\documentclass[12pt,a4paper,twoside]{article}
\usepackage{makeidx}
\usepackage{latexsym,amsmath,amssymb,amsfonts,amsthm}

\newtheorem{proposition}{Proposition}[section]

\newtheorem{corollary}{Corollary}[section]
\newtheorem{theorem}{Theorem}[section]
\newtheorem{remark}{Remark}[section]
\def\nabla{\bigtriangledown}

\makeindex \textheight 23.0cm
\begin{document}

\title{ A Method of Constructing Off--Diagonal \\
Solutions in Metric--Affine and String Gravity }
\author{Sergiu I. Vacaru\thanks{%
e--mail:\ vacaru@fisica.ist.utl.pt, sergiu$_{-}$vacaru@yahoo.com}\ ,
Evghenii Gaburov \thanks{
e--mail:\ eg35@leicester.ac.uk }\quad and Denis Gontsa \thanks{%
e--mail:\ d$_{-}$gontsa@yahoo.com } \quad \\
%EndAName
\\
{\small Centro Multidisciplinar de Astrofisica - CENTRA, Departamento de
Fisica,}\\
{\small Instituto Superior Tecnico, Av. Rovisco Pais 1, Lisboa, 1049-001,
Portugal,}\\
{\small {---}} \\
{\small $\dag$ \ Department of Physics and Astronomy, University}\\
{\small of Leicester, University Road, Leicester, LE1 7RH, UK} \\
{\small {---}} \\
{\small $\ddag$ \ Department of Physics, St. Petersbourg State University }\\
{\small P. O. Box 122, Petergoff, St. Petersbourg, 198904, Russia}}
\date{October 14, 2004 }
\maketitle

\begin{abstract}
The anholonomic frame method is generalized for non--Riemannian gravity
models defined by string corrections to the general relativity and
metric--affine gravity (MAG) theories. Such spacetime configurations are
modeled as metric--affine spaces provided with generic off--diagonal metrics
(which can not be diagonalized by coordinate transforms) and anholonomic
frames with associated nonlinear connection (N--connection) structure. We
investigate the field equations of MAG and string gravity with mixed
holonomic and anholonomic variables. There are proved the main theorems on
irreducible reduction to effective Einstein--Proca equations with respect to
anholonomic frames adapted to N--connections. String corrections induced by
the antisymmetric $H$--fields are considered. There are also proved the
theorems and criteria stating a new method of constructing exact solutions
with generic off--diagonal metric ansatz depending on 3-5 variables and
describing various type of locally anisotropic gravitational configurations
with torsion, nonmetricity and/or generalized Finsler--affine effective
geometry. We analyze solutions, generated in string gravity, when
generalized Finsler--affine metrics, torsion and nonmetricity interact with
three dimensional solitons.

\vskip5pt

Pacs:\ 04.50.+h, 04.20.JB, 02.40.-k,

MSC numbers: 83D05, 83C15, 83E15, 53B40, 53C07, 53C60
\end{abstract}

\tableofcontents

\vskip2.0cm

\section{Introduction}

Nowadays, there exists an interest to non--Riemannian descriptions of
gravity interactions derived in the low energy string theory \cite{sgr}
and/or certain noncommutative \cite{ncg} and quantum group generalizations %
\cite{majid} of gravity and field theory. Such effective models can be
expressed in terms of geometries with torsion and nonmetricity in the
framework of metric--affine gravity (MAG) \cite{mag} and a subclass of such
theories can be expressed as an effective Einstein--Proca gravity derived
via irreducible decompositions \cite{oveh}.

In a recent work \cite{vp1} we developed a unified scheme to the geometry of
anholonomic frames with associated nonlinear connection (N--connection)
structure for a large number of gauge and gravity models with locally
isotropic and anisotropic interactions and nontrivial torsion and
nonmetricity contributions and effective generalized
Finsler--Weyl--Riemann--Cartan geometries derived from MAG. The synthesis of
metric--affine and Finsler like theories was inspired by a number of exact
solutions parametrized by generic off--diagonal metrics and anholonomic
frames in Einstein, Einstein--Cartan, gauge and string gravity \cite{v1,v1a}%
. The resulting formalism admits inclusion of locally anisotropic spinor
interactions and extensions to noncommutative geometry and string/brane
gravity \cite{v2,vnces}. We concluded that the geometry of metric--affine
spaces enabled with an additional N--connection structure is sufficient not
only to model the bulk of physically important non--Riemannian geometries on
(pseudo) Riemannian spaces but also states the conditions when such
effective spaces with generic anisotropy can be defined as certain
generalized Finsler--affine geometric configurations constructed as exact
solutions of field equations. It was elaborated a detailed classification of
such spacetimes provided with N--connection structure.

If in the Ref. \cite{vp1} we paid attention to the geometrical
(pre--dynamical) aspects of the generalized Finsler--affine configurations
derived in MAG, the aim of this paper (the second partner) is to formulate a
variatonal formalism of deriving field equations on metric--affine spaces
provided with N--connection structure and to state the main theorems for
constructing exact off--diagonal solutions in such generalized
non--Riemannian gravity theories. We emphasize that generalized Finsler
metrics can be generated in string gravity connected to anholonomic
metric--affine configurations. In particular, we investigate how the
so--called Obukhov's equivalence theorem \cite{oveh} should be modified as
to include various type of Finsler--Lagrange--Hamilton--Cartan metrics, see
Refs. \cite{fin,ma,mhss}. The results of this paper consist a theoretical
background for constructing exact solutions in MAG and string gravity in the
third partner paper \cite{exsolmag} derived as exact solutions of
gravitational and matter field equations parametrized by generic
off--diagonal metrics (which can not be diagonalized by local coordinate
transforms) and anholonomic frames with associated N--connection structure.
Such solutions depending on 3-5 variables (generalizing to MAG the results
from \cite{v1,v1a,v2,vnces,vmethod}) differ substantially from those
elaborated in Refs. \cite{esolmag}; they define certain extensions to
nontrivial torsion and nonmetricity fields of certain generic off--diagonal
metrics in general relativity theory.

The plan of the paper is as follows: In Sec. 2 we outline the necessary
results on Finsler--affine geometry. Next, in Sec. 3, we formulate the field
equations on metric--affine spaces provided with N--connection structure. We
consider Lagrangians and derive geometrically the field equations of
Finsler--affine gravity. We prove the main theorems for the Einstein--Proca
systems distinguished by N--connection structure and analyze possible string
gravity corrections by $H$--fields from the bosonic string theory. There are
defined the restrictions on N--connection structures resulting in
Einstein--Cartan and Einstein gravity. Section 4 is devoted to extension of
the anholonomic frame method in MAG and string gravity. We formulate and
prove the main theorems stating the possibility of constructing exact
solutions parametrized by generic off--diagonal metrics, nontrivial torsion
and nonmetricity structures and possible sources of matter fields. In Sec. 5
we construct three classes of exact solutions. The first class of solutions
is stated for five subclasses of two dimensional generalized Finsler
geometries modeled in MAG with possible string corrections. The second class
of solutions is for MAG with effective variable and inhomogeneous
cosmological constant. The third class of solutions are for the string
Finsler--affine gravity (i. e. string gravity containing in certain limits
Finsler like metrics) with possible nonlinear three dimensional solitonic
interactions, Proca fields with almost vanishing masses, nontrivial torsions
and nonmetricity. In Sec. 6 we present the final remarks. In Appendices A, B
and C we give respectively the details on the proof of the Theorem 4.1
(stating the components of the Ricci tensor for generalized Finsler--affine
spaces), analyze the reduction of nonlinear solutions from five to four
dimensions and present a short characterization of five classes of
generalized Finsler--affine spaces.

Our basic notations and conventions are those from Ref. \cite{vp1} and
contain an interference of approaches elaborated in MAG and generalized
Finsler geometry. The spacetime is considered to be a manifold $V^{n+m}$ of
necessary smoothly class of dimension $n+m.$ The Greeck indices $\alpha
,\beta ,...$ split into subclasses like $\alpha =\left( i,a\right) ,$ $\beta
=\left( j,b\right) ...$ where the Latin indices $i,j,k,...$ run values $%
1,2,...n$ and $\ a,b,c,...$ run values $n+1,n+2,$ ..., $n+m.$ We follow the
Penrose convention on abstract indices \cite{pen} and use underlined indices
like $\underline{\alpha }=\left( \underline{i},\underline{a}\right) ,$ for
decompositions with respect to coordinate frames. The symbol ''\ $\doteqdot $%
'' will be used is some formulas will be introduced by definition and the
end of proofs will be stated by symbol $\blacksquare .$ The notations for
connections $\Gamma _{\ \beta \gamma }^{\alpha },$ metrics $g_{\alpha \beta
} $ and frames $e_{\alpha }$ and coframes $\vartheta ^{\beta },$ or another
geometrical and physical objects, are the standard ones from MAG if a
nonlinear connection (N--connection) structure is not emphasized on the
spacetime. If a N--connection and corresponding anholonomic frame structure
are prescribed, we use ''boldfaced'' symbols with possible splitting of the
objects and indices like $\mathbf{V}^{n+m},$ $\mathbf{\Gamma }_{\ \beta
\gamma }^{\alpha }=\left( L_{jk}^{i},L_{bk}^{a},C_{jc}^{i},C_{bc}^{a}\right)
,$\textbf{\ }$\mathbf{g}_{\alpha \beta }=\left( g_{ij},h_{ab}\right) ,$ $%
\mathbf{e}_{\alpha }=\left( e_{i},e_{a}\right) ,$ ...being distinguished by
N--connection (in brief, there are used the terms d--objects, d--tensor,
d--connection).

\section{Metric--Affine and Generalized Finsler Gravity}

In this section we recall some basic facts on metric--affine spaces provided
with nonlinear connection (N--connection) structure and generalized
Finsler--affine geometry \cite{vp1}.

The spacetime is modeled as a manifold $V^{n+m}$ of dimension $n+m,$ with $%
n\geq 2$ and $m\geq 1,$ admitting (co) vector/ tangent structures. It is
denoted by $\pi ^{T}:TV^{n+m}\rightarrow TV^{n}$ $\ $the differential of the
map $\pi :V^{n+m}\rightarrow V^{n}$ defined as a fiber--preserving morphism
of the tangent bundle $\left( TV^{n+m},\tau _{E},V^{n}\right) $ to $V^{n+m}$
and of tangent bundle $\left( TV^{n},\tau ,V^{n}\right) .$ We consider also
the kernel of the morphism $\pi ^{T}$ as a vector subbundle of the vector
bundle $\left( TV^{n+m},\tau _{E},V^{n+m}\right) .$ The kernel defines the
vertical subbundle over $V^{n+m},$ s denoted as $\left( vV^{n+m},\tau
_{V},V^{n+m}\right) .$ We parametrize the local coordinates of a point $u\in
V^{n+m}$ as $u^{\alpha }=\left( x^{i},y^{a}\right) ,$ where the values of
indices are $i,j,k,...=1,2,...,n$ and $a,b,c,...=n+1,n+2,...,n+m.$ The
inclusion mapping is written as $i:vV^{n+m}\rightarrow TV^{n+m}.$

A nonlinear connection (N--connection) $\mathbf{N}$ in a space $\left(
V^{n+m},\pi ,V^{n}\right) $ is a morphism of manifolds $N:TV^{n+m}%
\rightarrow vV^{n+m}$ defined by the splitting on the left of the exact
sequence
\begin{equation}
0\rightarrow vV^{n+m}\rightarrow TV^{n+m}/vV^{n+m}\rightarrow 0.
\label{eseq}
\end{equation}

The kernel of the morphism $\mathbf{N}$ \ is a subbundle of $\left(
TV^{n+m},\tau _{E},V^{n+m}\right) ,$ called the horizontal subspace and
denoted by $\left( hV^{n+m},\tau _{H},V^{n+m}\right) .$ Every tangent bundle
$(TV^{n+m},$ $\tau _{E},$ $V^{n+m})$ provided with a N--connection structure
is a Whitney sum of the vertical and horizontal subspaces (in brief, h- and
v-- subspaces), i. e.
\begin{equation}
TV^{n+m}=hV^{n+m}\oplus vV^{n+m}.  \label{wihit}
\end{equation}%
We note that the exact sequence (\ref{eseq}) defines the N--connection in a
global coordinate free form\ resulting in invariant splitting (\ref{wihit})
(see details in Refs. \cite{barthel,ma} stated for vector and tangent
bundles and generalizations on covector bundles, superspaces and
noncommutative spaces \cite{mhss} and \cite{v2}).

A N--connection structure prescribes a class of vielbein transforms%
\begin{eqnarray}
A_{\alpha }^{\ \underline{\alpha }}(u) &=&\mathbf{e}_{\alpha }^{\ \underline{%
\alpha }}=\left[
\begin{array}{cc}
e_{i}^{\ \underline{i}}(u) & N_{i}^{b}(u)e_{b}^{\ \underline{a}}(u) \\
0 & e_{a}^{\ \underline{a}}(u)%
\end{array}%
\right] ,  \label{vt1} \\
A_{\ \underline{\beta }}^{\beta }(u) &=&\mathbf{e}_{\ \underline{\beta }%
}^{\beta }=\left[
\begin{array}{cc}
e_{\ \underline{i}}^{i\ }(u) & -N_{k}^{b}(u)e_{\ \underline{i}}^{k\ }(u) \\
0 & e_{\ \underline{a}}^{a\ }(u)%
\end{array}%
\right] ,  \label{vt2}
\end{eqnarray}%
in particular case $e_{i}^{\ \underline{i}}=\delta _{i}^{\underline{i}}$ and
$e_{a}^{\ \underline{a}}=\delta _{a}^{\underline{a}}$ with $\delta _{i}^{%
\underline{i}}$ and $\delta _{a}^{\underline{a}}$ being the Kronecker
symbols, defining a global splitting of $\mathbf{V}^{n+m}$ into
''horizontal'' and ''vertical'' subspaces with the N--vielbein structure%
\begin{equation*}
\mathbf{e}_{\alpha }=\mathbf{e}_{\alpha }^{\ \underline{\alpha }}\partial _{%
\underline{\alpha }}\mbox{ and }\mathbf{\vartheta }_{\ }^{\beta }=\mathbf{e}%
_{\ \underline{\beta }}^{\beta }du^{\underline{\beta }}.
\end{equation*}%
We adopt the convention that for the spaces provided with N--connection
structure the geometrical objects can be denoted by ''boldfaced'' symbols if
it would be necessary to distinguish such objects from similar ones for
spaces without N--connection.

A N--connection $\mathbf{N}$ in a space $\mathbf{V}^{n+m}$ is parametrized
by its components $N_{i}^{a}(u)=N_{i}^{a}(x,y),$
\begin{equation*}
\mathbf{N}=N_{i}^{a}(u)d^{i}\otimes \partial _{a}
\end{equation*}%
and characterized by the N--connection curvature
\begin{equation*}
\mathbf{\Omega }=\frac{1}{2}\Omega _{ij}^{a}d^{i}\wedge d^{j}\otimes
\partial _{a},
\end{equation*}%
with N--connection curvature coefficients%
\begin{equation}
\Omega _{ij}^{a}=\delta _{\lbrack j}N_{i]}^{a}=\frac{\partial N_{i}^{a}}{%
\partial x^{j}}-\frac{\partial N_{j}^{a}}{\partial x^{i}}+N_{i}^{b}\frac{%
\partial N_{j}^{a}}{\partial y^{b}}-N_{j}^{b}\frac{\partial N_{i}^{a}}{%
\partial y^{b}}.  \label{ncurv}
\end{equation}%
On spaces provided with N--connection structure, we have to use
'N--elongated' operators like $\delta _{j}$ in (\ref{ncurv}) instead of
usual partial derivatives. They are defined by the vielbein configuration
induced by the N--connection, the N--elongated partial derivatives (in
brief, N--derivatives)
\begin{equation}
\mathbf{e}_{\alpha }\doteqdot \delta _{\alpha }=\left( \delta _{i},\partial
_{a}\right) \equiv \frac{\delta }{\delta u^{\alpha }}=\left( \frac{\delta }{%
\delta x^{i}}=\partial _{i}-N_{i}^{a}\left( u\right) \partial _{a},\frac{%
\partial }{\partial y^{a}}\right)  \label{dder}
\end{equation}%
and the N--elongated differentials (in brief, N--differentials)
\begin{equation}
\mathbf{\vartheta }_{\ }^{\beta }\doteqdot \delta \ ^{\beta }=\left(
d^{i},\delta ^{a}\right) \equiv \delta u^{\alpha }=\left( \delta
x^{i}=dx^{i},\delta y^{a}=dy^{a}+N_{i}^{a}\left( u\right) dx^{i}\right)
\label{ddif}
\end{equation}%
called also, respectively, the N--frame and N--coframe. There are used both
type of denotations $\mathbf{e}_{\alpha }\doteqdot \delta _{\alpha }$ and $%
\mathbf{\vartheta }_{\ }^{\beta }\doteqdot \delta \ ^{\alpha }$ in order to
preserve a connection to denotations from Refs. \cite{ma,v1,v1a,v2}. The
'boldfaced' symbols $\mathbf{e}_{\alpha }$ and $\mathbf{\vartheta }_{\
}^{\beta }$ will be considered in order to emphasize that they define
N--adapted vielbeins but the symbols $\delta _{\alpha }$ and $\delta \
^{\beta }$ will be used for the N--elongated partial derivatives and,
respectively, differentials.

The N--coframe (\ref{ddif}) satisfies the anholonomy relations
\begin{equation}
\left[ \delta _{\alpha },\delta _{\beta }\right] =\delta _{\alpha }\delta
_{\beta }-\delta _{\beta }\delta _{\alpha }=\mathbf{w}_{\ \alpha \beta
}^{\gamma }\left( u\right) \delta _{\gamma }  \label{anhr}
\end{equation}%
with nontrivial anholonomy coefficients $\mathbf{w}_{\beta \gamma }^{\alpha
}\left( u\right) $ computed as
\begin{equation}
\mathbf{w}_{~ji}^{a}=-\mathbf{w}_{~ij}^{a}=\Omega _{ij}^{a},\ \mathbf{w}%
_{~ia}^{b}=-\mathbf{w}_{~ai}^{b}=\partial _{a}N_{i}^{b}.  \label{anhc}
\end{equation}

The distinguished objects (by a N--connection on a spaces $\mathbf{V}^{n+m})$
are introduced in a coordinate free form as geometric objects adapted to the
splitting (\ref{wihit}). In brief, they are called d--objects, d--tensor,
d--connections, d--metrics....

There is an important class of linear connections adapted to the
N--connection structure:

A d--connection $\mathbf{D}$ on a space $\mathbf{V}^{n+m}$ is defined as a
linear connection $D$ conserving under a parallelism the global
decomposition (\ref{wihit}).

The N--adapted components $\mathbf{\Gamma }_{\beta \gamma }^{\alpha }$ of a
d-connection $\mathbf{D}_{\alpha }=(\delta _{\alpha }\rfloor \mathbf{D})$
are defined by the equations%
\begin{equation*}
\mathbf{D}_{\alpha }\delta _{\beta }=\mathbf{\Gamma }_{\ \alpha \beta
}^{\gamma }\delta _{\gamma },
\end{equation*}%
from which one immediately follows
\begin{equation}
\mathbf{\Gamma }_{\ \alpha \beta }^{\gamma }\left( u\right) =\left( \mathbf{D%
}_{\alpha }\delta _{\beta }\right) \rfloor \delta ^{\gamma }.  \label{dcon1}
\end{equation}%
The operations of h- and v-covariant derivations, $D_{k}^{[h]}=%
\{L_{jk}^{i},L_{bk\;}^{a}\}$ and $D_{c}^{[v]}=\{C_{jk}^{i},C_{bc}^{a}\}$ are
introduced as corresponding h- and v--parametrizations of (\ref{dcon1}),%
\begin{equation*}
L_{jk}^{i}=\left( \mathbf{D}_{k}\delta _{j}\right) \rfloor d^{i},\quad
L_{bk}^{a}=\left( \mathbf{D}_{k}\partial _{b}\right) \rfloor \delta
^{a},~C_{jc}^{i}=\left( \mathbf{D}_{c}\delta _{j}\right) \rfloor d^{i},\quad
C_{bc}^{a}=\left( \mathbf{D}_{c}\partial _{b}\right) \rfloor \delta ^{a}.
\end{equation*}%
The components $\mathbf{\Gamma }_{\ \alpha \beta }^{\gamma }=\left(
L_{jk}^{i},L_{bk}^{a},C_{jc}^{i},C_{bc}^{a}\right) $ completely define a
d--connection $\mathbf{D}$ in $\mathbf{V}^{n+m}.$

A metric structure $\mathbf{g}$ on a space $\mathbf{V}^{n+m}$ is defined as
a symmetric covariant tensor field of type $\left( 0,2\right) ,$ $g_{\alpha
\beta ,}$ being nondegenerate and of constant signature on $\mathbf{V}%
^{n+m}. $ A N--connection $\mathbf{N=}\{N_{\underline{i}}^{\underline{b}%
}\left( u\right) \}$ and a metric structure $\mathbf{g}=g_{\underline{\alpha
}\underline{\beta }}du^{\underline{\alpha }}\otimes du^{\underline{\beta }}$
on $\mathbf{V}^{n+m}$ are mutually compatible if there are satisfied the
conditions
\begin{equation*}
\mathbf{g}\left( \delta _{\underline{i}},\partial _{\underline{a}}\right) =0,%
\mbox{ or equivalently,
}g_{\underline{i}\underline{a}}\left( u\right) -N_{\underline{i}}^{%
\underline{b}}\left( u\right) h_{\underline{a}\underline{b}}\left( u\right)
=0,
\end{equation*}%
where $h_{\underline{a}\underline{b}}\doteqdot \mathbf{g}\left( \partial _{%
\underline{a}},\partial _{\underline{b}}\right) $ and $g_{\underline{i}%
\underline{a}}\doteqdot \mathbf{g}\left( \partial _{\underline{i}},\partial
_{\underline{a}}\right) \,$ resulting in
\begin{equation*}
N_{i}^{b}\left( u\right) =h^{ab}\left( u\right) g_{ia}\left( u\right)
\end{equation*}%
(the matrix $h^{ab}$ is inverse to $h_{ab};$ for simplicity, we do not
underline the indices in the last formula). In consequence, we define an
invariant h--v--decomposition of metric (in brief, a d--metric)%
\begin{equation*}
\mathbf{g}(X,Y)\mathbf{=}h\mathbf{g}(X,Y)+v\mathbf{g}(X,Y).
\end{equation*}%
With respect to a N--coframe (\ref{ddif}), the d--metric is written
\begin{equation}
\mathbf{g}=\mathbf{g}_{\alpha \beta }\left( u\right) \delta ^{\alpha
}\otimes \delta ^{\beta }=g_{ij}\left( u\right) d^{i}\otimes
d^{j}+h_{ab}\left( u\right) \delta ^{a}\otimes \delta ^{b},  \label{block2}
\end{equation}%
where $g_{ij}\doteqdot \mathbf{g}\left( \delta _{i},\delta _{j}\right) .$
The d--metric (\ref{block2}) can be equivalently written in
''off--diagonal'' with respect to a coordinate basis defined by usual local
differentials $du^{\alpha }=\left( dx^{i},dy^{a}\right) ,$
\begin{equation}
\underline{g}_{\alpha \beta }=\left[
\begin{array}{cc}
g_{ij}+N_{i}^{a}N_{j}^{b}h_{ab} & N_{j}^{e}h_{ae} \\
N_{i}^{e}h_{be} & h_{ab}%
\end{array}%
\right] .  \label{ansatz}
\end{equation}

A metric, for instance, parametrized in the form (\ref{ansatz})\ is generic
off--diagonal if it can not be diagonalized by any coordinate transforms.
The anholonomy coefficients (\ref{anhc}) do not vanish for the off--diagonal
form (\ref{ansatz}) and the equivalent d--metric (\ref{block2}).

The nonmetricity d--field
\begin{equation*}
\ \mathcal{Q}=\mathbf{Q}_{\alpha \beta }\mathbf{\vartheta }^{\alpha }\otimes
\mathbf{\vartheta }^{\beta }=\mathbf{Q}_{\alpha \beta }\delta \ ^{\alpha
}\otimes \delta ^{\beta }
\end{equation*}%
on a space $\mathbf{V}^{n+m}$ provided with N--connection structure is
defined by a d--tensor field with the coefficients
\begin{equation}
\mathbf{Q}_{\alpha \beta }\doteqdot -\mathbf{Dg}_{\alpha \beta }  \label{nmf}
\end{equation}%
where the covariant derivative $\mathbf{D}$ is for a d--connection (\ref%
{dcon1}) $\mathbf{\Gamma }_{\ \alpha }^{\gamma }=\mathbf{\Gamma }_{\ \alpha
\beta }^{\gamma }\mathbf{\vartheta }^{\beta }$ with $\mathbf{\Gamma }%
_{\alpha \beta }^{\gamma }=\left(
L_{jk}^{i},L_{bk}^{a},C_{jc}^{i},C_{bc}^{a}\right) .$

A linear connection $D_{X}$ is compatible\textbf{\ } with a d--metric $%
\mathbf{g}$ if%
\begin{equation}
D_{X}\mathbf{g}=0,  \label{mc}
\end{equation}%
i. e. if $Q_{\alpha \beta }\equiv 0.$ In a space provided with N--connection
structure, the metricity condition (\ref{mc}) may split into a set of
compatibility conditions on h- and v-- subspaces,
\begin{equation}
D^{[h]}(h\mathbf{g)}=0,D^{[v]}(h\mathbf{g)}=0,D^{[h]}(v\mathbf{g)}%
=0,D^{[v]}(v\mathbf{g)}=0.  \label{mca}
\end{equation}
For instance, if $D^{[v]}(h\mathbf{g)}=0$ and $D^{[h]}(v\mathbf{g)}=0,$ but,
in general, $D^{[h]}(h\mathbf{g)}\neq 0$ and $D^{[v]}(v\mathbf{g)}\neq 0$ we
have a nontrivial nonmetricity d--field $\mathbf{Q}_{\alpha \beta }=\mathbf{Q%
}_{\gamma \alpha \beta }\vartheta ^{\gamma }$ with irreducible
h--v--components $\mathbf{Q}_{\gamma \alpha \beta }=\left(
Q_{ijk},Q_{abc}\right) .$

In a metric--affine space, by acting on forms with a covariant derivative $%
D, $ we can also define another very important geometric objects (the
'gravitational field potentials', the torsion and, respectively, curvature;
see \cite{mag}):%
\begin{equation}
\ \mathbf{T}^{\alpha }\doteqdot \mathbf{D\vartheta }^{\alpha }=\delta
\mathbf{\vartheta }^{\alpha }+\mathbf{\Gamma }_{\ \beta }^{\gamma }\wedge
\mathbf{\vartheta }^{\beta }  \label{dt}
\end{equation}%
and
\begin{equation}
\ \mathbf{R}_{\ \beta }^{\alpha }\doteqdot \mathbf{D\Gamma }_{\ \beta
}^{\alpha }=\delta \mathbf{\Gamma }_{\ \beta }^{\alpha }-\mathbf{\Gamma }_{\
\beta }^{\gamma }\wedge \mathbf{\Gamma }_{\ \ \gamma }^{\alpha }  \label{dc}
\end{equation}%
For spaces provided with N--connection structures, we consider the same
formulas but for ''boldfaced'' symbols and change the usual differential $d$
$\ $into N-adapted operator $\delta .$

A general affine (linear) connection $D=\bigtriangledown +Z=\{\Gamma _{\beta
\alpha }^{\gamma }=\Gamma _{\bigtriangledown \beta \alpha }^{\gamma
}+Z_{\beta \alpha }^{\gamma }\}$
\begin{equation}
\Gamma _{\ \alpha }^{\gamma }=\Gamma _{\alpha \beta }^{\gamma }\vartheta
^{\beta },  \label{ac}
\end{equation}%
can always be decomposed into the Riemannian $\Gamma _{\bigtriangledown \
\beta }^{\alpha }$ and post--Riemannian $Z_{\ \beta }^{\alpha }$ parts \cite%
{mag,oveh},
\begin{equation}
\Gamma _{\ \beta }^{\alpha }=\Gamma _{\bigtriangledown \ \beta }^{\alpha
}+Z_{\ \beta }^{\alpha }.  \label{acc}
\end{equation}%
The distorsion 1-form $Z_{\ \beta }^{\alpha }$ from (\ref{acc}) is expressed
in terms of torsion and nonmetricity,%
\begin{equation}
Z_{\alpha \beta }=e_{\beta }\rfloor T_{\alpha }-e_{\alpha }\rfloor T_{\beta
}+\frac{1}{2}\left( e_{\alpha }\rfloor e_{\beta }\rfloor T_{\gamma }\right)
\vartheta ^{\gamma }+\left( e_{\alpha }\rfloor Q_{\beta \gamma }\right)
\vartheta ^{\gamma }-\left( e_{\beta }\rfloor Q_{\alpha \gamma }\right)
\vartheta ^{\gamma }+\frac{1}{2}Q_{\alpha \beta }  \label{dista}
\end{equation}%
where $T_{\alpha }$ is defined as (\ref{dt}) and $Q_{\alpha \beta }\doteqdot
-Dg_{\alpha \beta }.$ (We note that $Z_{\ \beta }^{\alpha }$ are $N_{\alpha
\beta }$ from Ref. \cite{oveh}, but in our works we use the symbol $N$ for
N--connections .) \ For $Q_{\beta \gamma }=0,$ we obtain from (\ref{dista})
just the distorsion for the Riemannian--Cartan geometry \cite{rcg}.

By substituting arbitrary (co) frames, metrics and linear connections into
N--adapted ones,
\begin{equation*}
e_{\alpha }\rightarrow \mathbf{e}_{\alpha },\vartheta ^{\beta }\rightarrow
\mathbf{\vartheta }^{\beta },g_{\alpha \beta }\rightarrow \mathbf{g}_{\alpha
\beta }=\left( g_{ij},h_{ab}\right) ,\Gamma _{\ \alpha }^{\gamma
}\rightarrow \mathbf{\Gamma }_{\ \alpha }^{\gamma },
\end{equation*}%
with $\mathbf{Q}_{\alpha \beta }=\mathbf{Q}_{\gamma \alpha \beta }\mathbf{%
\vartheta }^{\gamma }$ and $\mathbf{T}^{\alpha }$ as in (\ref{dt}), into
respective formulas (\ref{ac}), (\ref{acc}) and (\ref{dista}), \ we can
define an affine connection $\mathbf{D=\bigtriangledown +Z}=[\mathbf{\Gamma }%
_{\ \beta \alpha }^{\gamma }]$ with respect to N--adapted (co) frames,
\begin{equation}
\mathbf{\Gamma }_{\ \ \alpha }^{\gamma }=\mathbf{\Gamma }_{\ \alpha \beta
}^{\gamma }\mathbf{\vartheta }^{\beta },  \label{acn}
\end{equation}%
with
\begin{equation}
\mathbf{\Gamma }_{\ \beta }^{\alpha }=\mathbf{\Gamma }_{\bigtriangledown \
\beta }^{\alpha }+\mathbf{Z}_{\ \ \beta }^{\alpha },  \label{accn}
\end{equation}%
where
\begin{equation}
\mathbf{\Gamma }_{\ \gamma \alpha }^{\bigtriangledown }=\frac{1}{2}\left[
\mathbf{e}_{\gamma }\rfloor \ \delta \mathbf{\vartheta }_{\alpha }-\mathbf{e}%
_{\alpha }\rfloor \ \delta \mathbf{\vartheta }_{\gamma }-\left( \mathbf{e}%
_{\gamma }\rfloor \ \mathbf{e}_{\alpha }\rfloor \ \delta \mathbf{\vartheta }%
_{\beta }\right) \wedge \mathbf{\vartheta }^{\beta }\right] ,
\label{christa}
\end{equation}%
and
\begin{equation}
\mathbf{Z}_{\alpha \beta }=\mathbf{e}_{\beta }\rfloor \mathbf{T}_{\alpha }-%
\mathbf{e}_{\alpha }\rfloor \mathbf{T}_{\beta }+\frac{1}{2}\left( \mathbf{e}%
_{\alpha }\rfloor \mathbf{e}_{\beta }\rfloor \mathbf{T}_{\gamma }\right)
\mathbf{\vartheta }^{\gamma }+\left( \mathbf{e}_{\alpha }\rfloor \mathbf{Q}%
_{\beta \gamma }\right) \mathbf{\vartheta }^{\gamma }-\left( \mathbf{e}%
_{\beta }\rfloor \mathbf{Q}_{\alpha \gamma }\right) \mathbf{\vartheta }%
^{\gamma }+\frac{1}{2}\mathbf{Q}_{\alpha \beta }.  \label{distan}
\end{equation}%
The h-- and v--components of $\mathbf{\Gamma }_{\ \beta }^{\alpha }$ from (%
\ref{accn}) consists from the components of $\mathbf{\Gamma }%
_{\bigtriangledown \ \beta }^{\alpha }$ (considered for (\ref{christa})) and
of $\mathbf{Z}_{\alpha \beta }$ with $\mathbf{Z}_{\ \ \gamma \beta }^{\alpha
}=\left( Z_{jk}^{i},Z_{bk}^{a},Z_{jc}^{i},Z_{bc}^{a}\right) .$ We note that
for $\mathbf{Q}_{\alpha \beta }=0,$ the distorsion 1--form $\mathbf{Z}%
_{\alpha \beta }$ defines a Riemann--Cartan geometry adapted to the
N--connection structure.

A \ distinguished metric--affine space $\mathbf{V}^{n+m}$ is defined as a
usual metric--affine space additionally enabled with a N--connection
structure $\mathbf{N}=\{N_{i}^{a}\}$ inducing splitting into respective
irreducible horizontal and vertical subspaces of dimensions $n$ and $m.$
This space is provided with independent d--metric (\ref{block2}) and affine
d--connection (\ref{dcon1}) structures adapted to the N--connection.\

If a space $\mathbf{V}^{n+m}$ is provided with both N--connection $\mathbf{N}
$\ and d--metric $\mathbf{g}$ structures, there is a unique linear symmetric
and torsionless connection $\mathbf{\bigtriangledown },$ called the
Levi--Civita connection, being metric compatible such that $\bigtriangledown
_{\gamma }\mathbf{g}_{\alpha \beta }=0$ $\ $for $\mathbf{g}_{\alpha \beta
}=\left( g_{ij},h_{ab}\right) ,$ see (\ref{block2}), with the coefficients
\begin{equation*}
\mathbf{\Gamma }_{\alpha \beta \gamma }^{\bigtriangledown }=\mathbf{g}\left(
\delta _{\alpha },\bigtriangledown _{\gamma }\delta _{\beta }\right) =%
\mathbf{g}_{\alpha \tau }\mathbf{\Gamma }_{\bigtriangledown \beta \gamma
}^{\tau },\,
\end{equation*}%
computed as
\begin{equation}
\mathbf{\Gamma }_{\alpha \beta \gamma }^{\bigtriangledown }=\frac{1}{2}\left[
\delta _{\beta }\mathbf{g}_{\alpha \gamma }+\delta _{\gamma }\mathbf{g}%
_{\beta \alpha }-\delta _{\alpha }\mathbf{g}_{\gamma \beta }+\mathbf{g}%
_{\alpha \tau }\mathbf{w}_{\gamma \beta }^{\tau }+\mathbf{g}_{\beta \tau }%
\mathbf{w}_{\alpha \gamma }^{\tau }-\mathbf{g}_{\gamma \tau }\mathbf{w}%
_{\beta \alpha }^{\tau }\right]  \label{lcsym}
\end{equation}%
with respect to N--frames $\mathbf{e}_{\beta }\doteqdot \delta _{\beta }$ (%
\ref{dder}) and N--coframes $\mathbf{\vartheta }_{\ }^{\alpha }\doteqdot
\delta ^{\alpha }$ (\ref{ddif}).

We note that the Levi--Civita connection is not adapted to the N--connection
structure.\ Se, we can not state its coefficients in an irreducible form for
the h-- and v--subspaces. There is a type of d--connections which are
similar to the Levi--Civita connection but satisfying certain metricity
conditions adapted to the N--connection. They are introduced as metric
d--connections $\mathbf{D=}\left( D^{[h]},D^{[v]}\right) $ in a space \ $%
\mathbf{V}^{n+m}$ satisfying the metricity conditions if and only if
\begin{equation}
D_{k}^{[h]}g_{ij}=0,\ D_{a}^{[v]}g_{ij}=0,\ D_{k}^{[h]}h_{ab}=0,\
D_{a}^{[h]}h_{ab}=0.  \label{mcas}
\end{equation}

Let us consider an important example: The canonical d--connection $\widehat{%
\mathbf{D}}$\ \ $\mathbf{=}\left( \widehat{D}^{[h]},\widehat{D}^{[v]}\right)
,$ equivalently $\widehat{\mathbf{\Gamma }}_{\ \alpha }^{\gamma }=\widehat{%
\mathbf{\Gamma }}_{\ \alpha \beta }^{\gamma }\mathbf{\vartheta }^{\beta },$\
is defined by the h-- v--irreducible components $\widehat{\mathbf{\Gamma }}%
_{\ \alpha \beta }^{\gamma }=(\widehat{L}_{jk}^{i},\widehat{L}_{bk}^{a},$ $%
\widehat{C}_{jc}^{i},\widehat{C}_{bc}^{a}),$ where%
\begin{eqnarray}
\widehat{L}_{jk}^{i} &=&\frac{1}{2}g^{ir}\left( \frac{\delta g_{jk}}{\delta
x^{k}}+\frac{\delta g_{kr}}{\delta x^{j}}-\frac{\delta g_{jk}}{\delta x^{r}}%
\right) ,  \label{candcon} \\
\widehat{L}_{bk}^{a} &=&\frac{\partial N_{k}^{a}}{\partial y^{b}}+\frac{1}{2}%
h^{ac}\left( \frac{\delta h_{bc}}{\delta x^{k}}-\frac{\partial N_{k}^{d}}{%
\partial y^{b}}h_{dc}-\frac{\partial N_{k}^{d}}{\partial y^{c}}h_{db}\right)
,  \notag \\
\widehat{C}_{jc}^{i} &=&\frac{1}{2}g^{ik}\frac{\partial g_{jk}}{\partial
y^{c}},  \notag \\
\widehat{C}_{bc}^{a} &=&\frac{1}{2}h^{ad}\left( \frac{\partial h_{bd}}{%
\partial y^{c}}+\frac{\partial h_{cd}}{\partial y^{b}}-\frac{\partial h_{bc}%
}{\partial y^{d}}\right) .  \notag
\end{eqnarray}%
satisfying the torsionless conditions for the h--subspace and v--subspace,
respectively, $\widehat{T}_{jk}^{i}=\widehat{T}_{bc}^{a}=0.$

The components of the Levi--Civita connection $\mathbf{\Gamma }%
_{\bigtriangledown \beta \gamma }^{\tau }$ and the irreducible components of
the canonical d--connection \ $\widehat{\mathbf{\Gamma }}_{\ \beta \gamma
}^{\tau }$\ are related by formulas%
\begin{equation}
\mathbf{\Gamma }_{\bigtriangledown \beta \gamma }^{\tau }=\left( \widehat{L}%
_{jk}^{i},\widehat{L}_{bk}^{a}-\frac{\partial N_{k}^{a}}{\partial y^{b}},%
\widehat{C}_{jc}^{i}+\frac{1}{2}g^{ik}\Omega _{jk}^{a}h_{ca},\widehat{C}%
_{bc}^{a}\right) ,  \label{lcsyma}
\end{equation}%
where $\Omega _{jk}^{a}$\ \ is the N--connection curvature\ (\ref{ncurv}).

We can define and calculate the irreducible components of torsion and
curvature in a space $\mathbf{V}^{n+m}$ provided with additional
N--connection structure (these could be any metric--affine spaces \cite{mag}%
, or their particular, like Riemann--Cartan \cite{rcg}, cases with vanishing
nonmetricity and/or torsion, or any (co) vector / tangent bundles like in
Finsler geometry and generalizations).

The torsion $\mathbf{T}_{.\beta \gamma }^{\alpha
}=(T_{.jk}^{i},T_{ja}^{i},T_{.ij}^{a},T_{.bi}^{a},T_{.bc}^{a})$ of a
d--connection $\mathbf{\Gamma }_{\alpha \beta }^{\gamma
}=(L_{jk}^{i},L_{bk}^{a},C_{jc}^{i},C_{bc}^{a})$ (\ref{dcon1}) has
irreducible h- v--components (d--torsions)
\begin{eqnarray}
T_{.jk}^{i} &=&-T_{kj}^{i}=L_{jk}^{i}-L_{kj}^{i},\quad
T_{ja}^{i}=-T_{aj}^{i}=C_{.ja}^{i},\ T_{.ji}^{a}=-T_{.ij}^{a}=\frac{\delta
N_{i}^{a}}{\delta x^{j}}-\frac{\delta N_{j}^{a}}{\delta x^{i}}=\Omega
_{.ji}^{a},  \notag \\
T_{.bi}^{a} &=&-T_{.ib}^{a}=P_{.bi}^{a}=\frac{\partial N_{i}^{a}}{\partial
y^{b}}-L_{.bj}^{a},\
T_{.bc}^{a}=-T_{.cb}^{a}=S_{.bc}^{a}=C_{bc}^{a}-C_{cb}^{a}.\   \label{dtorsb}
\end{eqnarray}

We note that on (pseudo) Riemanian spacetimes the d--torsions can be induced
by the N--connection coefficients and reflect an anholonomic frame
structure. Such objects vanish when we transfer our considerations with
respect to holonomic bases for a trivial N--connection and zero ''vertical''
dimension.

The curvature $\mathbf{R}_{.\beta \gamma \tau }^{\alpha }=(R_{\
hjk}^{i},R_{\ bjk}^{a},P_{\ jka}^{i},P_{\ bka}^{c},S_{\ jbc}^{i},S_{\
bcd}^{a})$ of a d--con\-nec\-ti\-on $\mathbf{\Gamma }_{\alpha \beta
}^{\gamma }=(L_{jk}^{i},L_{bk}^{a},C_{jc}^{i},C_{bc}^{a})$ (\ref{dcon1}) has
irreducible h- v--components (d--curvatures)
\begin{eqnarray}
R_{\ hjk}^{i} &=&\frac{\delta L_{.hj}^{i}}{\delta x^{k}}-\frac{\delta
L_{.hk}^{i}}{\delta x^{j}}%
+L_{.hj}^{m}L_{mk}^{i}-L_{.hk}^{m}L_{mj}^{i}-C_{.ha}^{i}\Omega _{.jk}^{a},
\label{dcurv} \\
R_{\ bjk}^{a} &=&\frac{\delta L_{.bj}^{a}}{\delta x^{k}}-\frac{\delta
L_{.bk}^{a}}{\delta x^{j}}%
+L_{.bj}^{c}L_{.ck}^{a}-L_{.bk}^{c}L_{.cj}^{a}-C_{.bc}^{a}\ \Omega
_{.jk}^{c},  \notag \\
P_{\ jka}^{i} &=&\frac{\partial L_{.jk}^{i}}{\partial y^{k}}-\left( \frac{%
\partial C_{.ja}^{i}}{\partial x^{k}}%
+L_{.lk}^{i}C_{.ja}^{l}-L_{.jk}^{l}C_{.la}^{i}-L_{.ak}^{c}C_{.jc}^{i}\right)
+C_{.jb}^{i}P_{.ka}^{b},  \notag \\
P_{\ bka}^{c} &=&\frac{\partial L_{.bk}^{c}}{\partial y^{a}}-\left( \frac{%
\partial C_{.ba}^{c}}{\partial x^{k}}+L_{.dk}^{c%
\,}C_{.ba}^{d}-L_{.bk}^{d}C_{.da}^{c}-L_{.ak}^{d}C_{.bd}^{c}\right)
+C_{.bd}^{c}P_{.ka}^{d},  \notag \\
S_{\ jbc}^{i} &=&\frac{\partial C_{.jb}^{i}}{\partial y^{c}}-\frac{\partial
C_{.jc}^{i}}{\partial y^{b}}+C_{.jb}^{h}C_{.hc}^{i}-C_{.jc}^{h}C_{hb}^{i},
\notag \\
S_{\ bcd}^{a} &=&\frac{\partial C_{.bc}^{a}}{\partial y^{d}}-\frac{\partial
C_{.bd}^{a}}{\partial y^{c}}+C_{.bc}^{e}C_{.ed}^{a}-C_{.bd}^{e}C_{.ec}^{a}.
\notag
\end{eqnarray}

The components of the Ricci tensor
\begin{equation*}
\mathbf{R}_{\alpha \beta }=\mathbf{R}_{\ \alpha \beta \tau }^{\tau }
\end{equation*}%
with respect to a locally adapted frame (\ref{dder}) has four irreducible h-
v--components $\mathbf{R}_{\alpha \beta }=(R_{ij},R_{ia},R_{ai},S_{ab}),$
where%
\begin{eqnarray}
R_{ij} &=&R_{\ ijk}^{k},\quad R_{ia}=-\ ^{2}P_{ia}=-P_{\ ika}^{k},
\label{dricci} \\
R_{ai} &=&\ ^{1}P_{ai}=P_{\ aib}^{b},\quad S_{ab}=S_{\ abc}^{c}.  \notag
\end{eqnarray}%
We point out that because, in general, $^{1}P_{ai}\neq ~^{2}P_{ia}$ the
Ricci d--tensor is non symmetric.

Having defined a d--metric of type (\ref{block2}) in $\mathbf{V}^{n+m},$ we
can introduce the scalar curvature of a d--connection $\mathbf{D,}$
\begin{equation}
{\overleftarrow{\mathbf{R}}}=\mathbf{g}^{\alpha \beta }\mathbf{R}_{\alpha
\beta }=R+S,  \label{dscal}
\end{equation}%
where $R=g^{ij}R_{ij}$ and $S=h^{ab}S_{ab}$ and define the distinguished
form of the Einstein tensor (the Einstein d--tensor),
\begin{equation}
\mathbf{G}_{\alpha \beta }\doteqdot \mathbf{R}_{\alpha \beta }-\frac{1}{2}%
\mathbf{g}_{\alpha \beta }{\overleftarrow{\mathbf{R}}.}  \label{deinst}
\end{equation}

The introduced geometrical objects are extremely useful in definition of
field equations of MAG and string gravity with nontrivial N--connection
structure.

\section{ N--Connections and Field Equations}

The field equations of metric--affine gravity (in brief, MAG) \cite{mag,oveh}
can be reformulated with respect to frames and coframes consisting from
mixed holonomic and anholonomic components defined by the N--connection
structure. In this case, various type of (pseudo)\ Riemannian,
Riemann--Cartan and generalized Finsler metrics and additional torsion and
nonmetricity sructures with very general local anisotropy can be embedded
into MAG. It is known that in a metric--affine spacetime the curvature,
torsion and nonmetricity have correspondingly eleven, three and four
irreducible pieces. If the N--connection is defined in a metric--affine
spacetime, every irreducible component of curvature splits additionally into
six h- and v-- components (\ref{dcurv}), every irreducible component of
torsion splits additionally into five h- and v-- components (\ref{dtorsb})
and every irreducible component of nonmetricity splits additionally into two
h- and v-- components (defined by splitting of metrics into block ansatz (%
\ref{block2})).

\subsection{Lagrangians and field equations for Finsler--affine theories}

For an arbitrary d--connection $\mathbf{\Gamma }_{\ \beta }^{\alpha }$ in a
metric--affine space $\mathbf{V}^{n+m}$ provided with N--connecti\-on
structure (for simplicity, we can take $n+m=4)$ one holds the respective
decompositi\-ons for d--torsion and nonmetricity d--field,%
\begin{eqnarray}
^{(2)}\mathbf{T}^{\alpha } &\doteqdot &\frac{1}{3}\mathbf{\vartheta }%
^{\alpha }\wedge \mathbf{T,\mbox{ for }T}\doteqdot \mathbf{e}_{\alpha
}\rfloor \mathbf{T}^{\alpha },  \label{torsdec} \\
^{(3)}\mathbf{T}^{\alpha } &\doteqdot &\frac{1}{3}\ \ast \left( \mathbf{%
\vartheta }^{\alpha }\wedge \mathbf{P}\right) \mathbf{,\mbox{ for }P}%
\doteqdot \ast \left( \mathbf{T}^{\alpha }\wedge \mathbf{\vartheta }_{\alpha
}\right) ,  \notag \\
^{(1)}\mathbf{T}^{\alpha } &\doteqdot &\mathbf{T}^{\alpha }-^{(2)}\mathbf{T}%
^{\alpha }-^{(3)}\mathbf{T}^{\alpha }  \notag
\end{eqnarray}%
and
\begin{eqnarray}
^{(2)}\mathbf{Q}_{\alpha \beta } &\doteqdot &\frac{1}{3}\ast \left( \mathbf{%
\vartheta }_{\alpha }\wedge \mathbf{S}_{\beta }+\mathbf{\vartheta }_{\beta
}\wedge \mathbf{S}_{\alpha }\right) \mathbf{,\ }^{(4)}\mathbf{Q}_{\alpha
\beta }\doteqdot \mathbf{g}_{\alpha \beta }\mathbf{Q,}  \notag \\
^{(3)}\mathbf{Q}_{\alpha \beta } &\doteqdot &\frac{2}{9}\left[ \left(
\mathbf{\vartheta }_{\alpha }\mathbf{e}_{\beta }+\mathbf{\vartheta }_{\beta }%
\mathbf{e}_{\alpha }\right) \rfloor \mathbf{\Lambda }-\frac{1}{2}\mathbf{g}%
_{\alpha \beta }\mathbf{\Lambda }\right] ,  \label{nmdc} \\
^{(1)}\mathbf{Q}_{\alpha \beta } &\doteqdot &\mathbf{Q}_{\alpha \beta }-\
^{(2)}\mathbf{Q}_{\alpha \beta }-\ ^{(3)}\mathbf{Q}_{\alpha \beta }-\ ^{(4)}%
\mathbf{Q}_{\alpha \beta },  \notag
\end{eqnarray}%
where
\begin{eqnarray*}
\mathbf{Q} &\mathbf{\doteqdot }&\frac{1}{4}\mathbf{g}^{\alpha \beta }\mathbf{%
Q}_{\alpha \beta },\ \mathbf{\Lambda }\doteqdot \mathbf{\vartheta }^{\alpha }%
\mathbf{e}^{\beta }\rfloor \left( \mathbf{Q}_{\alpha \beta }-\mathbf{Qg}%
_{\alpha \beta }\right) , \\
\ \mathbf{\Theta }_{\alpha } &\doteqdot &\ast \left[ \left( \mathbf{Q}%
_{\alpha \beta }-\mathbf{Qg}_{\alpha \beta }\right) \wedge \mathbf{\vartheta
}^{\beta }\right] , \\
\mathbf{S}_{\alpha } &\mathbf{\doteqdot }&\ \mathbf{\Theta }_{\alpha }-\frac{%
1}{3}\mathbf{e}_{\alpha }\rfloor \left( \mathbf{\vartheta }^{\beta }\wedge
\mathbf{\Theta }_{\beta }\right)
\end{eqnarray*}%
and the Hodge dual ''$\ast $'' is such that $\mathbf{\eta }\doteqdot \ast 1$
is the volume 4--form and
\begin{equation*}
\mathbf{\eta }_{\alpha }\doteqdot \mathbf{e}_{\alpha }\rfloor \mathbf{\eta
=\ast \vartheta }_{\alpha },\ \mathbf{\eta }_{\alpha \beta }\doteqdot
\mathbf{e}_{\alpha }\rfloor \mathbf{\eta }_{\beta }\mathbf{=\ast }\left(
\mathbf{\vartheta }_{\alpha }\wedge \mathbf{\vartheta }_{\beta }\right) ,\
\mathbf{\eta }_{\alpha \beta \gamma }\doteqdot \mathbf{e}_{\gamma }\rfloor
\mathbf{\eta }_{\alpha \beta },\ \mathbf{\eta }_{\alpha \beta \gamma \tau
}\doteqdot \mathbf{e}_{\tau }\rfloor \mathbf{\eta }_{\alpha \beta \gamma }
\end{equation*}%
with $\mathbf{\eta }_{\alpha \beta \gamma \tau }$ being totally
antisymmetric. In higher dimensions, we have to consider $\mathbf{\eta }%
\doteqdot \ast 1$ as the volume $\left( n+m\right) $--form. For N--adapted
h- and v--constructions, we have to consider couples of 'volume' forms $%
\mathbf{\eta }\doteqdot \left( \eta ^{\lbrack g]}=\ast ^{\lbrack g]}1,\eta
^{\lbrack h]}=\ast ^{\lbrack h]}1\right) $ defined correspondingly by $%
\mathbf{g}_{\alpha \beta }=\left( g_{ij},h_{ab}\right) .$

With respect to N--adapted (co) frames $\mathbf{e}_{\beta }=\left( \delta
_{i},\partial _{a}\right) $ (\ref{dder}) and $\mathbf{\vartheta }^{\alpha
}=\left( d^{i},\delta ^{a}\right) $ (\ref{ddif}), the irreducible
decompositions (\ref{torsdec}) split into h- and v--components $^{(A)}%
\mathbf{T}^{\alpha }=\left( ^{(A)}\mathbf{T}^{i},\ ^{(A)}\mathbf{T}%
^{a}\right) $ for every $A=1,2,3,4.$ Because, by definition, $\mathbf{Q}%
_{\alpha \beta }\doteqdot \mathbf{Dg}_{\alpha \beta }$ and $\mathbf{g}%
_{\alpha \beta }=\left( g_{ij},h_{ab}\right) $ is a d--metric field, we
conclude that in a similar form can be decomposed the nonmetricity, $\mathbf{%
Q}_{\alpha \beta }=\left( Q_{ij},Q_{ab}\right).$ The symmetrizations in
formulas (\ref{nmdc}) hide splittings for $^{(1)}\mathbf{Q}_{\alpha \beta
},^{(2)}\mathbf{Q}_{\alpha \beta }$ and $^{(3)}\mathbf{Q}_{\alpha \beta }.$
Nevertheless, the h-- and v-- decompositions can be derived separately on
h-- and v--subspaces by distinguishing the interior product $\rfloor =\left(
\rfloor ^{\lbrack h]},\rfloor ^{\lbrack v]}\right) $ as to have $\mathbf{%
\eta }_{\alpha }=\left( \mathbf{\eta }_{i}=\delta _{i}\rfloor \mathbf{\eta }%
,\ \mathbf{\eta }_{a}=\partial _{a}\rfloor \mathbf{\eta }\right) $...and all
formulas after decompositions with respect to N--adapted frames (co
resulting into a separate relations in h-- and v--subspaces, when $^{(A)}%
\mathbf{Q}_{\alpha \beta }=\left( \ ^{(A)}\mathbf{Q}_{ij},\ ^{(A)}\mathbf{Q}%
_{ab}\right) $ for every $A=1,2,3,4.$

A generalized Finsler--affine theory is described by a Lagrangian%
\begin{equation*}
\mathcal{L}=\mathcal{L}_{GFA}+\mathcal{L}_{mat},
\end{equation*}%
where $\mathcal{L}_{mat}$ represents the Lagrangian of matter fields and%
\begin{eqnarray}
\mathcal{L}_{GFA} &=&\frac{1}{2\kappa }[-a_{0[Rh]}\mathbf{R}^{ij}\wedge \eta
_{ij}-a_{0[Rv]}\mathbf{R}^{ab}\wedge \eta _{ab}-a_{0[Ph]}\mathbf{P}%
^{ij}\wedge \eta _{ij}-a_{0[Pv]}\mathbf{P}^{ab}\wedge \eta _{ab}  \notag \\
&&-a_{0[Sh]}\mathbf{S}^{ij}\wedge \eta _{ij}-a_{0[Sv]}\mathbf{S}^{ab}\wedge
\eta _{ab}-2\lambda _{\lbrack h]}\eta _{\lbrack h]}-2\lambda _{\lbrack
v]}\eta _{\lbrack v]}  \label{actgfa} \\
&&+\mathbf{T}^{i}\wedge \ast ^{\lbrack h]}\left(
\sum\limits_{[A]=1}^{3}a_{[hA]}\ ^{[A]}\mathbf{T}_{i}\right) +\mathbf{T}%
^{a}\wedge \ast ^{\lbrack v]}\left( \sum\limits_{[A]=1}^{3}a_{[vA]}\ ^{[A]}%
\mathbf{T}_{a}\right)  \notag \\
&&+2\left( \sum\limits_{[I]=2}^{4}c_{[hI]}\ ^{[I]}\mathbf{Q}_{ij}\right)
\wedge \mathbf{\vartheta }^{i}\wedge \ast ^{\lbrack h]}\ \mathbf{T}%
^{j}+2\left( \sum\limits_{[I]=2}^{4}c_{[vI]}\ ^{[I]}\mathbf{Q}_{ab}\right)
\wedge \mathbf{\vartheta }^{a}\wedge \ast ^{\lbrack v]}\mathbf{T}^{b}  \notag
\\
&&+\mathbf{Q}_{ij}\wedge \left( \sum\limits_{\lbrack I]=1}^{4}b_{[hI]}\
^{[I]}\mathbf{Q}^{ij}\right) +\mathbf{Q}_{ab}\wedge \left(
\sum\limits_{\lbrack I]=1}^{4}b_{[vI]}\ ^{[I]}\mathbf{Q}^{ab}\right)  \notag
\\
&&+b_{[h5]}\left( ^{[3]}\mathbf{Q}_{ij}\wedge \mathbf{\vartheta }^{i}\right)
\wedge \ast ^{\lbrack h]}\left( ^{[4]}\mathbf{Q}^{kj}\wedge \mathbf{%
\vartheta }_{k}\right) +b_{[v5]}\left( ^{[3]}\mathbf{Q}_{ij}\wedge \mathbf{%
\vartheta }^{i}\right) \wedge \ast ^{\lbrack v]}\left( ^{[4]}\mathbf{Q}%
^{kj}\wedge \mathbf{\vartheta }_{k}\right) ]  \notag
\end{eqnarray}%
\begin{eqnarray*}
&&-\frac{1}{2\rho _{\lbrack Rh]}}\mathbf{R}^{ij}\wedge \ast ^{\lbrack
h]}\{\sum\limits_{[I]=1}^{6}w_{[RhI]}\ (^{[I]}\mathbf{R}_{ij}-\ ^{[I]}%
\mathbf{R}_{ji})+w_{[Rh7]}\mathbf{\vartheta }_{i}\wedge \lbrack \mathbf{e}%
_{k}\rfloor ^{\lbrack h]}\ (\ ^{[5]}\mathbf{R}_{\ \ j}^{k}-\ ^{[5]}\mathbf{R}%
_{j\ }^{\ k})] \\
&&+\sum\limits_{[I]=1}^{5}z_{[RhI]}\ (^{[I]}\mathbf{R}_{ij}+\ ^{[I]}\mathbf{R%
}_{ji})+z_{[Rh6]}\mathbf{\vartheta }_{k}\wedge \lbrack \mathbf{e}_{i}\rfloor
^{\lbrack h]}\ (\ ^{[2]}\mathbf{R}_{\ \ j}^{k}-\ ^{[2]}\mathbf{R}_{j\ }^{\
k})] \\
&&+\sum\limits_{[I]=7}^{9}z_{[RhI]}\mathbf{\vartheta }_{i}\wedge \lbrack
\mathbf{e}_{k}\rfloor ^{\lbrack h]}\ (\ ^{[I-4]}\mathbf{R}_{\ \ j}^{k}-\
^{[I-4]}\mathbf{R}_{j\ }^{\ k})]\}
\end{eqnarray*}%
\begin{eqnarray*}
&&-\frac{1}{2\rho _{\lbrack Rv]}}\mathbf{R}^{ab}\wedge \ast ^{\lbrack
v]}\{\sum\limits_{[I]=1}^{6}w_{[RvI]}\ (^{[I]}\mathbf{R}_{ab}-\ ^{[I]}%
\mathbf{R}_{ba})+w_{[Rv7]}\mathbf{\vartheta }_{a}\wedge \lbrack \mathbf{e}%
_{c}\rfloor ^{\lbrack v]}\ (\ ^{[5]}\mathbf{R}_{\ \ b}^{a}-\ ^{[5]}\mathbf{R}%
_{b\ }^{\ a})] \\
&&+\sum\limits_{[I]=1}^{5}z_{[RvI]}\ (^{[I]}\mathbf{R}_{ab}+\ ^{[I]}\mathbf{R%
}_{ba})+z_{[Rv6]}\mathbf{\vartheta }_{c}\wedge \lbrack \mathbf{e}_{a}\rfloor
^{\lbrack v]}\ (\ ^{[2]}\mathbf{R}_{\ \ b}^{c}-\ ^{[2]}\mathbf{R}_{b\ }^{\
c})] \\
&&+\sum\limits_{[I]=7}^{9}z_{[RvI]}\mathbf{\vartheta }_{a}\wedge \lbrack
\mathbf{e}_{c}\rfloor ^{\lbrack v]}\ (\ ^{[I-4]}\mathbf{R}_{\ \ b}^{c}-\
^{[I-4]}\mathbf{R}_{b\ }^{\ c})]\}
\end{eqnarray*}%
\begin{eqnarray*}
&&-\frac{1}{2\rho _{\lbrack Ph]}}\mathbf{P}^{ij}\wedge \ast ^{\lbrack
h]}\{\sum\limits_{[I]=1}^{6}w_{[PhI]}\ (^{[I]}\mathbf{P}_{ij}-\ ^{[I]}%
\mathbf{P}_{ji})+w_{[Ph7]}\mathbf{\vartheta }_{i}\wedge \lbrack \mathbf{e}%
_{k}\rfloor ^{\lbrack h]}\ (\ ^{[5]}\mathbf{P}_{\ \ j}^{k}-\ ^{[5]}\mathbf{P}%
_{j\ }^{\ k})] \\
&&+\sum\limits_{[I]=1}^{5}z_{[PhI]}\ (^{[I]}\mathbf{P}_{ij}+\ ^{[I]}\mathbf{P%
}_{ji})+z_{[Ph6]}\mathbf{\vartheta }_{k}\wedge \lbrack \mathbf{e}_{i}\rfloor
^{\lbrack h]}\ (\ ^{[2]}\mathbf{P}_{\ \ j}^{k}-\ ^{[2]}\mathbf{P}_{j\ }^{\
k})] \\
&&+\sum\limits_{[I]=7}^{9}z_{[PhI]}\mathbf{\vartheta }_{i}\wedge \lbrack
\mathbf{e}_{k}\rfloor ^{\lbrack h]}\ (\ ^{[I-4]}\mathbf{P}_{\ \ j}^{k}-\
^{[I-4]}\mathbf{P}_{j\ }^{\ k})]\}-
\end{eqnarray*}%
\begin{eqnarray*}
&&\frac{1}{2\rho _{\lbrack Pv]}}\mathbf{P}^{ab}\wedge \ast ^{\lbrack
v]}\{\sum\limits_{[I]=1}^{6}w_{[PvI]}\ (^{[I]}\mathbf{P}_{ab}-\ ^{[I]}%
\mathbf{P}_{ba})+w_{[Pv7]}\mathbf{\vartheta }_{a}\wedge \lbrack \mathbf{e}%
_{c}\rfloor ^{\lbrack v]}\ (\ ^{[5]}\mathbf{P}_{\ \ b}^{a}-\ ^{[5]}\mathbf{P}%
_{b\ }^{\ a})] \\
&&+\sum\limits_{[I]=1}^{5}z_{[PvI]}\ (^{[I]}\mathbf{P}_{ab}+\ ^{[I]}\mathbf{P%
}_{ba})+z_{[Pv6]}\mathbf{\vartheta }_{c}\wedge \lbrack \mathbf{e}_{a}\rfloor
^{\lbrack v]}\ (\ ^{[2]}\mathbf{P}_{\ \ b}^{c}-\ ^{[2]}\mathbf{P}_{b\ }^{\
c})] \\
&&+\sum\limits_{[I]=7}^{9}z_{[PvI]}\mathbf{\vartheta }_{a}\wedge \lbrack
\mathbf{e}_{c}\rfloor ^{\lbrack v]}\ (\ ^{[I-4]}\mathbf{P}_{\ \ b}^{c}-\
^{[I-4]}\mathbf{P}_{b\ }^{\ c})]\}
\end{eqnarray*}%
\begin{eqnarray*}
&&-\frac{1}{2\rho _{\lbrack Sh]}}\mathbf{S}^{ij}\wedge \ast ^{\lbrack
h]}\{\sum\limits_{[I]=1}^{6}w_{[ShI]}\ (^{[I]}\mathbf{S}_{ij}-\ ^{[I]}%
\mathbf{S}_{ji})+w_{[Sh7]}\mathbf{\vartheta }_{i}\wedge \lbrack \mathbf{e}%
_{k}\rfloor ^{\lbrack h]}\ (\ ^{[5]}\mathbf{S}_{\ \ j}^{k}-\ ^{[5]}\mathbf{S}%
_{j\ }^{\ k})] \\
&&+\sum\limits_{[I]=1}^{5}z_{[ShI]}\ (^{[I]}\mathbf{S}_{ij}+\ ^{[I]}\mathbf{S%
}_{ji})+z_{[Sh6]}\mathbf{\vartheta }_{k}\wedge \lbrack \mathbf{e}_{i}\rfloor
^{\lbrack h]}\ (\ ^{[2]}\mathbf{S}_{\ \ j}^{k}-\ ^{[2]}\mathbf{S}_{j\ }^{\
k})] \\
&&+\sum\limits_{[I]=7}^{9}z_{[ShI]}\mathbf{\vartheta }_{i}\wedge \lbrack
\mathbf{e}_{k}\rfloor ^{\lbrack h]}\ (\ ^{[I-4]}\mathbf{S}_{\ \ j}^{k}-\
^{[I-4]}\mathbf{S}_{j\ }^{\ k})]\}-
\end{eqnarray*}%
\begin{eqnarray*}
&&\frac{1}{2\rho _{\lbrack Sv]}}\mathbf{S}^{ab}\wedge \ast ^{\lbrack
v]}\{\sum\limits_{[I]=1}^{6}w_{[SvI]}\ (^{[I]}\mathbf{S}_{ab}-\ ^{[I]}%
\mathbf{S}_{ba})+w_{[Sv7]}\mathbf{\vartheta }_{a}\wedge \lbrack \mathbf{e}%
_{c}\rfloor ^{\lbrack v]}\ (\ ^{[5]}\mathbf{S}_{\ \ b}^{a}-\ ^{[5]}\mathbf{S}%
_{b\ }^{\ a})] \\
&&+\sum\limits_{[I]=1}^{5}z_{[SvI]}\ (^{[I]}\mathbf{S}_{ab}+\ ^{[I]}\mathbf{S%
}_{ba})+z_{[Sv6]}\mathbf{\vartheta }_{c}\wedge \lbrack \mathbf{e}_{a}\rfloor
^{\lbrack v]}\ (\ ^{[2]}\mathbf{S}_{\ \ b}^{c}-\ ^{[2]}\mathbf{S}_{b\ }^{\
c})] \\
&&+\sum\limits_{[I]=7}^{9}z_{[SvI]}\mathbf{\vartheta }_{a}\wedge \lbrack
\mathbf{e}_{c}\rfloor ^{\lbrack v]}\ (\ ^{[I-4]}\mathbf{S}_{\ \ b}^{c}-\
^{[I-4]}\mathbf{S}_{b\ }^{\ c})]\}.
\end{eqnarray*}%
Let us explain the denotations used in (\ref{actgfa}): The signature is
adapted in the form $\left( -+++\right) $ $\ $and there are considered two
Hodge duals, $\ast ^{\lbrack h]}$for h--subspace and $\ast ^{\lbrack v]}$for
v--subspace, and respectively two cosmological constants, $\lambda _{\lbrack
h]}$ and $\lambda _{\lbrack v]}.$ The strong gravity coupling constants $%
\rho _{\lbrack Rh]},\rho _{\lbrack Rv]},\rho _{\lbrack Ph]},....,$ the
constants $a_{0[Rh]},a_{0[Rv]},a_{0[Ph]},...,a_{[hA]},a_{[vA]},...$ $%
c_{[hI]},c_{[vI]},...$ are dimensionless and provided with labels $%
[R],[P],[h],[v],$ emphasizing $\ \ $that the constants are related, for
instance, to respective invariants of curvature, torsion, nonmetricity and
their h- and v--decompositions.

The action (\ref{actgfa}) describes all possible models of Einstein,
Einstein--Cartan and all type of Finsler--Lagrange--Cartan--Hamilton
gravities which can be modeled on metric affine spaces provided with
N--connection structure (i. e. with generic off--diagonal metrics) and
derived from quadratic MAG--type Lagrangians.

We can reduce the number of constants in $\mathcal{L}_{GFA}\rightarrow
\mathcal{L}_{GFA}^{\prime }$ if we select the limit resulting in the usual
quadratic MAG--Lagrangian \cite{mag} for trivial N--connection structure. In
this case, all constants for h-- and v-- decompositions coincide with those
from MAG without N--connection structure, for instance,%
\begin{equation*}
a_{0}=a_{0[Rh]}=a_{0[Rv]}=a_{0[Ph]}=...,\
a_{[A]}=a_{[hA]}=a_{[vA]}=...,...,c_{[I]}=c_{[hI]}=c_{[vI]},...
\end{equation*}%
The Lagrangian (\ref{actgfa}) can be reduced to a more simple one written in
terms of boldfaced symbols (emphasizing a nontrivial N--connection
structure) provided with Greek indices,

\begin{eqnarray}
\mathcal{L}_{GFA}^{\prime } &=&\frac{1}{2\kappa }[-a_{0}\mathbf{R}^{\alpha
\beta }\wedge \eta _{\alpha \beta }-2\lambda \eta +\mathbf{T}^{i}\wedge \ast
\left( \sum\limits_{\lbrack A]=1}^{3}a_{[A]}\ ^{[A]}\mathbf{T}_{i}\right)
\notag \\
&&+2\left( \sum\limits_{[I]=2}^{4}c_{[I]}\ ^{[I]}\mathbf{Q}_{\alpha \beta
}\right) \wedge \mathbf{\vartheta }^{\alpha }\wedge \ast \ \mathbf{T}^{\beta
}+\mathbf{Q}_{\alpha \beta }\wedge \left( \sum\limits_{\lbrack
I]=1}^{4}b_{[I]}\ ^{[I]}\mathbf{Q}^{\alpha \beta }\right)  \label{actfag1} \\
&&+\mathbf{Q}_{\alpha \beta }\wedge \left( \sum\limits_{\lbrack
I]=1}^{4}b_{[I]}\ ^{[I]}\mathbf{Q}^{\alpha \beta }\right) +b_{[5]}\left(
^{[3]}\mathbf{Q}_{\alpha \beta }\wedge \mathbf{\vartheta }^{\alpha }\right)
\wedge \ast \left( ^{\lbrack 4]}\mathbf{Q}^{\gamma \beta }\wedge \mathbf{%
\vartheta }_{\gamma }\right) ]  \notag \\
&&-\frac{1}{2\rho }\mathbf{R}^{\alpha \beta }\wedge \ast \lbrack
\sum\limits_{\lbrack I]=1}^{6}w_{[I]}\ \mathbf{\ }^{[I]}\mathbf{W}_{\alpha
\beta }+w_{[7]}\mathbf{\vartheta }_{\alpha }\wedge \left( \mathbf{e}_{\gamma
}\rfloor \ \mathbf{\ }^{[5]}\mathbf{W}_{\ \beta }^{\gamma }\right) \   \notag
\\
&&+\sum\limits_{[I]=1}^{5}z_{[I]}\ ^{[I]}\mathbf{Y}_{\alpha \beta }+z_{[6]}%
\mathbf{\vartheta }_{\gamma }\wedge \left( \mathbf{e}_{\alpha }\rfloor \
^{[2]}\mathbf{Y}_{\ \beta }^{\gamma }\right) +\sum\limits_{[I]=7}^{9}z_{[I]}%
\mathbf{\vartheta }_{\alpha }\wedge (\mathbf{e}_{\gamma }\rfloor \ \ ^{[I-4]}%
\mathbf{Y}_{\ \beta }^{\gamma })].  \notag
\end{eqnarray}%
where $\mathbf{\ }^{[I]}\mathbf{W}_{\alpha \beta }=$\ $\ ^{[I]}\mathbf{R}%
_{\alpha \beta }-\ ^{[I]}\mathbf{R}_{\beta \alpha }$ and $\ ^{[I]}\mathbf{Y}%
_{\alpha \beta }=$\ $^{[I]}\mathbf{R}_{\alpha \beta }+\ ^{[I]}\mathbf{R}%
_{\beta \alpha }.$ This action is just for the MAG quadratic theory but with
$\mathbf{e}_{\alpha }$ and $\mathbf{\vartheta }^{\beta }$ being adapted to
the N--connection structure as in (\ref{dder}) and (\ref{ddif}) with a
corresponding splitting of geometrical objects.

The field equations of a metric--affine space provided with N--connection
structure, $\mathbf{V}^{n+m}=\left[ N_{i}^{a},\mathbf{g}_{\alpha \beta
}=\left( g_{ij},h_{ab}\right) ,\mathbf{\Gamma }_{\alpha \beta }^{\gamma
}=\left( L_{jk}^{i},L_{bk}^{a},C_{jc}^{i},C_{bc}^{a}\right) \right] ,$ can
be obtained by the Noether procedure in its turn being N--adapted to (co)
frames $\mathbf{e}_{\alpha }$ and $\mathbf{\vartheta }^{\beta }.$ At the
first step, we parametrize the generalized Finsler--affine Lagrangian and
matter Lagrangian respectively as
\begin{equation*}
\mathcal{L}_{GFA}^{\prime }=\mathcal{L}_{[fa]}\left( N_{i}^{a},\mathbf{g}%
_{\alpha \beta },\mathbf{\vartheta }^{\gamma },\mathbf{Q}_{\alpha \beta },%
\mathbf{T}^{\alpha },\ \mathbf{R}_{\ \beta }^{\alpha }\right)
\end{equation*}%
and
\begin{equation*}
\mathcal{L}_{mat}=\mathcal{L}_{[m]}\left( N_{i}^{a},\mathbf{g}_{\alpha \beta
},\mathbf{\vartheta }^{\gamma },\mathbf{\Psi ,D\Psi }\right) ,
\end{equation*}%
where $\mathbf{T}^{\alpha }$ and$\ \mathbf{R}_{\ \beta }^{\alpha }$ are the
curvature of arbitrary d--connection $\mathbf{D}$ and $\mathbf{\Psi }$
represents the matter fields as a $p$--form. The action $\mathcal{S}$ on $%
\mathbf{V}^{n+m}$ is written
\begin{equation}
\mathcal{S}=\int \delta ^{n+m}u\sqrt{|\mathbf{g}_{\alpha \beta }|}\left[
\mathcal{L}_{[fa]}+\mathcal{L}_{[m]}\right]  \label{agmafgm}
\end{equation}%
which results in the matter and gravitational (generalized Finsler--affine
type) field equations.

\begin{theorem}
\ The Yang--Mills type field equations of the generalized Finsler--affine
gravity with matter derived by a variational procedure adapted to the
N--connection structure are defined by the system
\begin{eqnarray}
\mathbf{D}\left( \frac{\partial \mathcal{L}_{[m]}}{\partial \left( \mathbf{%
D\Psi }\right) }\right) -\left( -1\right) ^{p}\frac{\partial \mathcal{L}%
_{[m]}}{\partial \mathbf{\Psi }} &=&0,  \label{fefag} \\
\mathbf{D}\left( \frac{\partial \mathcal{L}_{[fa]}}{\partial \mathbf{Q}%
_{\alpha \beta }}\right) +2\frac{\partial \mathcal{L}_{[fa]}}{\partial
\mathbf{g}_{\alpha \beta }} &=&-\mathbf{\sigma }^{\alpha \beta },  \notag \\
\mathbf{D}\left( \frac{\partial \mathcal{L}_{[fa]}}{\partial \mathbf{T}%
^{\alpha }}\right) +2\frac{\partial \mathcal{L}_{[fa]}}{\partial \mathbf{%
\vartheta }^{\alpha }} &=&-\mathbf{\Sigma }_{\alpha },  \notag \\
\mathbf{D}\left( \frac{\partial \mathcal{L}_{[fa]}}{\partial \mathbf{R}_{\
\beta }^{\alpha }}\right) +\mathbf{\vartheta }^{\beta }\wedge \frac{\partial
\mathcal{L}_{[fa]}}{\partial \mathbf{T}^{\alpha }} &=&-\mathbf{\Delta }%
_{\alpha }^{\ \beta },  \notag
\end{eqnarray}%
where the material currents are defined
\begin{equation*}
\mathbf{\sigma }^{\alpha \beta }\doteqdot 2\frac{\mathbf{\delta }\mathcal{L}%
_{[m]}}{\mathbf{\delta g}_{\alpha \beta }},\ \mathbf{\Sigma }_{\alpha
}\doteqdot \frac{\mathbf{\delta }\mathcal{L}_{[m]}}{\mathbf{\delta \vartheta
}^{\alpha }},\ \mathbf{\Delta }_{\alpha }^{\ \beta }=\frac{\mathbf{\delta }%
\mathcal{L}_{[m]}}{\mathbf{\delta \Gamma }_{\ \beta }^{\alpha }}
\end{equation*}%
for variations ''boldfaced'' $\mathbf{\delta }\mathcal{L}_{[m]}/\mathbf{%
\delta }$\ computed with respect to N--adapted (co) frames.
\end{theorem}

The proof of this theorem consists from N--adapted variational calculus. The
equations (\ref{fefag}) transforms correspondingly into ''MATTER, ZEROTH,
FIRST, SECOND'' equations of MAG \cite{mag} for trivial N--connection
structures.

\begin{corollary}
The system (\ref{fefag}) has respectively the h-- and v--irreducible
components
\begin{equation*}
D^{[h]}\left( \frac{\partial \mathcal{L}_{[m]}}{\partial \left( D^{[h]}%
\mathbf{\Psi }\right) }\right) +D^{[v]}\left( \frac{\partial \mathcal{L}%
_{[m]}}{\partial \left( D^{[v]}\mathbf{\Psi }\right) }\right) -\left(
-1\right) ^{p}\frac{\partial \mathcal{L}_{[m]}}{\partial \mathbf{\Psi }}=0,
\end{equation*}%
\begin{eqnarray}
D^{[h]}\left( \frac{\partial \mathcal{L}_{[fa]}}{\partial Q_{ij}}\right)
+D^{[v]}\left( \frac{\partial \mathcal{L}_{[fa]}}{\partial Q_{ij}}\right) +2%
\frac{\partial \mathcal{L}_{[fa]}}{\partial g_{ij}} &=&-\sigma ^{ij},
\label{fefagd} \\
D^{[h]}\left( \frac{\partial \mathcal{L}_{[fa]}}{\partial Q_{ab}}\right)
+D^{[v]}\left( \frac{\partial \mathcal{L}_{[fa]}}{\partial Q_{ab}}\right) +2%
\frac{\partial \mathcal{L}_{[fa]}}{\partial g_{ab}} &=&-\sigma ^{ab},  \notag
\end{eqnarray}%
\begin{eqnarray*}
D^{[h]}\left( \frac{\partial \mathcal{L}_{[fa]}}{\partial T^{i}}\right)
+D^{[v]}\left( \frac{\partial \mathcal{L}_{[fa]}}{\partial T^{i}}\right) +2%
\frac{\partial \mathcal{L}_{[fa]}}{\partial \vartheta ^{i}} &=&-\Sigma _{i},
\\
D^{[h]}\left( \frac{\partial \mathcal{L}_{[fa]}}{\partial T^{a}}\right)
+D^{[v]}\left( \frac{\partial \mathcal{L}_{[fa]}}{\partial T^{a}}\right) +2%
\frac{\partial \mathcal{L}_{[fa]}}{\partial \vartheta ^{a}} &=&-\Sigma _{a},
\end{eqnarray*}%
\begin{eqnarray*}
D^{[h]}\left( \frac{\partial \mathcal{L}_{[fa]}}{\partial R_{\ j}^{i}}%
\right) +D^{[v]}\left( \frac{\partial \mathcal{L}_{[fa]}}{\partial R_{\
j}^{i}}\right) +\vartheta ^{j}\wedge \frac{\partial \mathcal{L}_{[fa]}}{%
\partial T^{i}} &=&-\mathbf{\Delta }_{i}^{\ j}, \\
D^{[h]}\left( \frac{\partial \mathcal{L}_{[fa]}}{\partial R_{\ b}^{a}}%
\right) +D^{[v]}\left( \frac{\partial \mathcal{L}_{[fa]}}{\partial R_{\
b}^{a}}\right) +\vartheta ^{b}\wedge \frac{\partial \mathcal{L}_{[fa]}}{%
\partial T^{a}} &=&-\mathbf{\Delta }_{a}^{\ b},
\end{eqnarray*}%
where \
\begin{eqnarray*}
\mathbf{\sigma }^{\alpha \beta } &=&\left( \sigma ^{ij},\sigma ^{ab}\right) %
\mbox{ for }\ \sigma ^{ij}\doteqdot 2\frac{\mathbf{\delta }\mathcal{L}_{[m]}%
}{\mathbf{\delta }g_{ij}},\ \sigma ^{ab}\doteqdot 2\frac{\mathbf{\delta }%
\mathcal{L}_{[m]}}{\mathbf{\delta }h_{ab}},\  \\
\mathbf{\Sigma }_{\alpha } &=&\left( \Sigma _{i},\Sigma _{a}\right)
\mbox{
for }\ \Sigma _{i}\doteqdot \frac{\mathbf{\delta }\mathcal{L}_{[m]}}{\mathbf{%
\delta }\vartheta ^{i}},\ \Sigma _{a}\doteqdot \frac{\mathbf{\delta }%
\mathcal{L}_{[m]}}{\mathbf{\delta }\vartheta ^{a}}, \\
\mathbf{\Delta }_{\alpha }^{\ \beta } &=&\left( \Delta _{i}^{\ j},\Delta
_{a}^{\ b}\right) \mbox{ for }\ \Delta _{i}^{\ j}=\frac{\mathbf{\delta }%
\mathcal{L}_{[m]}}{\mathbf{\delta }\Gamma _{\ j}^{i\ \ }},\ \Delta _{a}^{\
b}=\frac{\mathbf{\delta }\mathcal{L}_{[m]}}{\mathbf{\delta }\Gamma _{\
b}^{a\ \ }}\ .
\end{eqnarray*}
\end{corollary}

It should be noted that the complete h-- v--decomposition of the system (\ref%
{fefagd}) can be obtained if we represent the d--connection and curvature
forms as%
\begin{equation*}
\Gamma _{\ j}^{i\ \ }=L_{\ jk}^{i\ \ }dx^{j}+C_{\ ja}^{i\ \ }\delta y^{a}%
\mbox{ and }\Gamma _{\ b}^{a\ \ }=L_{\ bk}^{a\ \ }dx^{k}+C_{\ bc}^{a\ \
}\delta y^{c},
\end{equation*}%
see the d--connection components (\ref{dcon1}) and%
\begin{eqnarray*}
2R_{\ j}^{i} &=&R_{\ jkl}^{i}dx^{k}\wedge dx^{l}+P_{\ jka}^{i}dx^{k}\wedge
\delta y^{a}+S_{\ jba}^{i}\delta y^{b}\wedge \delta y^{a}, \\
2R_{\ f}^{e} &=&R_{\ fkl}^{e}dx^{k}\wedge dx^{l}+P_{\ fka}^{e}dx^{k}\wedge
\delta y^{a}+S_{\ fba}^{e}\delta y^{a}\wedge \delta y^{a},
\end{eqnarray*}%
see the d--curvature components (\ref{dcurv}).

\begin{remark}
\label{rfconf}For instance, a Finsler configuration can be modeled on a
metric affine space provided with N--connection structure, $\mathbf{V}^{n+m}=%
\left[ N_{i}^{a},\mathbf{g}_{\alpha \beta }=\left( g_{ij},h_{ab}\right) ,\
^{[F]}\widehat{\mathbf{\Gamma }}_{\alpha \beta }^{\gamma }\right] ,$ if $%
n=m, $ the ansatz for N--connection is of Cartan--Finsler type
\begin{equation*}
N_{j}^{a}\rightarrow \ ^{[F]}N_{j}^{i}=\frac{1}{8}\frac{\partial }{\partial
y^{j}}\left[ y^{l}y^{k}g_{[F]}^{ih}\left( \frac{\partial g_{hk}^{[F]}}{%
\partial x^{l}}+\frac{\partial g_{lh}^{[F]}}{\partial x^{k}}-\frac{\partial
g_{lk}^{[F]}}{\partial x^{h}}\right) \right] ,
\end{equation*}
the d--metric $\mathbf{g}_{\alpha \beta }=\mathbf{g}_{\alpha \beta }^{[F]}$
is defined by (\ref{block2}) with
\begin{equation*}
g_{ij}^{[F]}=g_{ij}=h_{ij}=\frac{1}{2}\partial ^{2}F/\partial y^{i}\partial
y^{j}
\end{equation*}
and $^{[F]}\widehat{\mathbf{\Gamma }}_{\alpha \beta }^{\gamma }$ is the
Finsler canonical d--connection computed as (\ref{candcon}). The data should
define an exact solution of the system of field equation (\ref{fefagd})
(equivalently of (\ref{fefag})).
\end{remark}

Similar Remarks hold true for all types of generalized Finsler--affine
spaces considered in Tables 1--11 from Ref. \cite{vp1}. We shall analyze the
possibility of modeling various type of locally anisotropic geometries by
the Einstein--Proca systems and in string gravity in next subsection.

\subsection{Effective Einstein--Proca systems and N--connections}

Any affine connection can always be decomposed into (pseudo) Riemannian, $%
\Gamma _{\bigtriangledown \ \beta }^{\alpha },$ and post--Riemannian, $Z_{\
\ \beta }^{\alpha },$ parts as $\Gamma _{\ \beta }^{\alpha }=\Gamma
_{\bigtriangledown \ \beta }^{\alpha }+Z_{\ \ \beta }^{\alpha },$ see
formulas (\ref{acc}) and (\ref{dista}) (or (\ref{accn}) and (\ref{distan})
if any N--connection structure is prescribed). This mean that it is possible
to split \ all quantities of a metric--affine theory into (pseudo)\
Riemannian and post--Riemannian pieces, for instance,
\begin{equation}
R_{\ \beta }^{\alpha }=R_{\bigtriangledown \ \beta }^{\alpha
}+\bigtriangledown Z_{\ \ \beta }^{\alpha }+Z_{\ \ \gamma }^{\alpha }\wedge
Z_{\ \ \beta }^{\gamma }.  \label{dist1}
\end{equation}%
Under certain assumptions one holds the Obukhov's equivalence theorem
according to which the field vacuum metric--affine gravity equations are
equivalent to Einstein's equations with an energy--momentum tensor
determined by a Proca field \cite{oveh,obet2}. We can generalize the
constructions and reformulate the equivalence theorem for generalized
Finsler--affine spaces and effective spaces provided with N--connection
structure.

\begin{theorem}
\label{teq}The system of effective field equations of MAG on spaces provided
with N--connection structure (\ref{fefag}) (equivalently, (\ref{fefagd})) \
for certain ansatz for torsion and nonmetricity fields (see (\ref{torsdec})
and (\ref{nmdc}))
\begin{eqnarray}
^{(1)}\mathbf{T}^{\alpha } &=&^{(2)}\mathbf{T}^{\alpha }=0,\ \ ^{(1)}\mathbf{%
Q}_{\alpha \beta }=\ ^{(2)}\mathbf{Q}_{\alpha \beta }=0,  \label{triplemag}
\\
\mathbf{Q} &=&k_{0}\mathbf{\phi ,\ \Lambda =}k_{1}\mathbf{\phi ,\ T=}k_{2}%
\mathbf{\phi ,}  \notag
\end{eqnarray}%
where $k_{0},k_{1},k_{2}=const$ and the Proca 1--form is $\mathbf{\phi =}%
\phi _{\alpha }\mathbf{\vartheta }^{\alpha }=\phi _{i}dx^{i}+\phi _{a}\delta
y^{a},$ reduces to the Einstein--Proca system of equations for the canonical
d--connection $\widehat{\mathbf{\Gamma }}_{\ \alpha \beta }^{\gamma }$ (\ref%
{candcon}) and massive d--field $\mathbf{\phi }_{\alpha },$%
\begin{eqnarray}
\frac{a_{0}}{2}\mathbf{\eta }_{\alpha \beta \gamma }\wedge \widehat{\mathbf{R%
}}^{\beta \gamma } &=&k\ \mathbf{\Sigma }_{\alpha },  \notag \\
\delta \left( \ast \mathbf{H}\right) +\mu ^{2}\mathbf{\phi }\mathbf{=0,} &&
\label{efmageq}
\end{eqnarray}%
where \ $\mathbf{H\doteqdot }\delta \mathbf{\phi ,}$ the mass$\mathbf{\ }\mu
=const$ and the energy--momentum is given by
\begin{equation*}
\mathbf{\Sigma }_{\alpha }=\mathbf{\Sigma }_{\alpha }^{[\phi ]}+\mathbf{%
\Sigma }_{\alpha }^{[\mathbf{m}]},
\end{equation*}%
\begin{equation*}
\mathbf{\Sigma }_{\alpha }^{[\phi ]}\mathbf{\doteqdot }\frac{z_{4}k_{0}^{2}}{%
2\rho }\{\left( \mathbf{e}_{\alpha }\rfloor \ \mathbf{H}\right) \wedge \ast
\mathbf{H-}\left( \mathbf{e}_{\alpha }\rfloor \ast \mathbf{H}\right) \wedge
\mathbf{H+}\mu ^{2}[\left( \mathbf{e}_{\alpha }\rfloor \ \mathbf{\phi }%
\right) \wedge \ast \mathbf{\phi -}\left( \mathbf{e}_{\alpha }\rfloor \ast
\mathbf{\phi }\right) \wedge \mathbf{\phi }]\}
\end{equation*}%
is the energy--momentum current of the Proca d--field and $\mathbf{\Sigma }%
_{\alpha }^{[\mathbf{\mu }]}$ is the energy--momentum current of the
additional matter d--fields satisfying the corresponding Euler--Largange
equations.
\end{theorem}

The proof of the Theorem is just the reformulation with respect to
N--adapted (co) frames (\ref{dder}) and (\ref{ddif}) of similar
considerations in Refs. \cite{oveh,obet2}. The constants $k_{0},k_{1}....$
are taken in terms of the gravitational coupling constants like in \cite{hm}
as to have connection to the usual MAG and Einstein theory for trivial
N--connection structures and for the dimension $m\rightarrow 0.$ We use the
triplet ansatz sector (\ref{triplemag}) \ of MAG theories \cite{oveh,obet2}.
It is a remarkable fact that the equivalence Theorem \ref{teq} holds also in
presence of arbitrary N--connections i. e. for all type of anholonomic
generalizations of the Einstein, Einstein--Cartan and Finsler--Lagrange and
Cartan--Hamilton geometries by introducing canonical d--connections (we can
also consider Berwald type d--connections).

\begin{corollary}
\ \label{corfag}In abstract index form, the effective field equations for
the generalized Finsler--affine gravity following from (\ref{efmageq})\ are
written%
\begin{eqnarray}
\widehat{\mathbf{R}}_{\alpha \beta }-\frac{1}{2}\mathbf{g}_{\alpha \beta }%
\overleftarrow{\mathbf{\hat{R}}} &=&\tilde{\kappa}\left( \mathbf{\Sigma }%
_{\alpha \beta }^{[\phi ]}+\mathbf{\Sigma }_{\alpha \beta }^{[\mathbf{m}%
]}\right) ,  \label{efeinst} \\
\widehat{\mathbf{D}}_{\nu }\mathbf{H}^{\nu \mu } &=&\mu ^{2}\mathbf{\phi }%
^{\mu },  \notag
\end{eqnarray}%
with $\mathbf{H}_{\nu \mu }\doteqdot \widehat{\mathbf{D}}_{\nu }\mathbf{\phi
}_{\mu }-\widehat{\mathbf{D}}_{\mu }\mathbf{\phi }_{\nu }+w_{\mu \nu
}^{\gamma }\mathbf{\phi }_{\gamma }$ being the field strengths of the
Abelian Proca field $\mathbf{\phi }^{\mu },\tilde{\kappa}=const,$ and
\begin{equation}
\mathbf{\Sigma }_{\alpha \beta }^{[\phi ]}=\mathbf{H}_{\alpha }^{\ \mu }%
\mathbf{H}_{\beta \mu }-\frac{1}{4}\mathbf{g}_{\alpha \beta }\mathbf{H}_{\mu
\nu }^{\ }\mathbf{H}^{\mu \nu }+\mu ^{2}\mathbf{\phi }_{\alpha }\mathbf{\phi
}_{\beta }-\frac{\mu ^{2}}{2}\mathbf{g}_{\alpha \beta }\mathbf{\phi }_{\mu }%
\mathbf{\phi }^{\mu }.  \label{sourcef}
\end{equation}
\end{corollary}

The Ricci d--tensor $\widehat{\mathbf{R}}_{\alpha \beta }$ and scalar $%
\overleftarrow{\mathbf{\hat{R}}}$ from (\ref{efeinst}) can be decomposed in
irreversible h-- and v--invariant components like (\ref{dricci}) and (\ref%
{dscal}),%
\begin{eqnarray}
\widehat{R}_{ij}-\frac{1}{2}g_{ij}\left( \widehat{R}+\widehat{S}\right) &=&%
\tilde{\kappa}\left( \mathbf{\Sigma }_{ij}^{[\phi ]}+\mathbf{\Sigma }_{ij}^{[%
\mathbf{m}]}\right) ,  \label{ep1} \\
\widehat{S}_{ab}-\frac{1}{2}h_{ab}\left( \widehat{R}+\widehat{S}\right) &=&%
\tilde{\kappa}\left( \mathbf{\Sigma }_{ab}^{[\phi ]}+\mathbf{\Sigma }_{ab}^{[%
\mathbf{m}]}\right) ,  \label{ep2} \\
^{1}P_{ai} &=&\tilde{\kappa}\left( \mathbf{\Sigma }_{ai}^{[\phi ]}+\mathbf{%
\Sigma }_{ai}^{[\mathbf{m}]}\right) ,  \label{ep3} \\
\ -^{2}P_{ia} &=&\tilde{\kappa}\left( \mathbf{\Sigma }_{ia}^{[\phi ]}+%
\mathbf{\Sigma }_{ia}^{[\mathbf{m}]}\right) .  \label{ep4}
\end{eqnarray}%
The constants are those from \cite{oveh} being related to the constants from
(\ref{actfag1}),%
\begin{equation*}
\mu ^{2}=\frac{1}{z_{k}\kappa }\left( -4\beta _{4}+\frac{k_{1}}{2k_{0}}\beta
_{5}+\frac{k_{2}}{k_{0}}\gamma _{4}\right) ,
\end{equation*}%
where
\begin{eqnarray*}
k_{0} &=&4\alpha _{2}\beta _{3}-3(\gamma _{3})^{2}\neq 0,\ k_{1}=9\left(
\frac{1}{2}\alpha _{5}\beta _{5}-\gamma _{3}\gamma _{4}\right) ,\
k_{2}=3\left( 4\beta _{3}\gamma _{4}-\frac{3}{2}\beta _{5}\gamma _{3}\right)
, \\
\alpha _{2} &=&a_{2}-2a_{0},\ \beta _{3}=b_{3}+\frac{a_{0}}{8},\ \beta
_{4}=b_{4}-\frac{3a_{0}}{8},\ \gamma _{3}=c_{3}+a_{0},\ \gamma
_{4}=c_{4}+a_{0}.
\end{eqnarray*}%
If
\begin{equation}
\beta _{4}\rightarrow \frac{1}{4k_{0}}\left( \frac{1}{2}\beta
_{5}k_{1}+k_{2}\gamma _{4}\right) ,  \label{procvan}
\end{equation}%
the mass of Proca field $\mu ^{2}\rightarrow 0.$ The system becomes like the
Einstein--Maxwell one with the source (\ref{sourcef}) defined by the
antisymmetric field $\mathbf{H}_{\mu \nu }^{\ }$ in its turn being
determined by a solution of $\widehat{\mathbf{D}}_{\nu }\widehat{\mathbf{D}}%
^{\nu }\mathbf{\phi }_{\alpha }=0$ (a wave like equation in a curved space
provided with N--connection). Even in this case the nonmetricity and torsion
can be nontrivial, for instance, oscillating (see (\ref{triplemag})).

We note that according the Remark \ref{rfconf}, the system (\ref{efeinst})
defines, for instance, a Finsler configuration if the d--metric $\mathbf{g}%
_{\alpha \beta },$ the d--connection $\widehat{\mathbf{D}}_{\nu }$ and the
N--connection are of Finsler type (or contains as imbeddings such objects).

\subsection{Einstein--Cartan gravity and N--connections}

The Einstein--Cartan gravity contains gravitational configurations with
nontrivial N--con\-nec\-ti\-on structure. The simplest model with local
anisotropy is to write on a space $\mathbf{V}^{n+m}$ the Einstein equations
for the canonical d--connection $\widehat{\mathbf{\Gamma }}_{\ \alpha \beta
}^{\gamma }$ (\ref{candcon}) introduced in the Einstein d--tensor (\ref%
{deinst}),%
\begin{equation*}
\widehat{\mathbf{R}}_{\alpha \beta }-\frac{1}{2}\mathbf{g}_{\alpha \beta }%
\overleftarrow{\mathbf{\hat{R}}}=\kappa \mathbf{\Sigma }_{\alpha \beta }^{[%
\mathbf{m}]},
\end{equation*}%
or in terms of differential forms,
\begin{equation}
\mathbf{\eta }_{\alpha \beta \gamma }\wedge \widehat{\mathbf{R}}^{\beta
\gamma }=\kappa \mathbf{\Sigma }_{\alpha }^{[\mathbf{m}]}  \label{einst1}
\end{equation}%
which is a particular case of equations (\ref{efmageq}). The model contains
nontrivial d--torsions, $\widehat{\mathbf{T}}_{\ \alpha \beta }^{\gamma },$
computed by introducing the components of (\ref{candcon}) into formulas (\ref%
{dtorsb}). We can consider that specific distributions of ''spin
dust/fluid'' of Weyssenhoff and Raabe type, or any generalizations, adapted
to the N--connection structure, can constitute the source of certain
algebraic equations for torsion (see details in Refs. \cite{rcg}) or even to
consider generalizations for dynamical equations for torsion like in gauge
gravity theories \cite{ggrav}. A more special case is defined by the
theories when the d--torsions $\widehat{\mathbf{T}}_{\ \alpha \beta
}^{\gamma }$ are induced by specific frame effects of N--connection
structures. Such models contain all possible distorsions to generalized
Finsler--Lagrange--Cartan spacetimes of the Einstein gravity and emphasize
the conditions when such generalizations to locally anisotropic gravity
preserve the local Lorentz invariance or even model Finsler like
configurations in the framework of general relativity.

Let us express the 1--form of the canonical d--connection $\widehat{\mathbf{%
\Gamma }}_{\ \alpha }^{\gamma }$ as the deformation of the Levi--Civita
connection $\mathbf{\Gamma }_{\bigtriangledown \ \alpha }^{\gamma },$%
\begin{equation}
\widehat{\mathbf{\Gamma }}_{\ \alpha }^{\gamma }=\mathbf{\Gamma }%
_{\bigtriangledown \ \alpha }^{\gamma }+\widehat{\mathbf{Z}}_{\ \alpha
}^{\gamma }  \label{dist2}
\end{equation}%
where%
\begin{equation}
\widehat{\mathbf{Z}}_{\alpha \beta }=\mathbf{e}_{\beta }\rfloor \widehat{%
\mathbf{T}}_{\alpha }-\mathbf{e}_{\alpha }\rfloor \widehat{\mathbf{T}}%
_{\beta }+\frac{1}{2}\left( \mathbf{e}_{\alpha }\rfloor \mathbf{e}_{\beta
}\rfloor \widehat{\mathbf{T}}_{\gamma }\right) \mathbf{\vartheta }^{\gamma }
\label{aux53}
\end{equation}%
being a particular case of formulas (\ref{accn}) and (\ref{distan}) when
nonmetricity vanishes, $\mathbf{Q}_{\alpha \beta }=0.$ This induces a
distorsion of the curvature tensor like (\ref{dist1}) but for d--objects,
expressing (\ref{einst1}) in the form
\begin{equation}
\mathbf{\eta }_{\alpha \beta \gamma }\wedge \mathbf{R}_{\bigtriangledown
}^{\beta \gamma }+\mathbf{\eta }_{\alpha \beta \gamma }\wedge \mathbf{Z}%
_{\bigtriangledown \ }^{\beta \gamma }=\kappa \mathbf{\Sigma }_{\alpha }^{[%
\mathbf{m}]}  \label{einst1a}
\end{equation}%
where%
\begin{equation*}
\mathbf{Z}_{\bigtriangledown \ \gamma }^{\beta }=\bigtriangledown \mathbf{Z}%
_{~\ \gamma }^{\beta }+\mathbf{Z}_{~\ \alpha }^{\beta }\wedge \mathbf{Z}_{~\
\gamma }^{\alpha }.
\end{equation*}

\begin{theorem}
The Einstein equations (\ref{einst1}) for the canonical d--connection $%
\widehat{\mathbf{\Gamma }}_{\ \alpha }^{\gamma }$ constructed for a
d--metric field $\mathbf{g}_{\alpha \beta }=[g_{ij},h_{ab}]$ (\ref{block2})
and N--connection $N_{i}^{a}$ is equivalent to the gravitational field
equations for the Einstein--Cartan theory with torsion $\widehat{\mathbf{T}}%
_{\ \alpha }^{\gamma }$ defined by the N--connection, see formulas (\ref%
{dtorsb}).
\end{theorem}

\textbf{Proof:} The proof is trivial and follows from decomposition (\ref%
{dist2}).

\begin{remark}
Every type of generalized Finsler--Lagrange geometries is characterized by a
corresponding N-- and d--connection and d--metric structures, see Tables
1--11 in Ref. \cite{vp1}. For the canonical d--connection such locally
anisotropic geometries can be modeled on Riemann--Cartan manifolds as
solutions of (\ref{einst1}) for a prescribed type of d--torsions (\ref%
{dtorsb}).
\end{remark}

\begin{corollary}
\label{corcond1}A generalized Finsler geometry can be modeled in a (pseudo)
Riemann spacetime by a d--metric $\mathbf{g}_{\alpha \beta }=[g_{ij},h_{ab}]$
(\ref{block2}), equivalently by generic off--diagonal metric (\ref{ansatz}),
satisfying the Einstein equations for the Levi--Civita connection,
\begin{equation}
\mathbf{\eta }_{\alpha \beta \gamma }\wedge \mathbf{R}_{\bigtriangledown
}^{\beta \gamma }=\kappa \mathbf{\Sigma }_{\alpha }^{[\mathbf{m}]}
\label{einst2}
\end{equation}%
if and only if
\begin{equation}
\mathbf{\eta }_{\alpha \beta \gamma }\wedge \mathbf{Z}_{\bigtriangledown \
}^{\beta \gamma }=0.  \label{cond1}
\end{equation}
\end{corollary}

The proof follows from equations (\ref{einst1a}). We emphasize that the
conditions (\ref{cond1}) are imposed for the deformations of the Ricci
tensors computed from distorsions of the Levi--Civita connection to the
cannonical d--connection. In general, a solution $\mathbf{g}_{\alpha \beta
}=[g_{ij},h_{ab}]$ of the Einstein equations (\ref{einst2}) can be
characterized alternatively by d--connections and N--connections as follows
from relation (\ref{lcsyma}). The alternative geometric description contains
nontrivial torsion fields. The simplest such anholonomic configurations can
be defined by the condition of vanishing of N--connection curvature (\ref%
{ncurv}), $\Omega _{ij}^{a}=0,\,$\ but even in such cases there are
nontrivial anholonomy coefficients, see (\ref{anhc}), $\mathbf{w}_{~ia}^{b}=-%
\mathbf{w}_{~ai}^{b}=\partial _{a}N_{i}^{b},$ and nonvanishing d--torsions (%
\ref{dtorsb}),
\begin{equation*}
\widehat{T}_{ja}^{i}=-\widehat{T}_{aj}^{i}=\widehat{C}_{.ja}^{i}\mbox{ and }%
\widehat{T}_{.bi}^{a}=-\widehat{T}_{.ib}^{a}=\widehat{P}_{.bi}^{a}=\frac{%
\partial N_{i}^{a}}{\partial y^{b}}-\widehat{L}_{.bj}^{a},
\end{equation*}%
being induced by off--diagonal terms in the metric (\ref{ansatz}).

\subsection{String gravity and N--connections}

The subjects concerning generalized Finsler (super) geometry, spinors and
(super) strings are analyzed in details in Refs. \cite{v2}. Here, we
consider the simplest examples when Finsler like geometries can be modeled
in string gravity and related to certain metric--affine structures.

For instance, in the sigma model for bosonic string (see, \cite{sgr}), the
background connection is taken to be not the Levi--Civita one, but a certain
deformation by the strength (torsion) tensor
\begin{equation*}
H_{\mu \nu \rho }\doteqdot \delta _{\mu }B_{\nu \rho }+\delta _{\rho }B_{\mu
\nu }+\delta _{\nu }B_{\rho \mu }
\end{equation*}%
of an antisymmetric field $B_{\nu \rho },$ defined as
\begin{equation*}
\mathcal{D}_{\mu }=\bigtriangledown _{\mu }+\frac{1}{2}H_{\mu \nu }^{\quad
\rho }.
\end{equation*}%
We consider the $H$--field defined by using N--elongated operators (\ref%
{dder}) in order to compute the coefficients with respect to anholonomic
frames.

The condition of the Weyl invariance to hold in two dimensions in the lowest
nontrivial approximation in string constant $\alpha ^{\prime },$ see \cite%
{v2}, turn out to be
\begin{eqnarray*}
R_{\mu \nu } &=&-\frac{1}{4}H_{\mu }^{\ \nu \rho }H_{\nu \lambda \rho
}+2\bigtriangledown _{\mu }\bigtriangledown _{\nu }\Phi , \\
\bigtriangledown _{\lambda }H_{\ \mu \nu }^{\lambda } &=&2\left(
\bigtriangledown _{\lambda }\Phi \right) H_{\ \mu \nu }^{\lambda }, \\
\left( \bigtriangledown \Phi \right) ^{2} &=&\bigtriangledown _{\lambda
}\bigtriangledown ^{\lambda }\Phi +\frac{1}{4}R+\frac{1}{48}H_{\mu \nu \rho
}H^{\mu \nu \rho }.
\end{eqnarray*}%
where $\Phi $ is the dilaton field. For trivial dilaton configurations, $%
\Phi =0,$ we may write
\begin{eqnarray*}
R_{\mu \nu } &=&-\frac{1}{4}H_{\mu }^{\ \nu \rho }H_{\nu \lambda \rho }, \\
\bigtriangledown _{\lambda }H_{\ \mu \nu }^{\lambda } &=&0.
\end{eqnarray*}%
In Refs. \cite{v2} we analyzed string gravity models derived from
superstring effective actions, for instance, from the 4D Neveu-Schwarz
action. In this paper we consider, for simplicity, a model with zero dilaton
field but with nontrivial $H$--field related to the d--torsions induced by
the N--connection and canonical d--connection.

A class of Finsler like metrics can be derived from the bosonic string
theory if $\mathbf{H}_{\nu \lambda \rho }$ and $\mathbf{B}_{\nu \rho } $ are
related to the d--torsions components, for instance, with $\widehat{\mathbf{T%
}}_{\ \alpha \beta }^{\gamma }.$ Really, we can take an ansatz
\begin{equation*}
\mathbf{B}_{\nu \rho }=\left[ B_{ij},B_{ia},B_{ab}\right]
\end{equation*}%
and consider that
\begin{equation}
\mathbf{H}_{\nu \lambda \rho }=\widehat{\mathbf{Z}}_{\ \nu \lambda \rho }+%
\widehat{\mathbf{H}}_{\nu \lambda \rho }  \label{aux51a}
\end{equation}%
where $\widehat{\mathbf{Z}}_{\ \nu \lambda \rho }$ is the distorsion of the
Levi--Civita connection induced by $\widehat{\mathbf{T}}_{\ \alpha \beta
}^{\gamma },$ see (\ref{aux53}). In this case the induced by N--connection
torsion structure is related to the antisymmetric $H$--field and
correspondingly to the $B$--field from string theory. The equations
\begin{equation}
\bigtriangledown ^{\nu }\mathbf{H}_{\nu \lambda \rho }=\bigtriangledown
^{\nu }(\widehat{\mathbf{Z}}_{\ \nu \lambda \rho }+\widehat{\mathbf{H}}_{\nu
\lambda \rho })=0  \label{aux51}
\end{equation}%
impose certain dynamical restrictions to the N--connection coefficients $%
N_{i}^{a}$ and d--metric $\mathbf{g}_{\alpha \beta }=[g_{ij},h_{ab}]$ \
contained in $\widehat{\mathbf{T}}_{\ \alpha \beta }^{\gamma }.$ If on the
background space it is prescribed the cannonical d--connection $\widehat{%
\mathbf{D}}$, we can state a model with (\ref{aux51}) redefined as
\begin{equation}
\widehat{\mathbf{D}}^{\nu }\mathbf{H}_{\nu \lambda \rho }=\widehat{\mathbf{D}%
}^{\nu }(\widehat{\mathbf{Z}}_{\ \nu \lambda \rho }+\widehat{\mathbf{H}}%
_{\nu \lambda \rho })=0,  \label{aux51b}
\end{equation}%
where $\widehat{\mathbf{H}}_{\nu \lambda \rho }$ are computed for stated
values of $\widehat{\mathbf{T}}_{\ \alpha \beta }^{\gamma }.$ For trivial
N--connections when $\widehat{\mathbf{Z}}_{\ \nu \lambda \rho }\rightarrow 0$
and $\widehat{\mathbf{D}}^{\nu }\rightarrow \bigtriangledown ^{\nu },$ the $%
\widehat{\mathbf{H}}_{\nu \lambda \rho }$ transforms into usual $H$--fields.

\begin{proposition}
The dynamics of generalized Finsler--affine string gravity is defined by the
system of field equations%
\begin{eqnarray}
\widehat{\mathbf{R}}_{\alpha \beta }-\frac{1}{2}\mathbf{g}_{\alpha \beta }%
\overleftarrow{\mathbf{\hat{R}}} &=&\tilde{\kappa}\left( \mathbf{\Sigma }%
_{\alpha \beta }^{[\phi ]}+\mathbf{\Sigma }_{\alpha \beta }^{[\mathbf{m}]}+%
\mathbf{\Sigma }_{\alpha \beta }^{[\mathbf{T}]}\right) ,  \label{fagfe} \\
\widehat{\mathbf{D}}_{\nu }\mathbf{H}^{\nu \mu } &=&\mu ^{2}\mathbf{\phi }%
^{\mu },  \notag \\
\widehat{\mathbf{D}}^{\nu }(\widehat{\mathbf{Z}}_{\ \nu \lambda \rho }+%
\widehat{\mathbf{H}}_{\nu \lambda \rho }) &=&0  \notag
\end{eqnarray}%
with $\mathbf{H}_{\nu \mu }\doteqdot \widehat{\mathbf{D}}_{\nu }\mathbf{\phi
}_{\mu }-\widehat{\mathbf{D}}_{\mu }\mathbf{\phi }_{\nu }+w_{\mu \nu
}^{\gamma }\mathbf{\phi }_{\gamma }$ being the field strengths of the
Abelian Proca field $\mathbf{\phi }^{\mu },\tilde{\kappa}=const,$
\begin{equation*}
\mathbf{\Sigma }_{\alpha \beta }^{[\phi ]}=\mathbf{H}_{\alpha }^{\ \mu }%
\mathbf{H}_{\beta \mu }-\frac{1}{4}\mathbf{g}_{\alpha \beta }\mathbf{H}_{\mu
\nu }^{\ }\mathbf{H}^{\mu \nu }+\mu ^{2}\mathbf{\phi }_{\alpha }\mathbf{\phi
}_{\beta }-\frac{\mu ^{2}}{2}\mathbf{g}_{\alpha \beta }\mathbf{\phi }_{\mu }%
\mathbf{\phi }^{\mu },
\end{equation*}%
and%
\begin{equation*}
\mathbf{\Sigma }_{\alpha \beta }^{[\mathbf{T}]}=\mathbf{\Sigma }_{\alpha
\beta }^{[\mathbf{T}]}\left( \widehat{\mathbf{T}},\Phi \right)
\end{equation*}%
contains contributions of $\widehat{\mathbf{T}}$ and $\Phi $ fields.
\end{proposition}

\textbf{Proof:} It follows as an extension of the Corollary \ref{corfag} to
sources induced by string corrections. The system (\ref{fagfe}) should be
completed by the field equations for the matter fields present in $\mathbf{%
\Sigma }_{\alpha \beta }^{[\mathbf{m}]}.$

Finally, we note that the equations (\ref{fagfe}) reduce to equations of
type (\ref{einst1a}) (for Riemann--Cartan configurations with zero
nonmetricity),
\begin{equation*}
\mathbf{\eta }_{\alpha \beta \gamma }\wedge \mathbf{R}_{\bigtriangledown
}^{\beta \gamma }+\mathbf{\eta }_{\alpha \beta \gamma }\wedge \mathbf{Z}%
_{\bigtriangledown \ }^{\beta \gamma }=\kappa \mathbf{\Sigma }_{\alpha }^{[%
\mathbf{T}]},
\end{equation*}%
and to equations of type (\ref{einst2}) and (\ref{cond1}) (for (pseudo)
Riemannian configurations)
\begin{eqnarray}
\mathbf{\eta }_{\alpha \beta \gamma }\wedge \mathbf{R}_{\bigtriangledown
}^{\beta \gamma } &=&\kappa \mathbf{\Sigma }_{\alpha }^{[\mathbf{T}]},
\label{cond2} \\
\mathbf{\eta }_{\alpha \beta \gamma }\wedge \mathbf{Z}_{\bigtriangledown \
}^{\beta \gamma } &=&0  \notag
\end{eqnarray}%
with sources defined by torsion (related to N--connection) from string
theory.

\section{The Anholonomic Frame Method \newline
in MAG and String Gravity}

In a series of papers, see Refs. \cite{v1,v1a,vnces,vmethod}, the
anholonomic frame method of constructing exact solutions with generic
off--diagonal metrics (depending on 2-4 variables) in general relativity,
gauge gravity and certain extra dimension generalizations was elaborated. In
this section, we develop the method in MAG and string gravity with
applications to different models of \ five dimensional (in brief, 5D)\
generalized Finsler--affine spaces.

We consider a metric--affine\ space provided with N--connection structure $%
\mathbf{N}=[N_{i}^{4}(u^{\alpha }),$ $N_{i}^{5}(u^{\alpha })]$ where the
local coordinates are labeled $u^{\alpha }=(x^{i},y^{4}=v,y^{5}),$ for $%
i=1,2,3.$ We state the general condition when exact solutions of the field
equations of the generalized Finsler--affine string gravity depending on
holonomic variables $x^{i}$ and on one anholonomic (equivalently,
anisotropic) variable $y^{4}=v$ can be constructed in explicit form. Every
coordinate from a set $u^{\alpha }$ can may be time like, 3D space like, or
extra dimensional. For simplicity, the partial derivatives are denoted $%
a^{\times }=\partial a/\partial x^{1},a^{\bullet }=\partial a/\partial
x^{2},a^{\prime }=\partial a/\partial x^{3},a^{\ast }=\partial a/\partial v.$

The 5D metric
\begin{equation}
\mathbf{g}=\mathbf{g}_{\alpha \beta }\left( x^{i},v\right) du^{\alpha
}\otimes du^{\beta }  \label{metric5}
\end{equation}%
has the metric coefficients $\mathbf{g}_{\alpha \beta }$ parametrized with
respect to the coordinate dual basis by an off--diagonal matrix (ansatz) {\
%%\footnotesize
\begin{equation}
\left[
\begin{array}{ccccc}
g_{1}+w_{1}^{\ 2}h_{4}+n_{1}^{\ 2}h_{5} & w_{1}w_{2}h_{4}+n_{1}n_{2}h_{5} &
w_{1}w_{3}h_{4}+n_{1}n_{3}h_{5} & w_{1}h_{4} & n_{1}h_{5} \\
w_{1}w_{2}h_{4}+n_{1}n_{2}h_{5} & g_{2}+w_{2}^{\ 2}h_{4}+n_{2}^{\ 2}h_{5} &
w_{2}w_{3}h_{4}+n_{2}n_{3}h_{5} & w_{2}h_{4} & n_{2}h_{5} \\
w_{1}w_{3}h_{4}+n_{1}n_{3}h_{5} & w_{2}w_{3}h_{4}+n_{2}n_{3}h_{5} &
g_{3}+w_{3}^{\ 2}h_{4}+n_{3}^{\ 2}h_{5} & w_{3}h_{4} & n_{3}h_{5} \\
w_{1}h_{4} & w_{2}h_{4} & w_{3}h_{4} & h_{4} & 0 \\
n_{1}h_{5} & n_{2}h_{5} & n_{3}h_{5} & 0 & h_{5}%
\end{array}%
\right] ,  \label{ansatz5}
\end{equation}%
} with the coefficients being some necessary smoothly class functions of
type
\begin{eqnarray}
g_{1} &=&\pm 1,g_{2,3}=g_{2,3}(x^{2},x^{3}),h_{4,5}=h_{4,5}(x^{i},v),  \notag
\\
w_{i} &=&w_{i}(x^{i},v),n_{i}=n_{i}(x^{i},v),  \notag
\end{eqnarray}%
where the $N$--coefficients from (\ref{dder}) and (\ref{ddif}) are
parametrized $N_{i}^{4}=w_{i}$ and $N_{i}^{5}=n_{i}.$

\begin{theorem}
\label{t5dr}The nontrivial components of the 5D Ricci d--tensors (\ref%
{dricci}), $\widehat{\mathbf{R}}_{\alpha \beta }=(\widehat{R}_{ij},\widehat{R%
}_{ia},$ $\widehat{R}_{ai},\widehat{S}_{ab}),$ for the d--metric (\ref%
{block2}) and canonical d--connection $\widehat{\mathbf{\Gamma }}_{\ \alpha
\beta }^{\gamma }$(\ref{candcon}) both defined by the ansatz (\ref{ansatz5}%
), computed with respect to anholonomic frames (\ref{dder}) and (\ref{ddif}%
), consist from h- and v--irreducible components:
\begin{eqnarray}
R_{2}^{2}=R_{3}^{3}=-\frac{1}{2g_{2}g_{3}}[g_{3}^{\bullet \bullet }-\frac{%
g_{2}^{\bullet }g_{3}^{\bullet }}{2g_{2}}-\frac{(g_{3}^{\bullet })^{2}}{%
2g_{3}}+g_{2}^{^{\prime \prime }}-\frac{g_{2}^{^{\prime }}g_{3}^{^{\prime }}%
}{2g_{3}}-\frac{(g_{2}^{^{\prime }})^{2}}{2g_{2}}], &&  \label{ricci1a} \\
S_{4}^{4}=S_{5}^{5}=-\frac{1}{2h_{4}h_{5}}\left[ h_{5}^{\ast \ast
}-h_{5}^{\ast }\left( \ln \sqrt{|h_{4}h_{5}|}\right) ]^{\ast }\right] , &&
\label{ricci2a} \\
R_{4i}=-w_{i}\frac{\beta }{2h_{5}}-\frac{\alpha _{i}}{2h_{5}}, &&
\label{ricci3a} \\
R_{5i}=-\frac{h_{5}}{2h_{4}}\left[ n_{i}^{\ast \ast }+\gamma n_{i}^{\ast }%
\right] , &&  \label{ricci4a}
\end{eqnarray}%
where
\begin{eqnarray}
\alpha _{i} &=&\partial _{i}{h_{5}^{\ast }}-h_{5}^{\ast }\partial _{i}\ln
\sqrt{|h_{4}h_{5}|},\beta =h_{5}^{\ast \ast }-h_{5}^{\ast }[\ln \sqrt{%
|h_{4}h_{5}|}]^{\ast },\gamma =3h_{5}^{\ast }/2h_{5}-h_{4}^{\ast }/h_{4}
\label{abc} \\
h_{4}^{\ast } &\neq &0,\text{ }h_{5}^{\ast }\neq 0\
\mbox{ cases
with
vanishing }\text{ }h_{4}^{\ast }\mbox{ and/or }\text{ }h_{5}^{\ast }%
\mbox{
should be analyzed additionally}.  \notag
\end{eqnarray}
\end{theorem}

The proof of Theorem \ref{t5dr} is given in Appendix \ref{appa}.

We can generalize the ansatz (\ref{ansatz5}) by introducing a conformal
factor $\omega (x^{i},v)$ and additional deformations of the metric via
coefficients $\zeta _{\hat{\imath}}(x^{i},v)$ (here, the indices with 'hat'
take values like $\hat{{i}}=1,2,3,5),$ i. e. for metrics of type
\begin{equation}
\mathbf{g}^{[\omega ]}=\omega ^{2}(x^{i},v)\hat{\mathbf{g}}_{\alpha \beta
}\left( x^{i},v\right) du^{\alpha }\otimes du^{\beta },  \label{cmetric}
\end{equation}%
were the coefficients $\hat{\mathbf{g}}_{\alpha \beta }$ are parametrized by
the ansatz {\scriptsize
\begin{equation}
\left[
\begin{array}{ccccc}
g_{1}+(w_{1}^{\ 2}+\zeta _{1}^{\ 2})h_{4}+n_{1}^{\ 2}h_{5} &
(w_{1}w_{2}+\zeta _{1}\zeta _{2})h_{4}+n_{1}n_{2}h_{5} & (w_{1}w_{3}+\zeta
_{1}\zeta _{3})h_{4}+n_{1}n_{3}h_{5} & (w_{1}+\zeta _{1})h_{4} & n_{1}h_{5}
\\
(w_{1}w_{2}+\zeta _{1}\zeta _{2})h_{4}+n_{1}n_{2}h_{5} & g_{2}+(w_{2}^{\
2}+\zeta _{2}^{\ 2})h_{4}+n_{2}^{\ 2}h_{5} & (w_{2}w_{3}++\zeta _{2}\zeta
_{3})h_{4}+n_{2}n_{3}h_{5} & (w_{2}+\zeta _{2})h_{4} & n_{2}h_{5} \\
(w_{1}w_{3}+\zeta _{1}\zeta _{3})h_{4}+n_{1}n_{3}h_{5} & (w_{2}w_{3}++\zeta
_{2}\zeta _{3})h_{4}+n_{2}n_{3}h_{5} & g_{3}+(w_{3}^{\ 2}+\zeta _{3}^{\
2})h_{4}+n_{3}^{\ 2}h_{5} & (w_{3}+\zeta _{3})h_{4} & n_{3}h_{5} \\
(w_{1}+\zeta _{1})h_{4} & (w_{2}+\zeta _{2})h_{4} & (w_{3}+\zeta _{3})h_{4}
& h_{4} & 0 \\
n_{1}h_{5} & n_{2}h_{5} & n_{3}h_{5} & 0 & h_{5}+\zeta _{5}h_{4}%
\end{array}%
\right] .
\end{equation}%
}\label{ansatzc} Such 5D metrics have a second order anisotropy \cite%
{v2,mhss} when the $N$--coefficients are paramet\-ri\-zed in the first order
anisotropy like $N_{i}^{4}=w_{i}$ and $N_{i}^{5}=n_{i}$ (with three
anholonomic, $x^{i},$ and two anholonomic, $y^{4}$ and $y^{5},$ coordinates)
and in the second order anisotropy (on the second 'shell', \ with four
holonomic, $(x^{i},y^{5}),$ and one anholonomic,$y^{4},$ coordinates) with $%
N_{\hat{{i}}}^{5}=\zeta _{\hat{{i}}},$ in this work we state, for
simplicity, $\zeta _{{5}}=0.$ For trivial values $\omega =1$ and $\zeta _{%
\hat{\imath}}=0,$ the metric (\ref{cmetric}) transforms into (\ref{metric5}).

The Theorem \ref{t5dr} can be extended as to include the ansatz (\ref%
{cmetric}):

\begin{theorem}
\label{t5dra}The nontrivial components of the 5D Ricci d--tensors (\ref%
{dricci}), $\widehat{\mathbf{R}}_{\alpha \beta }=(\widehat{R}_{ij},\widehat{R%
}_{ia},$ $\widehat{R}_{ai},\widehat{S}_{ab}),$ for the metric (\ref{block2})
and canonical d--connection $\widehat{\mathbf{\Gamma }}_{\ \alpha \beta
}^{\gamma }$ (\ref{candcon}) defined by the ansatz (\ref{ansatzc}), computed
with respect to the anholonomic frames (\ref{dder}) and (\ref{ddif}), are
given by the same formulas (\ref{ricci1a})--(\ref{ricci4a}) if there are
satisfied the conditions
\begin{equation}
\hat{{\delta }}_{i}h_{4}=0\mbox{\ and\  }\hat{{\delta }}_{i}\omega =0
\label{conf1}
\end{equation}%
for $\hat{{\delta }}_{i}=\partial _{i}-\left( w_{i}+\zeta _{i}\right)
\partial _{4}+n_{i}\partial _{5}$ when the values $\zeta _{\widetilde{i}%
}=\left( \zeta _{{i}},\zeta _{{5}}=0\right) $ are to be defined as any
solutions of (\ref{conf1}).
\end{theorem}

The proof of Theorem \ref{t5dra} consists from a straightforward calculation
of the components of the Ricci tensor (\ref{dricci}) like in Appendix \ref%
{appa}. The simplest way to do this is to compute the deformations by the
conformal factor of the coefficients of the canonical connection (\ref%
{candcon}) and then to use the calculus for Theorem \ref{t5dr}. \ Such
deformations induce corresponding deformations of the Ricci tensor (\ref%
{dricci}). \ The condition that we have the same values of the Ricci tensor
for the (\ref{ansatz}) and (\ref{ansatzc}) results in equations (\ref{conf1}%
) and (\ref{confeq}) which are compatible, for instance, if for instance, if
\begin{equation}
\omega ^{q_{1}/q_{2}}=h_{4}~(q_{1}\mbox{ and }q_{2}\mbox{ are
integers}),  \label{confq}
\end{equation}%
and $\zeta _{{i}}$ satisfy the equations \
\begin{equation}
\partial _{i}\omega -(w_{i}+\zeta _{{i}})\omega ^{\ast }=0.  \label{confeq}
\end{equation}%
\ There are also different possibilities to satisfy the condition (\ref%
{conf1}). For instance, if $\omega =\omega _{1}$ $\omega _{2},$ we can
consider that $h_{4}=\omega _{1}^{q_{1}/q_{2}}$ $\omega _{2}^{q_{3}/q_{4}}$ $%
\ $for some integers $q_{1},q_{2},q_{3}$ and $q_{4}\blacksquare $

There are some important consequences of the Theorems \ref{t5dr} and \ref%
{t5dra}:

\begin{corollary}
\label{ceint}The non--trivial components of the Einstein tensor [see (\ref%
{deinst}) for the canonical d--connection] $\widehat{\mathbf{G}}_{\ \beta
}^{\alpha }=\widehat{\mathbf{R}}_{\ \beta }^{\alpha }-\frac{1}{2}%
\overleftarrow{\mathbf{\hat{R}}}\delta _{\beta }^{\alpha }$ for the ansatz (%
\ref{ansatz5}) and (\ref{ansatzc}) given with respect to the N--adapted (co)
frames are
\begin{equation}
G_{1}^{1}=-\left( R_{2}^{2}+S_{4}^{4}\right)
,G_{2}^{2}=G_{3}^{3}=-S_{4}^{4},G_{4}^{4}=G_{5}^{5}=-R_{2}^{2}.
\label{einstdiag}
\end{equation}
\end{corollary}

The relations (\ref{einstdiag}) can be derived following the formulas for
the Ricci tensor (\ref{ricci1a})--(\ref{ricci4a}). They impose the condition
that the dynamics of such gravitational fields is defined by two independent
components $R_{2}^{2}$ and $S_{4}^{4}$ and result in

\begin{corollary}
\label{cors}The system of effective 5D Einstein--Proca equations on spaces
provided with N--connection structure (\ref{efeinst}) (equivalently, the
system (\ref{ep1})--(\ref{ep4})\_is compatible for the generic off--diagonal
ansatz (\ref{ansatz5}) and (\ref{ansatzc}) if the energy--momentum tensor $%
\mathbf{\Upsilon }_{\alpha \beta }=\tilde{\kappa}(\mathbf{\Sigma }_{\alpha
\beta }^{[\phi ]}+\mathbf{\Sigma }_{\alpha \beta }^{[\mathbf{m}]})$ of the
Proca and matter fields given with respect to N-- frames is diagonal and
satisfies the conditions
\begin{equation}
\Upsilon _{2}^{2}=\Upsilon _{3}^{3}=\Upsilon _{2}(x^{2},x^{3},v),\ \Upsilon
_{4}^{4}=\Upsilon _{5}^{5}=\Upsilon _{4}(x^{2},x^{3}),\mbox{ and }\Upsilon
_{1}=\Upsilon _{2}+\Upsilon _{4}.  \label{emcond}
\end{equation}
\end{corollary}

\begin{remark}
\label{remconds}Instead of the energy--momentum tensor $\mathbf{\Upsilon }%
_{\alpha \beta }=\tilde{\kappa}(\mathbf{\Sigma }_{\alpha \beta }^{[\phi ]}+%
\mathbf{\Sigma }_{\alpha \beta }^{[\mathbf{m}]})$ for the Proca and matter
fields we can consider any source, for instance, with string corrections,
when $\mathbf{\Upsilon }_{\alpha \beta }^{[str]}=\tilde{\kappa}\left(
\mathbf{\Sigma }_{\alpha \beta }^{[\phi ]}+\mathbf{\Sigma }_{\alpha \beta
}^{[\mathbf{m}]}+\mathbf{\Sigma }_{\alpha \beta }^{[\mathbf{T}]}\right) $
like in (\ref{fagfe}) satisfying the conditions (\ref{emcond}).
\end{remark}

If the conditions of the Corollary \ref{cors}, or Remark \ref{remconds}, are
satisfied, the h- and v-- irreducible components of the 5D Einstein--Proca
equations (\ref{ep1}) and (\ref{ep4}), or of the string gravity equations (%
\ref{fagfe}), for the ansatz (\ref{ansatz5}) and (\ref{ansatzc}) transform
into the system%
\begin{eqnarray}
R_{2}^{2} &=&R_{3}^{3}=-\frac{1}{2g_{2}g_{3}}[g_{3}^{\bullet \bullet }-\frac{%
g_{2}^{\bullet }g_{3}^{\bullet }}{2g_{2}}-\frac{(g_{3}^{\bullet })^{2}}{%
2g_{3}}+g_{2}^{^{\prime \prime }}-\frac{g_{2}^{^{\prime }}g_{3}^{^{\prime }}%
}{2g_{3}}-\frac{(g_{2}^{^{\prime }})^{2}}{2g_{2}}]=-\Upsilon
_{4}(x^{2},x^{3}),  \label{ep1a} \\
S_{4}^{4} &=&S_{5}^{5}=-\frac{1}{2h_{4}h_{5}}\left[ h_{5}^{\ast \ast
}-h_{5}^{\ast }\left( \ln \sqrt{|h_{4}h_{5}|}\right) ^{\ast }]\right]
=-\Upsilon _{2}(x^{2},x^{3},v).  \label{ep2a} \\
R_{4i} &=&-w_{i}\frac{\beta }{2h_{5}}-\frac{\alpha _{i}}{2h_{5}}=0,
\label{ep3a} \\
R_{5i} &=&-\frac{h_{5}}{2h_{4}}\left[ n_{i}^{\ast \ast }+\gamma n_{i}^{\ast }%
\right] =0.  \label{ep4a}
\end{eqnarray}

A very surprising result is that we are able to construct exact solutions of
the 5D Einstein--Proca equations with anholonomic variables and generic
off--diagonal metrics:

\begin{theorem}
\label{texs}The system of second order nonlinear partial differential
equations (\ref{ep1a})--(\ref{ep4a}) and (\ref{confeq}) can be solved in
general form if there are given certain values of functions $%
g_{2}(x^{2},x^{3})$ (or, inversely, $g_{3}(x^{2},x^{3})),\ h_{4}\left(
x^{i},v\right) $ (or, inversely, $h_{5}\left( x^{i},v\right) ),$ $\omega
\left( x^{i},v\right) $ and of sources $\Upsilon _{2}(x^{2},x^{3},v)$ and $%
\Upsilon _{4}(x^{2},x^{3}).$
\end{theorem}

We outline the main steps of constructing exact solutions and proving this
Theorem.

\begin{itemize}
\item The general solution of equation (\ref{ep1a}) can be written in the
form
\begin{equation}
\varpi =g_{[0]}\exp [a_{2}\widetilde{x}^{2}\left( x^{2},x^{3}\right) +a_{3}%
\widetilde{x}^{3}\left( x^{2},x^{3}\right) ],  \label{solricci1a}
\end{equation}%
were $g_{[0]},a_{2}$ and $a_{3}$ are some constants and the functions $%
\widetilde{x}^{2,3}\left( x^{2},x^{3}\right) $ define any coordinate
transforms $x^{2,3}\rightarrow \widetilde{x}^{2,3}$ for which the 2D line
element becomes conformally flat, i. e.
\begin{equation}
g_{2}(x^{2},x^{3})(dx^{2})^{2}+g_{3}(x^{2},x^{3})(dx^{3})^{2}\rightarrow
\varpi (x^{2},x^{3})\left[ (d\widetilde{x}^{2})^{2}+\epsilon (d\widetilde{x}%
^{3})^{2}\right] ,  \label{con10}
\end{equation}%
where $\epsilon =\pm 1$ for a corresponding signature. In coordinates $%
\widetilde{x}^{2,3},$ the equation (\ref{ep1a}) transform into%
\begin{equation*}
\varpi \left( \ddot{\varpi}+\varpi ^{\prime \prime }\right) -\dot{\varpi}%
-\varpi ^{\prime }=2\varpi ^{2}\Upsilon _{4}(\tilde{x}^{2},\tilde{x}^{3})
\end{equation*}%
or%
\begin{equation}
\ddot{\psi}+\psi ^{\prime \prime }=2\Upsilon _{4}(\tilde{x}^{2},\tilde{x}%
^{3}),  \label{auxeq01}
\end{equation}%
for $\psi =\ln |\varpi |.$ The integrals of (\ref{auxeq01}) depends on the
source $\Upsilon _{4}.$ As a particular case we can consider that $\Upsilon
_{4}=0.$ There are three alternative possibilities to generate solutions of (%
\ref{ep1a}). For instance, we can prescribe that $g_{2}=g_{3}$ and get the
equation (\ref{auxeq01}) for $\psi =\ln |g_{2}|=\ln |g_{3}|.$ If we suppose
that $g_{2}^{^{\prime }}=0,$ for a given $g_{2}(x^{2}),$ we obtain from (\ref%
{ep1a})%
\begin{equation*}
g_{3}^{\bullet \bullet }-\frac{g_{2}^{\bullet }g_{3}^{\bullet }}{2g_{2}}-%
\frac{(g_{3}^{\bullet })^{2}}{2g_{3}}=2g_{2}g_{3}\Upsilon _{4}(x^{2},x^{3})
\end{equation*}%
which can be integrated explicitly for given values of $\Upsilon _{4}.$
Similarly, we can generate solutions for a prescribed $g_{3}(x^{3})$ in the
equation
\begin{equation*}
g_{2}^{^{\prime \prime }}-\frac{g_{2}^{^{\prime }}g_{3}^{^{\prime }}}{2g_{3}}%
-\frac{(g_{2}^{^{\prime }})^{2}}{2g_{2}}=2g_{2}g_{3}\Upsilon
_{4}(x^{2},x^{3}).
\end{equation*}%
We note that a transform (\ref{con10}) is always possible for 2D metrics and
the explicit form of solutions depends on chosen system of 2D coordinates
and on the signature $\epsilon =\pm 1.$ In the simplest case with $\Upsilon
_{4}=0$ the equation (\ref{ep1a}) is solved by arbitrary two functions $%
g_{2}(x^{3})$ and $g_{3}(x^{2}).$

\item For $\Upsilon _{2}=0,$ the equation (\ref{ep2a}) relates two functions
$h_{4}\left( x^{i},v\right) $ and $h_{5}\left( x^{i},v\right) $ following
two possibilities:

a) to compute
\begin{eqnarray}
\sqrt{|h_{5}|} &=&h_{5[1]}\left( x^{i}\right) +h_{5[2]}\left( x^{i}\right)
\int \sqrt{|h_{4}\left( x^{i},v\right) |}dv,~h_{4}^{\ast }\left(
x^{i},v\right) \neq 0;  \notag \\
&=&h_{5[1]}\left( x^{i}\right) +h_{5[2]}\left( x^{i}\right) v,\ h_{4}^{\ast
}\left( x^{i},v\right) =0,  \label{p2}
\end{eqnarray}%
for some functions $h_{5[1,2]}\left( x^{i}\right) $ stated by boundary
conditions;

b) or, inversely, to compute $h_{4}$ for a given $h_{5}\left( x^{i},v\right)
,h_{5}^{\ast }\neq 0,$%
\begin{equation}
\sqrt{|h_{4}|}=h_{[0]}\left( x^{i}\right) (\sqrt{|h_{5}\left( x^{i},v\right)
|})^{\ast },  \label{p1}
\end{equation}%
with $h_{[0]}\left( x^{i}\right) $ given by boundary conditions. We note
that the sourceless equation (\ref{ep2a}) is satisfied by arbitrary pairs of
coefficients $h_{4}\left( x^{i},v\right) $ and $h_{5[0]}\left( x^{i}\right)
. $ Solutions with $\Upsilon _{2}\neq 0$ can be found by ansatz of type
\begin{equation}
h_{5}[\Upsilon _{2}]=h_{5},h_{4}[\Upsilon _{2}]=\varsigma _{4}\left(
x^{i},v\right) h_{4},  \label{auxf02}
\end{equation}%
where $h_{4}$ and $h_{5}$ are related by formula (\ref{p2}), or (\ref{p1}).
Substituting (\ref{auxf02}), we obtain%
\begin{equation}
\varsigma _{4}\left( x^{i},v\right) =\varsigma _{4[0]}\left( x^{i}\right)
-\int \Upsilon _{2}(x^{2},x^{3},v)\frac{h_{4}h_{5}}{4h_{5}^{\ast }}dv,
\label{auxf02a}
\end{equation}%
where $\varsigma _{4[0]}\left( x^{i}\right) $ are arbitrary functions.

\item The exact solutions of (\ref{ep3a}) for $\beta \neq 0$ are defined
from an algebraic equation, $w_{i}\beta +\alpha _{i}=0,$ where the
coefficients $\beta $ and $\alpha _{i}$ are computed as in formulas (\ref%
{abc}) by using the solutions for (\ref{ep1a}) and (\ref{ep2a}). The general
solution is
\begin{equation}
w_{k}=\partial _{k}\ln [\sqrt{|h_{4}h_{5}|}/|h_{5}^{\ast }|]/\partial
_{v}\ln [\sqrt{|h_{4}h_{5}|}/|h_{5}^{\ast }|],  \label{w}
\end{equation}%
with $\partial _{v}=\partial /\partial v$ and $h_{5}^{\ast }\neq 0.$ If $%
h_{5}^{\ast }=0,$ or even $h_{5}^{\ast }\neq 0$ but $\beta =0,$ the
coefficients $w_{k}$ could be arbitrary functions on $\left( x^{i},v\right)
. $ \ For the vacuum Einstein equations this is a degenerated case imposing
the the compatibility conditions $\beta =\alpha _{i}=0,$ which are
satisfied, for instance, if the $h_{4}$ and $h_{5}$ are related as in the
formula (\ref{p1}) but with $h_{[0]}\left( x^{i}\right) =const.$

\item Having defined $h_{4}$ and $h_{5}$ and computed $\gamma $ from (\ref%
{abc}) we can solve the equation (\ref{ep4a}) by integrating on variable ''$%
v $'' the equation $n_{i}^{\ast \ast }+\gamma n_{i}^{\ast }=0.$ The exact
solution is
\begin{eqnarray}
n_{k} &=&n_{k[1]}\left( x^{i}\right) +n_{k[2]}\left( x^{i}\right) \int
[h_{4}/(\sqrt{|h_{5}|})^{3}]dv,~h_{5}^{\ast }\neq 0;  \notag \\
&=&n_{k[1]}\left( x^{i}\right) +n_{k[2]}\left( x^{i}\right) \int
h_{4}dv,\qquad ~h_{5}^{\ast }=0;  \label{n} \\
&=&n_{k[1]}\left( x^{i}\right) +n_{k[2]}\left( x^{i}\right) \int [1/(\sqrt{%
|h_{5}|})^{3}]dv,~h_{4}^{\ast }=0,  \notag
\end{eqnarray}%
for some functions $n_{k[1,2]}\left( x^{i}\right) $ stated by boundary
conditions.
\end{itemize}

The exact solution of (\ref{confeq}) is given by some arbitrary functions $%
\zeta _{i}=\zeta _{i}\left( x^{i},v\right) $ if \ both $\partial _{i}\omega
=0$ and $\omega ^{\ast }=0,$ we chose $\zeta _{i}=0$ for $\omega =const,$
and
\begin{eqnarray}
\zeta _{i} &=&-w_{i}+(\omega ^{\ast })^{-1}\partial _{i}\omega ,\quad \omega
^{\ast }\neq 0,  \label{confsol} \\
&=&(\omega ^{\ast })^{-1}\partial _{i}\omega ,\quad \omega ^{\ast }\neq 0,%
\mbox{ for vacuum solutions}.  \notag
\end{eqnarray}%
\bigskip

The Theorem \ref{texs} states a general method of constructing exact
solutions in MAG, of the Einstein--Proca equations and various string
gravity generalizations with generic off--diagonal metrics. Such solutions
are with associated N--connection structure. This method can be also applied
in order to generate, for instance, certain Finsler or Lagrange
configurations as v-irreducible components. The 5D ansatz can not be used to
generate standard Finsler or Lagrange geometries because the dimension of
such spaces can not be an odd number. Nevertheless, the anholonomic frame
method can be applied in order to generate 4D exact solutions containing
Finsler--Lagrange configurations, see Appendix \ref{ss4d}.

\ Summarizing the results for the nondegenerated cases when $h_{4}^{\ast
}\neq 0$ and $h_{5}^{\ast }\neq 0$ and (for simplicity, for a trivial
conformal factor $\omega ),$ we derive an explicit result for 5D exact
solutions with local coordinates $u^{\alpha }=\left( x^{i},y^{a}\right) $
when $x^{i}=\left( x^{1},x^{\widehat{i}}\right) ,x^{\widehat{i}}=\left(
x^{2},x^{3}\right) ,y^{a}=\left( y^{4}=v,y^{a}\right) $ and arbitrary
signatures $\epsilon _{\alpha }=\left( \epsilon _{1},\epsilon _{2},\epsilon
_{3},\epsilon _{4},\epsilon _{5}\right) $ (where $\epsilon _{\alpha }=\pm
1): $

\begin{corollary}
\label{corgsol1}Any off--diagonal metric
\begin{eqnarray}
\delta s^{2} &=&\epsilon _{1}(dx^{1})^{2}+\epsilon _{\widehat{k}}g_{\widehat{%
k}}\left( x^{\widehat{i}}\right) (dx^{\widehat{k}})^{2}+  \notag \\
&&\epsilon _{4}h_{0}^{2}(x^{i})\left[ f^{\ast }\left( x^{i},v\right) \right]
^{2}|\varsigma _{\Upsilon }\left( x^{i},v\right) |\left( \delta v\right)
^{2}+\epsilon _{5}f^{2}\left( x^{i},v\right) \left( \delta y^{5}\right) ^{2},
\notag \\
\delta v &=&dv+w_{k}\left( x^{i},v\right) dx^{k},\ \delta
y^{5}=dy^{5}+n_{k}\left( x^{i},v\right) dx^{k},  \label{gensol1}
\end{eqnarray}%
with coefficients of necessary smooth class, where\ \ $g_{\widehat{k}}\left(
x^{\widehat{i}}\right) $ is a solution of the 2D equation (\ref{ep1a}) for a
given source $\Upsilon _{4}\left( x^{\widehat{i}}\right) ,$%
\begin{equation*}
\varsigma _{\Upsilon }\left( x^{i},v\right) =\varsigma _{4}\left(
x^{i},v\right) =\varsigma _{4[0]}\left( x^{i}\right) -\frac{\epsilon _{4}}{16%
}h_{0}^{2}(x^{i})\int \Upsilon _{2}(x^{\widehat{k}},v)[f^{2}\left(
x^{i},v\right) ]^{2}dv,
\end{equation*}%
and the N--connection coefficients $N_{i}^{4}=w_{i}(x^{k},v)$ and $%
N_{i}^{5}=n_{i}(x^{k},v)$ are
\begin{equation}
w_{i}=-\frac{\partial _{i}\varsigma _{\Upsilon }\left( x^{k},v\right) }{%
\varsigma _{\Upsilon }^{\ast }\left( x^{k},v\right) }  \label{gensol1w}
\end{equation}%
and
\begin{equation}
n_{k}=n_{k[1]}\left( x^{i}\right) +n_{k[2]}\left( x^{i}\right) \int \frac{%
\left[ f^{\ast }\left( x^{i},v\right) \right] ^{2}}{\left[ f\left(
x^{i},v\right) \right] ^{2}}\varsigma _{\Upsilon }\left( x^{i},v\right) dv,
\label{gensol1n}
\end{equation}%
define an exact solution of the system of Einstein equations with holonomic
and anholonomic variables (\ref{ep1a})--(\ref{ep4a}) for arbitrary
nontrivial functions $f\left( x^{i},v\right) $ (with $f^{\ast }\neq 0),$ $%
h_{0}^{2}(x^{i})$, $\varsigma _{4[0]}\left( x^{i}\right) ,n_{k[1]}\left(
x^{i}\right) $ and $\ n_{k[2]}\left( x^{i}\right) ,$ and sources $\Upsilon
_{2}(x^{\widehat{k}},v),\Upsilon _{4}\left( x^{\widehat{i}}\right) $ and any
integration constants and signatures $\epsilon _{\alpha }=\pm 1$ to be
defined by certain boundary conditions and physical considerations.
\end{corollary}

Any metric (\ref{gensol1}) with $h_{4}^{\ast }\neq 0$ and $h_{5}^{\ast }\neq
0$ has the property to be generated by a function of four variables $f\left(
x^{i},v\right) $ with emphasized dependence on the anisotropic coordinate $%
v, $ because $f^{\ast }\doteqdot \partial _{v}f\neq 0$ and by arbitrary
sources $\Upsilon _{2}(x^{\widehat{k}},v),\Upsilon _{4}\left( x^{\widehat{i}%
}\right) .$ The rest of arbitrary functions not depending on $v$ have been
obtained in result of integradion of partial differential equations. This
fix a specific class of metrics generated by using the relation (\ref{p1})
and the first formula in (\ref{n}). We can generate also a different class
of solutions with $h_{4}^{\ast }=0$ by considering the second formula in (%
\ref{p2}) and respective formulas in (\ref{n}). The ''degenerated'' cases
with $h_{4}^{\ast }=0$ but $h_{5}^{\ast }\neq 0$ and inversely, $h_{4}^{\ast
}\neq 0$ but $h_{5}^{\ast }=0$ are more special and request a proper
explicit construction of solutions. Nevertheless, such type of solutions are
also generic off--diagonal and they could be of substantial interest.

The sourceless case with vanishing $\Upsilon _{2}$ and $\Upsilon _{4}$ is
defined following

\begin{remark}
Any off--diagonal metric (\ref{gensol1}) with $\varsigma _{\Upsilon }=1,$ $%
h_{0}^{2}(x^{i})=$ $h_{0}^{2}=const,$ $w_{i}=0$ and $n_{k}$ computed as in (%
\ref{gensol1n}) but for $\varsigma _{\Upsilon }=1,$ defines a vacuum
solution of 5D Einstein equations for the canonical d--connection (\ref%
{candcon}) computed for the ansatz (\ref{gensol1}).
\end{remark}

By imposing additional constraints on arbitrary functions from $%
N_{i}^{5}=n_{i}$ and $N_{i}^{5}=w_{i},$ we can select off--diagonal
gravitational configurations with distorsions of the Levi--Civita connection
resulting in canonical d--connections with the same solutions of the vacuum
Einstein equations. For instance, we can model Finsler like geometries in
general relativity, see Corollary \ref{corcond1}. Under similar conditions
the ansatz (\ref{ansatz5}) was used for constructing exact off--diagonal
solutions in the 5D Einstein gravity, see Refs. \cite{v1,v1a,v2}.

Let us consider the procedure of selecting solutions with off--diagonal
metrics from an ansatz (\ref{gensol1}) with trivial N--connection curvature
(such metrics consists a simplest subclass which can be restricted to
(pseudo) Riemannian ones). The corresponding nontrivial coefficients the
N--connection curvature \ (\ref{ncurv}) are computed
\begin{equation*}
\Omega _{ij}^{4}=\partial _{i}w_{j}-\partial _{j}w_{i}+w_{i}w_{j}^{\ast
}-w_{j}w_{i}^{\ast }\mbox{ and }\Omega _{ij}^{5}=\partial _{i}n_{j}-\partial
_{j}n_{i}+w_{i}n_{j}^{\ast }-w_{j}n_{i}^{\ast }.
\end{equation*}%
So, there are imposed six constraints, $\Omega _{ij}^{4}=\Omega _{ij}^{5}=0,$
for $i,j...=1,2,4$ on six functions $w_{i}$ and $n_{i}$ computed
respectively as (\ref{gensol1n}) and (\ref{gensol1n}) which can be satisfied
by a corresponding subclass of functions $f\left( x^{i},v\right) $ (with $%
f^{\ast }\neq 0),$ $h_{0}^{2}(x^{i}),$ $\varsigma _{4[0]}\left( x^{i}\right)
,n_{k[1]}\left( x^{i}\right) ,$ $\ n_{k[2]}\left( x^{i}\right) $ and $%
\Upsilon _{2}(x^{\widehat{k}},v),\Upsilon _{4}\left( x^{\widehat{i}}\right) $
(in general, we have to solve certain first order partial derivative
equations with may be reduced to algebraic relations by corresponding
parametrizations). For instance, in the vacuum case when $w_{j}=0,$ we
obtain $\Omega _{ij}^{5}=\partial _{i}n_{j}-\partial _{j}n_{i}.$ The
simplest example when condition $\Omega _{\widehat{i}\widehat{j}%
}^{5}=\partial _{\widehat{i}}n_{\widehat{j}}-\partial _{\widehat{j}}n_{%
\widehat{i}}=0,$ with $\widehat{i},\widehat{j}=2,3$ (reducing the metric (%
\ref{gensol1}) to a 4D one trivially embedded into 5D) is satisfied is to
take $n_{3[1]}=n_{3[2]}=0$ in (\ref{gensol1n}) and consider that $f=f\left(
x^{2},v\right) $ with $n_{2[1]}=n_{2[1]}\left( x^{2}\right) $ and $%
n_{2[2]}=n_{2[2]}\left( x^{2}\right) ,$ i. e. by eliminating the dependence
of the coefficients on $x^{3}.$ This also results in a generic off--diagonal
solution, because the anholonomy coefficients (\ref{anhc}) are not trivial,
for instance, $w_{24}^{5}=n_{2}^{\ast }$ and $w_{14}^{5}=n_{1}^{\ast }.$

Another interesting remark is that even we have reduced the canonical
d--connection to the Levi--Civita one [with respect to N--adapted (co)
frames; this imposes the metric to be (pseudo) Riemannian] by selecting the
arbitrary functions as to have $\Omega _{ij}^{a}=0,$ one could be
nonvanishing d--torsion components like $T_{41}^{5}=P_{41}^{5}$ and $%
T_{41}^{5}=P_{41}^{5}$ in (\ref{dtorsb}). Such objects, as well the
anholonomy coefficients $w_{24}^{5}$ and $w_{14}^{5}$ (which can be also
considered as torsion like objects) are constructed by taking certain
''scarps'' from the coefficients of off--diagonal metrics and anholonomic
frames. They are induced by the frame anholonomy (like ''torsions'' in
rotating anholonomic systems of reference for the Newton gravtity and
mechanics with constraints) and vanish if we transfer the constructions with
respect to any holonomic basis.

The above presented results are for generic 5D off--diagonal metrics,
anholonomic transforms and nonlinear field equations. Reductions to a lower
dimensional theory are not trivial in such cases. We emphasize some specific
points of this procedure in the Appendix \ref{ss4d} (see details in \cite%
{vmethod}).

\section{Exact Solutions}

\label{exsol} There were found a set of exact solutions in MAG \cite%
{esolmag,obet2,oveh} describing various configuration of Einstein--Maxwell
of dilaton gravity emerging from low energy string theory, soliton and
multipole solutions and generalized Plebanski--Demianski solutions,
colliding waves and static black hole metrics. In this section we are going
to look for some classes of 4D and 5D solutions of the Einstein--Proca
equations in MAG related to string gravity modeling generalized
Finsler--affine geometries and extending to such spacetimes some our
previous results \cite{v1,v1a,v2}.

\subsection{Finsler--Lagrange metrics in string and metric--affine gravity}

\label{solmlrc}

As we discussed in section 2, the generalized Finsler--Lagrange spaces can
be modeled in metric--affine spacetimes provided with N--connection
structure. In this subsection, we show how such two dimensional Finsler like
spaces with d--metrics depending on one anisotropic coordinate $y^{4}=v$
(denoted as $\mathbf{F}^{2}=\left[ V^{2},F\left( x^{2},x^{3},y\right) \right]
,$ $\mathbf{L}^{2}=\left[ V^{2},L\left( x^{2},x^{3},y\right) \right] $ and $%
\mathbf{GL}^{2}=\left[ V^{2},g_{ij}\left( x^{2},x^{3},y\right) \right] $
according to Ref. \cite{vp1}) can be modeled by corresponding diad
transforms on spacetimes with 5D (or 4D) d--metrics being exact solutions of
the field equations for the generalized Finsler--affine string gravity (\ref%
{fagfe}) (as a particular case we can consider the Einstein--Proca system (%
\ref{ep1a})--(\ref{ep4a}) and (\ref{confeq})). For every particular case of
locally anisotropic spacetime, for instance, outlined in Appendix C, see
Table \ref{tablegs}, the quadratic form $\tilde{g}_{ij},$ d--metric $\mathbf{%
\tilde{g}}_{\alpha \beta }=\left[ \tilde{g}_{ij},\tilde{g}_{ij}\right] $ and
N--connection $\tilde{N}_{j}^{a}$ one holds

\begin{theorem}
Any 2D locally anisotropic structure given by \ $\mathbf{\tilde{g}}_{\alpha
\beta }$ and $\tilde{N}_{j}^{a}$ can be modeled on the space of exact
solutions of the 5D (or 4D) the generalized Finsler--affine string gravity
system defined by the ansatz (\ref{ansatzc}) (or (\ref{ansatzc4})).
\end{theorem}

We give the proof via an explicit construction. Let us consider
\begin{equation*}
\mathbf{g}_{\alpha \beta }=[g_{ij},h_{ab}]=\left[ \omega g_{2}\left(
x^{2},x^{3}\right) ,\omega g_{3}\left( x^{2},x^{3}\right) ,\omega
h_{4}\left( x^{2},x^{3},v\right) ,\omega h_{5}\left( x^{2},x^{3},v\right) %
\right]
\end{equation*}%
for $\omega =\omega \left( x^{2},x^{3},v\right) $ and
\begin{equation*}
N_{i}^{a}=\left[ N_{i}^{4}=w_{i}\left( x^{2},x^{3},v\right)
,N_{i}^{5}=n_{i}\left( x^{2},x^{3},v\right) \right] ,
\end{equation*}%
where indices are running the values $a=4,5$ and $i=2,3$ define an exact 4D
solution of the equations (\ref{fagfe}) (or, in the particular case, of the
system (\ref{ep1a})--(\ref{ep4a}), for simplicity, we put $\omega \left(
x^{2},x^{3},v\right) =1).$ We can relate the data $\left( \mathbf{g}_{\alpha
\beta },N_{i}^{a}\right) $ to any data $\left( \mathbf{\tilde{g}}_{\alpha
\beta },\tilde{N}_{j}^{a}\right) $ via nondegenerate diadic transforms $%
e_{i}^{i^{\prime }}=e_{i}^{i^{\prime }}\left( x^{2},x^{3},v\right) ,\
l_{a}^{i^{\prime }}=l_{a}^{i^{\prime }}\left( x^{2},x^{3},v\right) $ and $%
q_{a}^{i^{\prime }}=q_{a}^{i^{\prime }}\left( x^{2},x^{3},v\right) $ (and
theirs inverse matrices)%
\begin{equation}
g_{ij}=e_{i}^{i^{\prime }}e_{j}^{j^{\prime }}\tilde{g}_{i^{\prime }j^{\prime
}},\ h_{ab}=l_{a}^{i^{\prime }}l_{b}^{i^{\prime }}\tilde{g}_{i^{\prime
}j^{\prime }},N_{i^{\prime }}^{a}=q_{a^{\prime }}^{a}\tilde{N}_{j^{\prime
}}^{a^{\prime }}\ .  \label{aux03}
\end{equation}%
Such transforms may be associated to certain tetradic transforms of the
N--elongated (co) frames ((\ref{ddif})) (\ref{dder}). If for the given data $%
\left( \mathbf{g}_{\alpha \beta },N_{i}^{a}\right) $ and $\left( \mathbf{%
\tilde{g}}_{\alpha \beta },\tilde{N}_{j}^{a}\right) $ in (\ref{aux03}), we
can solve the corresponding systems of quadratic algebraic equations \ and
define nondegenerate matrices $\left( e_{i}^{i^{\prime }}\right) ,\left(
l_{a}^{i^{\prime }}\right) $ and $\left( q_{i^{\prime }}^{a}\right) ,$ we
argue that the 2D locally anisotropic spacetime $\left( \mathbf{\tilde{g}}%
_{\alpha \beta },\tilde{N}_{j}^{a}\right) $ (really, it is a 4D spacetime
with generic off--diagonal metric and associated N--connection structure)
can be modeled on by a class of exact solutions of effective Einstein--Proca
equations for MAG.$\blacksquare $

The d--metric with respect to transformed N--adapted diads is \ written in
the form
\begin{equation}
\mathbf{g=}\tilde{g}_{i^{\prime }j^{\prime }}\mathbf{e}^{i^{\prime }}\otimes
\mathbf{e}^{j^{\prime }}+\tilde{g}_{i^{\prime }j^{\prime }}\mathbf{\tilde{e}}%
^{i^{\prime }}\otimes \mathbf{\tilde{e}}^{j^{\prime }}  \label{blok2b}
\end{equation}%
where
\begin{equation*}
\mathbf{e}^{i^{\prime }}=e_{i}^{i^{\prime }}dx^{i},\ \mathbf{\tilde{e}}%
^{i^{\prime }}=l_{a}^{i^{\prime }}\mathbf{\tilde{e}}^{a},\ \mathbf{\tilde{e}}%
^{a}=dy^{a}+\tilde{N}_{j^{\prime }}^{a}\mathbf{\tilde{e}}_{[N]}^{j^{\prime
}},\ \mathbf{\tilde{e}}_{[N]}^{j^{\prime }}=q_{i}^{j^{\prime }}dx^{i}.
\end{equation*}%
The d--metric (\ref{blok2b}) has the coefficients corresponding to
generalized Finsler--Lagange spaces and emphasizes that any quadratic form $%
\tilde{g}_{i^{\prime }j^{\prime }}$ from Table \ref{tablegs} can be related
via an exact solution $\left( g_{ij},h_{ab},N_{i^{\prime }}^{a}\right) .$

We note that we can define particular cases of imbeddings with $%
h_{ab}=l_{a}^{i^{\prime }}l_{b}^{i^{\prime }}\tilde{g}_{i^{\prime }j^{\prime
}}$ and $N_{j^{\prime }}^{a}=q_{i^{\prime }}^{a}\tilde{N}_{j^{\prime
}}^{i^{\prime }}$ for a prescribed value of $g_{ij}=\tilde{g}_{i^{\prime
}j^{\prime }}$ and try to model only the quadratic form $\tilde{h}%
_{i^{\prime }j^{\prime }}$ in MAG. Similar considerations were presented for
particular cases of modeling Finsler structures and generalizations in
Einstein and Einstein--Cartan spaces \cite{v1,v2}, see the conditions (\ref%
{cond2}).

\subsection{Solutions in MAG with effective variable cosmological constant}

A class of 4D solutions in MAG with local anisotropy can be derived from (%
\ref{efeinst}) $\ $\ for $\mathbf{\Sigma }_{\alpha \beta }^{[\mathbf{m}]}=0$
and almost vanishing mass $\mu \rightarrow 0$ of the Proca field in the
source $\mathbf{\Sigma }_{\alpha \beta }^{[\mathbf{\phi }]}.$ This holds in
absence of matter fields and when the constant in the action for the
Finsler--affine gravity are subjected to the condition (\ref{procvan}). We
consider that $\phi _{\mu }=\left( \phi _{\widehat{i}}\left( x^{\widehat{k}%
}\right) ,\phi _{a}=0\right) ,$ where $\widehat{i},\widehat{k},...=2,3$ and $%
a,b,...=4,5,$ with respect to a N--adapted coframe (\ref{ddif}) and choose a
metric ansatz of type (\ref{ansatz4}) with $g_{2}=1$ and $g_{3}=-1$ which
select a flat h--subspace imbedded into a general anholonomic 4D background
with nontrivial $h_{ab}$ and N--connection structure $N_{i}^{a}.$ The
h--covariant derivatives are $\widehat{D}^{[h]}$ $\phi _{\widehat{i}}=\left(
\partial _{2}\phi _{\widehat{i}},\partial _{3}\phi _{\widehat{i}}\right) $
because the coefficients $\widehat{L}_{\ jk}^{i}$ and $\widehat{C}_{\
ja}^{i} $ are zero in (\ref{candcon}) and any contraction with $\phi _{a}=0$
results in zero values. In this case the Proca equations, $\widehat{\mathbf{D%
}}_{\nu }\mathbf{H}^{\nu \mu }=\mu ^{2}\mathbf{\phi }^{\mu },$ transform in
a Maxwell like equation,%
\begin{equation}
\partial _{2}(\partial _{2}\phi _{\widehat{i}})-\partial _{3}(\partial
_{3}\phi _{\widehat{i}})=0,  \label{waveeq}
\end{equation}%
for the potential $\phi _{\widehat{i}},$ with the dynamics in the
h--subspace distinguished by a N--connection structure to be defined latter.
We note that $\phi _{i}$ is not an electromagnetic field, but a component of
the metric--affine gravity related to nonmetricity and torsion. The relation
$\mathbf{Q}=k_{0}\mathbf{\phi ,\ \Lambda =}k_{1}\mathbf{\phi ,\ T=}k_{2}%
\mathbf{\phi }$ from (\ref{triplemag}) transforms into $Q_{\widehat{i}%
}=k_{0}\phi _{\widehat{i}},\Lambda _{\widehat{i}}=k_{1}\phi _{\widehat{i}%
},T_{i}=k_{2}\phi _{\widehat{i}},$ and vanishing $Q_{a},\Lambda _{a}$ and $%
T_{a},$ defined, for instance, by a wave solution of (\ref{waveeq}),
\begin{equation}
\phi _{\widehat{i}}=\phi _{\lbrack 0]\widehat{i}}\cos \left( \varrho
_{i}x^{i}+\varphi _{\lbrack 0]}\right)  \label{wavesoltf}
\end{equation}%
for any constants $\phi _{\lbrack 0]2,3},$ $\varphi _{\lbrack 0]}$ and $%
\left( \varrho _{2}\right) ^{2}-\left( \varrho _{3}\right) ^{2}=0.$ In this
simplified model we have related plane waves of nonmetricity and torsion
propagating on an anholonomic background provided with N--connection. Such
nonmetricity and torsion do not vanish even $\mu \rightarrow 0$ and the
Proca field is approximated by a massless vector field defined in the
h--subspace.

The energy--momentum tensor $\mathbf{\Sigma }_{\alpha \beta }^{[\mathbf{\phi
}]}$ for the massless field (\ref{wavesoltf}) is defined by a nontrivial
value%
\begin{equation*}
H_{23}=\partial _{2}\phi _{3}-\partial _{3}\phi _{2}=\varepsilon
_{23}\lambda _{\lbrack h]}\sin \left( \varrho _{i}x^{i}+\varphi _{\lbrack
0]}\right)
\end{equation*}%
with antisymmetric $\varepsilon _{23},\varepsilon _{23}=1,$ and constant $%
\lambda _{\lbrack h]}$ taken for a normalization $\varepsilon _{23}\lambda
_{\lbrack h]}=$ $\varrho _{2}\phi _{\lbrack 0]3}-\varrho _{3}\phi _{\lbrack
0]2}.$ This tensor is diagonal with respect to N--adapted (co) frames, $%
\mathbf{\Sigma }_{\alpha }^{[\mathbf{\phi }]\beta }=\{\Upsilon _{2},\Upsilon
_{2},0,0\}$ with
\begin{equation}
\Upsilon _{2}\left( x^{2},x^{3}\right) =-\lambda _{\lbrack h]}^{2}\sin
^{2}\left( \varrho _{i}x^{i}+\varphi _{\lbrack 0]}\right) .  \label{source01}
\end{equation}%
So, we have the case from (\ref{einstdiag4}) and (\ref{emcond4}) with $%
\Upsilon _{2}\left( x^{2},x^{2},v\right) \rightarrow \Upsilon _{2}\left(
x^{2},x^{2}\right) $ and $\Upsilon _{4},$ i. e.
\begin{equation}
G_{2}^{2}=G_{3}^{3}=-S_{4}^{4}=\Upsilon _{2}\left( x^{2},x^{2}\right)
\mbox{
and }G_{4}^{4}=G_{4}^{4}=-R_{2}^{2}=0.  \label{aux62}
\end{equation}%
There are satisfied the compatibility conditions from Corollary \ref{cors}.
For the above stated ansatz for the d--metric and $\phi $--field, the system
(\ref{efeinst}) $\ $reduces to a particular case of (\ref{ep1a})--(\ref{ep4a}%
), when the first equation is trivially satisfied by $g_{2}=1$ and $g_{3}=-1$
but the second one is
\begin{equation}
S_{4}^{4}=S_{5}^{5}=-\frac{1}{2h_{4}h_{5}}\left[ h_{5}^{\ast \ast
}-h_{5}^{\ast }\left( \ln \sqrt{|h_{4}h_{5}|}\right) ^{\ast }]\right]
=\lambda _{\lbrack h]}^{2}\sin ^{2}\left( \varrho _{i}x^{i}+\varphi
_{\lbrack 0]}\right) .  \label{aux04sol}
\end{equation}%
The right part of this equation is like a ''cosmological constant'', being
nontrivial in the h--subspace and polarized by a nonmetricity and torsion
wave (we can state $x^{2}=t$ and choose the signature $\left( -+---\right)
). $

The exact solution of (\ref{aux04sol}) exists according the Theorem \ref%
{texs} (see formulas (\ref{p2})--(\ref{auxf02a})). Taking any $%
h_{4}=h_{4}[\lambda _{\lbrack h]}=0]$ and $h_{5}=h_{5}[\lambda _{\lbrack
h]}=0]$ solving the equation with $\lambda _{\lbrack h]}=0,$ for instance,
like in (\ref{p1}), we can express the general solution with nontrivial
source like
\begin{equation*}
h_{5}[\lambda _{\lbrack h]}]=h_{5},\ h_{4}[\lambda _{\lbrack h]}]=\varsigma
_{\lbrack \lambda ]}\left( x^{i},v\right) h_{4},
\end{equation*}%
where (for an explicit source (\ref{source01}) in (\ref{auxf02a}))%
\begin{equation*}
\varsigma _{\lbrack \lambda ]}\left( t,x^{3},v\right) =\varsigma
_{4[0]}\left( t,x^{3}\right) -\frac{\lambda _{\lbrack h]}^{2}}{4}\sin
^{2}\left( \varrho _{2}t+\varrho _{3}x^{3}+\varphi _{\lbrack 0]}\right) \int
\frac{h_{4}h_{5}}{h_{5}^{\ast }}dv,
\end{equation*}%
where $\varsigma _{4[0]}\left( t,x^{3}\right) =1$ if we want to have $%
\varsigma _{\lbrack \lambda ]}$ for $\lambda _{\lbrack h]}^{2}\rightarrow 0$%
. A particular class of 4D off--diagonal exact solutions with $h_{4,5}^{\ast
}\neq 0$ (see the Corollary \ref{corgsol1} with $x^{2}=t$ stated to be the
time like coordinate and $x^{1}$ considered as the extra 5th dimensional one
to be eliminated for reductions 5D$\rightarrow $4D) is parametrized by the
generic off--diagonal metric
\begin{eqnarray}
\delta s^{2} &=&(dt)^{2}-\left( dx^{3}\right) -h_{0}^{2}(t,x^{3})\left[
f^{\ast }\left( t,x^{3},v\right) \right] ^{2}|\varsigma _{\lbrack \lambda
]}\left( t,x^{3},v\right) |\left( \delta v\right) ^{2}-f^{2}\left(
t,x^{3},v\right) \left( \delta y^{5}\right) ^{2},  \notag \\
\delta v &=&dv+w_{\widehat{k}}\left( t,x^{3},v\right) dx^{\widehat{k}},\
\delta y^{5}=dy^{5}+n_{\widehat{k}}\left( t,x^{3},v\right) dx^{\widehat{k}},
\label{dm5}
\end{eqnarray}%
with coefficients of necessary smooth class, where\ \ $g_{\widehat{k}}\left(
x^{\widehat{i}}\right) $ is a solution of the 2D equation (\ref{ep1a}) for a
given source $\Upsilon _{4}\left( x^{\widehat{i}}\right) ,$%
\begin{equation*}
\varsigma _{\lbrack \lambda ]}\left( t,x^{3},v\right) =1+\frac{\lambda
_{\lbrack h]}^{2}}{16}h_{0}^{2}(t,x^{3})\sin ^{2}\left( \varrho
_{2}t+\varrho _{3}x^{3}+\varphi _{\lbrack 0]}\right) f^{2}\left(
t,x^{3},v\right) ,
\end{equation*}%
and the N--connection coefficients $N_{\widehat{i}}^{4}=w_{\widehat{i}%
}(t,x^{3},v)$ and $N_{\widehat{i}}^{5}=n_{\widehat{i}}(t,x^{3},v)$ are
\begin{equation*}
w_{2,3}=-\frac{\partial _{2,3}\varsigma _{\lbrack \lambda ]}\left(
t,x^{3},v\right) }{\varsigma _{\lbrack \lambda ]}^{\ast }\left(
t,x^{3},v\right) }
\end{equation*}%
and
\begin{equation*}
n_{2,3}\left( t,x^{3},v\right) =n_{2,3[1]}\left( t,x^{3}\right)
+n_{2,3[2]}\left( t,x^{3}\right) \int \frac{\left[ f^{\ast }\left(
t,x^{3},v\right) \right] ^{2}}{\left[ f\left( t,x^{3},v\right) \right] ^{2}}%
\varsigma _{\lbrack \lambda ]}\left( t,x^{3},v\right) dv,
\end{equation*}%
define an exact 4D solution of the system of Einstein--Proca equations (\ref%
{ep1})--(\ref{ep4}) for vanishing mass $\mu \rightarrow 0,$ with holonomic
and anholonomic variables and 1-form field
\begin{equation*}
\phi _{\mu }=\left[ \phi _{\widehat{i}}=\phi _{\lbrack 0]\widehat{i}}\cos
\left( \varrho _{2}t+\varrho _{3}x^{3}+\varphi _{\lbrack 0]}\right) ,\phi
_{4}=0,\phi _{0}=0\right]
\end{equation*}%
for arbitrary nontrivial functions $f\left( t,x^{3},v\right) $ (with $%
f^{\ast }\neq 0),$ $h_{0}^{2}(t,x^{3})$, $n_{k[1,2]}\left( t,x^{3}\right) $
and sources $\Upsilon _{2}\left( t,x^{3}\right) =-\lambda _{\lbrack
h]}^{2}\sin ^{2}\left( \varrho _{2}t+\varrho _{3}x^{3}+\varphi _{\lbrack
0]}\right) $ and $\Upsilon _{4}=0$ and any integration constants to be
defined by certain boundary conditions and additional physical arguments.
For instance, we can consider ellipsoidal symmetries for the set of space
coordinates $\left( x^{3},y^{4}=v,y^{5}\right) $ considered on possibility
to be ellipsoidal ones, or even with topologically nontrivial configurations
like torus, with toroidal coordinates. Such exact solutions emphasize
anisotropic dependences on coordinate $v$ and do not depend on $y^{5}.$

\subsection{3D solitons in string Finsler--affine gravity}

The d--metric (\ref{dm5}) can be extended as to define a class of exact
solutions of generalized Finsler affine string gravity (\ref{fagfe}), for
certain particular cases describing 3D solitonic configurations.

We start with the the well known ansatz in string theory (see, for instance, %
\cite{kir}) for the $H$--field (\ref{aux51a}) when
\begin{equation}
\mathbf{H}_{\nu \lambda \rho }=\widehat{\mathbf{Z}}_{\ \nu \lambda \rho }+%
\widehat{\mathbf{H}}_{\nu \lambda \rho }=\lambda _{\lbrack H]}\sqrt{|\mathbf{%
g}_{\alpha \beta }|}\varepsilon _{\nu \lambda \rho }  \label{ans61}
\end{equation}%
where $\varepsilon _{\nu \lambda \rho }$ is completely antisymmetric and $%
\lambda _{\lbrack H]}=const,$ which satisfies the field equations for $%
\mathbf{H}_{\nu \lambda \rho },$ see (\ref{aux51b}). \ The ansatz (\ref%
{ans61}) is chosen for a locally anisotropic background with $\widehat{%
\mathbf{Z}}_{\ \nu \lambda \rho }$ defined by the d--torsions for the
canonical d--connection. So, the values $\widehat{\mathbf{H}}_{\nu \lambda
\rho }$ are constrained to solve the equations (\ref{ans61}) for a fixed
value of the cosmological constant $\lambda _{\lbrack H]}$ effectively
modeling some corrections from string gravity. In this case, the source (\ref%
{source01}) is modified to
\begin{equation*}
\mathbf{\Sigma }_{\alpha }^{[\mathbf{\phi }]\beta }+\mathbf{\Sigma }_{\alpha
}^{[\mathbf{H}]\beta }=\{\Upsilon _{2}+\frac{\lambda _{\lbrack H]}^{2}}{4}%
,\Upsilon _{2}+\frac{\lambda _{\lbrack H]}^{2}}{4},\frac{\lambda _{\lbrack
H]}^{2}}{4},\frac{\lambda _{\lbrack H]}^{2}}{4}\}
\end{equation*}%
and the equations (\ref{aux62}) became more general,
\begin{equation}
G_{2}^{2}=G_{3}^{3}=-S_{4}^{4}=\Upsilon _{2}\left( x^{2},x^{2}\right) +\frac{%
\lambda _{\lbrack H]}^{2}}{4}\mbox{
and }G_{4}^{4}=G_{4}^{4}=-R_{2}^{2}=\frac{\lambda _{\lbrack H]}^{2}}{4},
\label{aux63}
\end{equation}%
or, in component form%
\begin{eqnarray}
R_{2}^{2} &=&R_{3}^{3}=-\frac{1}{2g_{2}g_{3}}[g_{3}^{\bullet \bullet }-\frac{%
g_{2}^{\bullet }g_{3}^{\bullet }}{2g_{2}}-\frac{(g_{3}^{\bullet })^{2}}{%
2g_{3}}+g_{2}^{^{\prime \prime }}-\frac{g_{2}^{^{\prime }}g_{3}^{^{\prime }}%
}{2g_{3}}-\frac{(g_{2}^{^{\prime }})^{2}}{2g_{2}}]=-\frac{\lambda _{\lbrack
H]}^{2}}{4},  \label{eq64a} \\
S_{4}^{4} &=&S_{5}^{5}=-\frac{1}{2h_{4}h_{5}}\left[ h_{5}^{\ast \ast
}-h_{5}^{\ast }\left( \ln \sqrt{|h_{4}h_{5}|}\right) ^{\ast }]\right] =-%
\frac{\lambda _{\lbrack H]}^{2}}{4}+\lambda _{\lbrack h]}^{2}\sin ^{2}\left(
\varrho _{i}x^{i}+\varphi _{\lbrack 0]}\right) .  \label{eq64b}
\end{eqnarray}%
The solution of (\ref{eq64a})\ can be found as in the case for (\ref{auxeq01}%
), when $\psi =\ln |g_{2}|=\ln |g_{3}|$ is a solution of
\begin{equation}
\ddot{\psi}+\psi ^{\prime \prime }=-\frac{\lambda _{\lbrack H]}^{2}}{2},
\label{aux73}
\end{equation}%
where, for simplicity we choose the h--variables $x^{2}=\tilde{x}^{2}$ and $%
x^{3}=\tilde{x}^{3}.$

The solution of (\ref{eq64b}) can be constructed similarly to the equation (%
\ref{aux04sol}) but for a modified source (see Theorem \ref{texs} and
formulas (\ref{p2})--(\ref{auxf02a})). Taking any $h_{4}=h_{4}[\lambda
_{\lbrack h]}=0,\lambda _{\lbrack H]}=0]$ and $h_{5}=h_{5}[\lambda _{\lbrack
h]}=0,\lambda _{\lbrack H]}=0]$ solving the equation with $\lambda _{\lbrack
h]}=0$ and $\lambda _{\lbrack H]}=0$ like in (\ref{p1}), we can express the
general solution with nontrivial source like
\begin{equation*}
h_{5}[\lambda _{\lbrack h]},\lambda _{\lbrack H]}]=h_{5},\ h_{4}[\lambda
_{\lbrack h]},\lambda _{\lbrack H]}]=\varsigma _{\lbrack \lambda ,H]}\left(
x^{i},v\right) h_{4},
\end{equation*}%
where (for an explicit source from (\ref{eq64b}) in (\ref{auxf02a}))%
\begin{equation*}
\varsigma _{\lbrack \lambda ,H]}\left( t,x^{3},v\right) =\varsigma
_{4[0]}\left( t,x^{3}\right) -\frac{1}{4}\left[ \lambda _{\lbrack
h]}^{2}\sin ^{2}\left( \varrho _{2}t+\varrho _{3}x^{3}+\varphi _{\lbrack
0]}\right) -\frac{\lambda _{\lbrack H]}^{2}}{4}\right] \int \frac{h_{4}h_{5}%
}{h_{5}^{\ast }}dv,
\end{equation*}%
where $\varsigma _{4[0]}\left( t,x^{3}\right) =1$ if we want to have $%
\varsigma _{\lbrack \lambda ]}$ for $\lambda _{\lbrack h]}^{2},\lambda
_{\lbrack H]}^{2}\rightarrow 0$.

We define a class of 4D off--diagonal exact solutions of the system (\ref%
{fagfe}) with $h_{4,5}^{\ast }\neq 0$ (see the Corollary \ref{corgsol1} with
$x^{2}=t$ stated to be the time like coordinate and $x^{1}$ considered as
the extra 5th dimensional one to be eliminated for reductions 5D$\rightarrow
$4D) is parametrized by the generic off--diagonal metric
\begin{eqnarray}
\delta s^{2} &=&e^{\psi (t,x^{3})}(dt)^{2}-e^{\psi (t,x^{3})}\left(
dx^{3}\right) -f^{2}\left( t,x^{3},v\right) \left( \delta y^{5}\right) ^{2}
\notag \\
&&-h_{0}^{2}(t,x^{3})\left[ f^{\ast }\left( t,x^{3},v\right) \right]
^{2}|\varsigma _{\lbrack \lambda ,H]}\left( t,x^{3},v\right) |\left( \delta
v\right) ^{2},  \label{eq65} \\
\delta v &=&dv+w_{\widehat{k}}\left( t,x^{3},v\right) dx^{\widehat{k}},\
\delta y^{5}=dy^{5}+n_{\widehat{k}}\left( t,x^{3},v\right) dx^{\widehat{k}},
\notag
\end{eqnarray}%
with coefficients of necessary smooth class, where\ \ $g_{\widehat{k}}\left(
x^{\widehat{i}}\right) $ is a solution of the 2D equation (\ref{ep1a}) for a
given source $\Upsilon _{4}\left( x^{\widehat{i}}\right) ,$%
\begin{equation*}
\varsigma _{\lbrack \lambda ,H]}\left( t,x^{3},v\right) =1+\frac{%
h_{0}^{2}(t,x^{3})}{16}\left[ \lambda _{\lbrack h]}^{2}\sin ^{2}\left(
\varrho _{2}t+\varrho _{3}x^{3}+\varphi _{\lbrack 0]}\right) -\frac{\lambda
_{\lbrack H]}^{2}}{4}\right] f^{2}\left( t,x^{3},v\right) ,
\end{equation*}%
and the N--connection coefficients $N_{\widehat{i}}^{4}=w_{\widehat{i}%
}(t,x^{3},v)$ and $N_{\widehat{i}}^{5}=n_{\widehat{i}}(t,x^{3},v)$ are
\begin{equation*}
w_{2,3}=-\frac{\partial _{2,3}\varsigma _{\lbrack \lambda ,H]}\left(
t,x^{3},v\right) }{\varsigma _{\lbrack \lambda ,H]}^{\ast }\left(
t,x^{3},v\right) }
\end{equation*}%
and
\begin{equation*}
n_{2,3}\left( t,x^{3},v\right) =n_{2,3[1]}\left( t,x^{3}\right)
+n_{2,3[2]}\left( t,x^{3}\right) \int \frac{\left[ f^{\ast }\left(
t,x^{3},v\right) \right] ^{2}}{\left[ f\left( t,x^{3},v\right) \right] ^{2}}%
\varsigma _{\lbrack \lambda ,H]}\left( t,x^{3},v\right) dv,
\end{equation*}%
define an exact 4D solution of the system of generalized Finsler--affine
gravity equations (\ref{fagfe}) for vanishing Proca mass $\mu \rightarrow 0,$
with holonomic and anholonomic variables, 1-form field
\begin{equation}
\phi _{\mu }=\left[ \phi _{\widehat{i}}=\phi _{\lbrack 0]\widehat{i}%
}(t,x^{3})\cos \left( \varrho _{2}t+\varrho _{3}x^{3}+\varphi _{\lbrack
0]}\right) ,\phi _{4}=0,\phi _{0}=0\right]   \label{aux71}
\end{equation}%
and nontrivial effective $H$--field $\mathbf{H}_{\nu \lambda \rho }=\lambda
_{\lbrack H]}\sqrt{|\mathbf{g}_{\alpha \beta }|}\varepsilon _{\nu \lambda
\rho }$ for arbitrary nontrivial functions $f\left( t,x^{3},v\right) $ (with
$f^{\ast }\neq 0),$ $h_{0}^{2}(t,x^{3})$, $n_{k[1,2]}\left( t,x^{3}\right) $
and sources
\begin{equation*}
\Upsilon _{2}\left( t,x^{3}\right) =\lambda _{\lbrack H]}^{2}/4-\lambda
_{\lbrack h]}^{2}(t,x^{3})\sin ^{2}\left( \varrho _{2}t+\varrho
_{3}x^{3}+\varphi _{\lbrack 0]}\right) \mbox{ and }\Upsilon _{4}=\lambda
_{\lbrack H]}^{2}/4
\end{equation*}%
and any integration constants to be defined by certain boundary conditions
and additional physical arguments. The function $\phi _{\lbrack 0]\widehat{i}%
}(t,x^{3})$ in (\ref{aux71}) is taken to solve the equation
\begin{equation}
\partial _{2}[e^{-\psi \left( t,x^{3}\right) }\partial _{2}\phi
_{k}]-\partial _{3}[e^{-\psi \left( t,x^{3}\right) }\partial _{3}\phi
_{k}]=L_{ki}^{j}\partial ^{i}\phi _{j}-L_{ij}^{i}\partial ^{j}\phi _{k}
\label{eq74}
\end{equation}%
where $L_{ki}^{j}$ are computed for the d--metric  (\ref{eq65}) following
the formulas (\ref{candcon}). For $\psi =0,$ we obtain just the plane wave
equation (\ref{waveeq}) when $\phi _{\lbrack 0]\widehat{i}}$ and $\lambda
_{\lbrack h]}^{2}(t,x^{3})$ reduce to constant values. We do not fix here
any value of $\psi \left( t,x^{3}\right) $ solving (\ref{aux73}) in order to
define explicitly a particular solution of (\ref{eq74}). We note that for
any value of $\psi \left( t,x^{3}\right) $ we can solve the inhomogeneous
wave equation  (\ref{eq74}) by using solutions of the homogeneous case.

For simplicity, we do not present here the explicit value of $\sqrt{|\mathbf{%
g}_{\alpha \beta }|}$ computed for the d--metric (\ref{eq65}) as well the
values for distorsions $\widehat{\mathbf{Z}}_{\ \nu \lambda \rho },$ defined
by d--torsions of the canonical d--connection, see formulas (\ref{aux53})
and (\ref{dtorsb}) (the formulas are very cumbersome and do not reflect
additional physical properties). Having defined $\widehat{\mathbf{Z}}_{\ \nu
\lambda \rho },$ we can compute
\begin{equation*}
\widehat{\mathbf{H}}_{\nu \lambda \rho }=\lambda _{\lbrack H]}\sqrt{|\mathbf{%
g}_{\alpha \beta }|}\varepsilon _{\nu \lambda \rho }-\widehat{\mathbf{Z}}_{\
\nu \lambda \rho }.
\end{equation*}%
We note that the torsion $\widehat{\mathbf{T}}_{\ \lambda \rho }^{\nu }$
contained in $\widehat{\mathbf{Z}}_{\ \nu \lambda \rho },$ related to string
corrections by the $H$--field, is different from the torsion $\mathbf{T=}%
k_{2}\mathbf{\phi }$ and nontrivial nonmetricity $\mathbf{Q}=k_{0}\mathbf{%
\phi ,\ \Lambda =}k_{1}\mathbf{\phi ,\ }$from the metric--affine part of the
theory, see (\ref{triplemag}).

We can choose the function $f\left( t,x^{3},v\right) $ from (\ref{eq65}), or
(\ref{dm5}), as it would be a solution of the Kadomtsev--Petviashvili (KdP)
equation \cite{kdp}, i. e. to satisfy
\begin{equation*}
f^{\bullet \bullet }+\epsilon \left( f^{\prime}+6ff^{\ast }+f^{\ast \ast
\ast }\right) ^{\ast }=0,\ \epsilon =\pm 1,
\end{equation*}%
or, for another locally anisotropic background, to satisfy the $(2+1)$%
--dimensional sine--Gordon (SG) equation,
\begin{equation*}
-f^{\bullet \bullet }+f%
%TCIMACRO{\U{b4}}%
%BeginExpansion
{\acute{}}%
%EndExpansion
%TCIMACRO{\U{b4}}%
%BeginExpansion
{\acute{}}%
%EndExpansion
+f^{\ast \ast }=\sin f,
\end{equation*}%
see Refs. \cite{soliton} on gravitational solitons and theory of solitons.
In this case, we define a nonlinear model of graviational plane wave and 3D
solitons in the framework of the MAG with string corrections by $H$--field.
Such solutions generalized those considered in Refs. \cite{v1} for 4D and 5D
gravity.

We can also consider that $F/L=f^{2}\left( t,x^{3},v\right) $ is just the
generation function for a 2D model of Finsler/Lagrange geometry (being of
any solitonic or another type nature). In this case, the geometric
background is characterized by this type locally anisotropic configurations
(for Finsler metrics we shall impose corresponding homogeneity conditions on
coordinates).

\section{ Final Remarks}

In this paper we have investigated the dynamical aspects of metric--affine
gravity (MAG) with certain additional string corrections defined by the
antisymmetric $H$--field when the metric structure is generic off--diagonal
and the spacetime is provided with an anholonomic frame structure with
associated nonlinear connection (N--connection). We analyzed the
corresponding class of Lagrangians and derived the field equations of MAG
and string gravity with mixed holonomic and anholonomic variables. The main
motivation for this work is to determine the place and significance of such
models of gravity which contain as exact solutions certain classes of
metrics and connections modeling Finsler like geometries even in the limits
to the general relativity theory.

The work supports the results of Refs. \cite{v1,v1a} where various classes
of exact solutions in Einstein, Einstein--Cartan, gauge and string gravity
modeling Finsler--Lagrange configurations were constructed. We provide an
irreducible decomposition techniques (in our case with additional
N--connection splitting) and study the dynamics of MAG fields generating the
locally anisotropic geometries and interactions classified in Ref. \cite{vp1}%
. There are proved the main theorems on irreducible reduction to effective
Einstein--Proca equations with string corrections and formulated a new
method of constructing exact solutions.

As explicit examples of the new type of locally anisotropic configurations
in MAG and string gravity, we have elaborated three new classes of exact
solutions depending on 3-4 variables possessing nontrivial torsion and
nonmetricity fields, describing plane wave and three dimensional soliton
interactions and induced generalized Finsler--affine effective
configurations.

Finally, it seems worthwhile to note that such Finsler like configurations
do not violates the postulates of the general relativity theory in the
corresponding limits to the four dimensional Einstein theory because such
metrics transform into exact solutions of this theory. The anisotropies are
modeled by certain anholonomic frame constraints on a (pseudo) Riemannian
spacetime. In this case the restrictions imposed on physical applications of
the Finsler geometry, derived from experimental data on possible limits for
brocken local Lorentz invariance (see, for instance, Ref. \cite{will}), do
not hold.

\appendix

\section{Proof of Theorem \ref{t5dr}}

\bigskip \label{appa}We give some details on straightforward calculations
outlined in Ref. \cite{vmethod} for (pseudo) Riemannian and Riemann--Cartan
spaces. In brief, the proof of Theorem \ref{t5dr} is to be performed in this
way:\ Introducing $N_{i}^{4}=w_{i}$ and $N_{i}^{5}=n_{i}$ in (\ref{dder})
and (\ref{ddif}) and re--writting (\ref{ansatz5}) into a diagonal (in our
case) block form (\ref{block2}), we compute the h- and v--irreducible
components of the cannonical d--connection (\ref{candcon}). The next step is
to compute d--curvatures (\ref{dcurv}) and by contracting of indices to
define the components of the Ricci d--tensor (\ref{dricci}) which results in
(\ref{ricci1a})--(\ref{ricci4a}). We emphasize that such computations can
not be performed directly by applying any Tensor, Maple of Mathematica
macros because, in our case, we consider canonical d--connections instead of
the\ Levi--Civita connection \cite{cartanmacros}. We give the details of
such calculus related to N--adapted anholonomic frames.

The five dimensional (5D) local coordinates are $x^{i}$ and $y^{a}=\left(
v,y\right) ,$ i. e. $y^{4}=v,$ $y^{5}=y,$ were indices $i,j,k...=1,2,3$ and $%
a,b,c,...=4,5.$ Our reductions to 4D will be considered by excluding
dependencies on the variable $x^{1}$ and for trivial embeddings of 4D
off--diagonal ansatz into 5D ones. The signatures of metrics coud be
arbitrary ones. In general, the spacetime could be with torsion, but we
shall always be interested to define the limits to (pseudo) Riemannian
spaces.

The d--metric (\ref{block2}) for an ansatz (\ref{ansatz5}) with $%
g_{1}=const, $ is written
\begin{eqnarray}
\delta s^{2}&=&g_{1}{(dx^{1})}^{2}+g_{2}\left((x^{2},x^{3}\right) {(dx^{2})}%
^{2}+g_{3}\left( x^{k}\right){(dx^{3})}^{2}+h_{4}\left( x^{k},v\right) {%
(\delta v)}^{2}+h_{5}\left( x^{k},v\right){(\delta y)}^{2},  \notag \\
\delta v &=&dv+w_{i}\left( x^{k},v\right) dx^{i},\ \delta y=dy+n_{i}\left(
x^{k},v\right) dx^{i}  \label{ans4d}
\end{eqnarray}%
when the generic off--diagonal metric (\ref{metric5}) is associated to a
N--connection structure $N_{i}^{a}$ with $N_{i}^{4}=w_{i}\left(
x^{k},v\right) $ and $N_{i}^{5}=n_{i}\left( x^{k},v\right) .$ We note that
the metric (\ref{ans4d}) does not depend on variable $y^{5}=y,$ but
emphasize the dependence on ''anisotropic'' variable $y^{4}=v.$

If we regroup (\ref{ans4d}) with respect to true differentials $du^{\alpha
}=\left( dx^{i},dy^{a}\right) $ we obtain just the ansatz (\ref{ansatz5}).
It is a cumbersome task to perform tensor calculations (for instance, of
curvature and Ricci tensors) with such generic off--diagonal ansatz but the
formulas simplify substantially with respect to N--adapted frames of type(%
\ref{dder}) and (\ref{ddif}) and for effectively diagonalized metrics like (%
\ref{ans4d}).

So, the \ metric (\ref{metric5}) transform in a diagonal one with respect to
the pentads (frames, funfbeins)%
\begin{equation}
e^{i}=dx^{i},e^{4}=\delta v=dv+w_{i}\left( x^{k},v\right)
dx^{i},e^{5}=\delta y=dy+n_{i}\left( x^{k},v\right) dx^{i}  \label{ddif4a}
\end{equation}%
or
\begin{equation*}
\delta u^{\alpha }=\left( dx^{i},\delta y^{a}=dy^{a}+N_{i}^{a}dx^{i}\right)
\end{equation*}%
being dual to the N--elongated partial derivative operators,
\begin{eqnarray}
e_{1} &=&\delta _{1}=\frac{\partial }{\partial x^{1}}-N_{1}^{a}\frac{%
\partial }{\partial y^{a}}=\frac{\partial }{\partial x^{1}}-w_{1}\frac{%
\partial }{\partial v}-n_{1}\frac{\partial }{\partial y},  \label{dder4a} \\
e_{2} &=&\delta _{2}=\frac{\partial }{\partial x^{2}}-N_{2}^{a}\frac{%
\partial }{\partial y^{a}}=\frac{\partial }{\partial x^{2}}-w_{2}\frac{%
\partial }{\partial v}-n_{2}\frac{\partial }{\partial y},  \notag \\
e_{3} &=&\delta _{3}=\frac{\partial }{\partial x^{3}}-N_{3}^{a}\frac{%
\partial }{\partial y^{a}}=\frac{\partial }{\partial x^{3}}-w_{3}\frac{%
\partial }{\partial v}-n_{3}\frac{\partial }{\partial y},  \notag \\
e_{4} &=&\frac{\partial }{\partial y^{4}}=\frac{\partial }{\partial v},\
e_{5}=\frac{\partial }{\partial y^{5}}=\frac{\partial }{\partial y}  \notag
\end{eqnarray}%
when $\delta _{\alpha }=\frac{\delta }{\partial u^{\alpha }}=\left( \frac{%
\delta }{\partial x^{i}}=\frac{\partial }{\partial x^{i}}-N_{i}^{a}\frac{%
\partial }{\partial y^{a}},\frac{\partial }{\partial y^{b}}\right) .$

The N--elongated partial derivatives of a function $f\left( u^{\alpha
}\right) =f\left( x^{i},y^{a}\right) =f\left( x,r,v,y\right) $ are computed
in the formSo the N--elongated derivatives are%
\begin{equation*}
\delta _{2}f=\frac{\delta f}{\partial u^{2}}=\frac{\delta f}{\partial x^{2}}=%
\frac{\delta f}{\partial x}=\frac{\partial f}{\partial x}-N_{2}^{a}\frac{%
\partial f}{\partial y^{a}}=\frac{\partial f}{\partial x}-w_{2}\frac{%
\partial f}{\partial v}-n_{2}\frac{\partial f}{\partial y}=f^{\bullet
}-w_{2}\ f^{\prime }-n_{2}\ f^{\ast }
\end{equation*}%
where
\begin{equation*}
f^{\bullet }=\frac{\partial f}{\partial x^{2}}=\frac{\partial f}{\partial x}%
,\ f^{\prime }=\frac{\partial f}{\partial x^{3}}=\frac{\partial f}{\partial r%
},\ f^{\ast }=\frac{\partial f}{\partial y^{4}}=\frac{\partial f}{\partial v}%
.
\end{equation*}%
The N--elongated differential is
\begin{equation*}
\delta f=\frac{\delta f}{\partial u^{\alpha }}\delta u^{\alpha }.
\end{equation*}%
The N--elongated differential calculus should be applied if we work with
respect to N--adapted frames.

\subsection{Calculation of N--connection curvature}

We compute the coefficients (\ref{ncurv}) for the d--metric (\ref{ans4d})
(equivalently, the ansatz (\ref{ansatz5})) defining the curvature of
N--connection $N_{i}^{a},$ by substituting $N_{i}^{4}=w_{i}\left(
x^{k},v\right) $ and $N_{i}^{5}=n_{i}\left( x^{k},v\right) ,$ where $i=2,3$
and $a=4,5.$ The result for nontrivial values is
\begin{eqnarray}
\Omega _{23}^{4} &=&-\Omega _{23}^{4}=w_{2}^{\prime }-w_{3}^{\cdot
}-w_{3}w_{2}^{\ast }-w_{2}w_{3}^{\ast },  \label{omega} \\
\Omega _{23}^{5} &=&-\Omega _{23}^{5}=n_{2}^{\prime }-n_{3}^{\cdot
}-w_{3}n_{2}^{\ast }-w_{2}n_{3}^{\ast }.  \notag
\end{eqnarray}%
The canonical d--connection $\widehat{\mathbf{\Gamma }}_{\ \alpha \beta
}^{\gamma }=\left( \widehat{L}_{jk}^{i},\widehat{L}_{bk}^{a},\widehat{C}%
_{jc}^{i},\widehat{C}_{bc}^{a}\right) $ (\ref{candcon}) defines the
covariant derivative $\widehat{\mathbf{D}},$ satisfying the metricity
conditions $\widehat{\mathbf{D}}_{\alpha }\mathbf{g}_{\gamma \delta }=0$ for
$\mathbf{g}_{\gamma \delta }$ being the metric (\ref{ans4d}) with the
coefficients written with respect to N--adapted frames. $\widehat{\mathbf{%
\Gamma }}_{\ \alpha \beta }^{\gamma }$ has nontrivial d--torsions.

We compute the Einstein tensors for the canonica d--connection $\widehat{%
\mathbf{\Gamma }}_{\ \alpha \beta }^{\gamma }$ defined by the ansatz (\ref%
{ans4d}) with respect to N--adapted frames (\ref{ddif4a}) and (\ref{dder4a}%
). This results in exactly integrable vacuum Einstein equations and certain
type of sources. Such solutions could be with nontrivial torsion for
different classes of linear connections from Riemann--Cartan and generalized
Finsler geometries. So, the anholonomic frame method offers certain
possibilites to be extended to in string gravity where the torsion could be
not zero. But we can always select the limit to Levi--Civita connections, i.
e. to (pseudo) Riemannian spaces by considering additional constraints, see
Corollary \ref{corcond1} and/or conditions (\ref{cond2}).

\subsection{Calculation of the canonical d--connection}

We compute the coefficients (\ref{candcon}) for the d--metric (\ref{ans4d})
(equivalently, the ansatz (\ref{ansatz5})) when $g_{jk}=\{g_{j}\}$ and $%
h_{bc}=\{h_{b}\}$ are diagonal and $g_{ik}$ depend only on $x^{2}$ and $%
x^{3} $ but not on $y^{a}.$

We have
\begin{eqnarray}
\delta _{k}g_{ij} &=&\partial _{k}g_{ij}-w_{k}g_{ij}^{\ast }=\partial
_{k}g_{ij},\ \delta _{k}h_{b}=\partial _{k}h_{b}-w_{k}h_{b}^{\ast }
\label{aux01} \\
\delta _{k}w_{i} &=&\partial _{k}w_{i}-w_{k}w_{i}^{\ast },\ \delta
_{k}n_{i}=\partial _{k}n_{i}-w_{k}n_{i}^{\ast }  \notag
\end{eqnarray}%
resulting in formulas
\begin{equation*}
\widehat{L}_{jk}^{i}=\frac{1}{2}g^{ir}\left( \frac{\delta g_{jk}}{\delta
x^{k}}+\frac{\delta g_{kr}}{\delta x^{j}}-\frac{\delta g_{jk}}{\delta x^{r}}%
\right) =\frac{1}{2}g^{ir}\left( \frac{\partial g_{jk}}{\delta x^{k}}+\frac{%
\partial g_{kr}}{\delta x^{j}}-\frac{\partial g_{jk}}{\delta x^{r}}\right)
\end{equation*}%
The nontrivial values of $\widehat{L}_{jk}^{i}$ are%
\begin{eqnarray}
\widehat{L}_{22}^{2} &=&\frac{g_{2}^{\bullet }}{2g_{2}}=\alpha _{2}^{\bullet
},\ \widehat{L}_{23}^{2}=\frac{g_{2}^{\prime }}{2g_{2}}=\alpha
_{2}^{^{\prime }},\ \widehat{L}_{33}^{2}=-\frac{g_{3}^{\bullet }}{2g_{2}}
\label{aux02} \\
\widehat{L}_{22}^{3} &=&-\frac{g_{2}^{\prime }}{2g_{3}},\ \widehat{L}%
_{23}^{3}=\frac{g_{3}^{\bullet }}{2g_{3}}=\alpha _{3}^{\bullet },\ \
\widehat{L}_{33}^{3}=\frac{g_{3}^{\prime }}{2g_{3}}=\alpha _{3}^{^{\prime }}.
\notag
\end{eqnarray}%
In a similar form we compute the components%
\begin{equation*}
\widehat{L}_{bk}^{a}=\partial _{b}N_{k}^{a}+\frac{1}{2}h^{ac}\left( \partial
_{k}h_{bc}-N_{k}^{d}\frac{\partial h_{bc}}{\partial y^{d}}-h_{dc}\partial
_{b}N_{k}^{d}-h_{db}\partial _{c}N_{k}^{d}\right)
\end{equation*}%
having nontrivial values
\begin{eqnarray}
\widehat{L}_{42}^{4} &=&\frac{1}{2h_{4}}\left( h_{4}^{\bullet
}-w_{2}h_{4}^{\ast }\right) =\delta _{2}\ln \sqrt{|h_{4}|}\doteqdot \delta
_{2}\beta _{4},  \label{aux02a} \\
\widehat{L}_{43}^{4} &=&\frac{1}{2h_{4}}\left( h_{4}^{^{\prime
}}-w_{3}h_{4}^{\ast }\right) =\delta _{3}\ln \sqrt{|h_{4}|}\doteqdot \delta
_{3}\beta _{4}  \notag
\end{eqnarray}%
\begin{equation}
\widehat{L}_{5k}^{4}=-\frac{h_{5}}{2h_{4}}n_{k}^{\ast },\ \widehat{L}%
_{bk}^{5}=\partial _{b}n_{k}+\frac{1}{2h_{5}}\left( \partial
_{k}h_{b5}-w_{k}h_{b5}^{\ast }-h_{5}\partial _{b}n_{k}\right) ,
\label{aux02b}
\end{equation}

\begin{eqnarray}
\widehat{L}_{4k}^{5} &=&n_{k}^{\ast }+\frac{1}{2h_{5}}\left(
-h_{5}n_{k}^{\ast }\right) =\frac{1}{2}n_{k}^{\ast },  \label{aux02c} \\
\widehat{L}_{5k}^{5} &=&\frac{1}{2h_{5}}\left( \partial
_{k}h_{5}-w_{k}h_{5}^{\ast }\right) =\delta _{k}\ln \sqrt{|h_{4}|}=\delta
_{k}\beta _{4}.  \notag
\end{eqnarray}

We note that
\begin{equation}
\widehat{C}_{jc}^{i}=\frac{1}{2}g^{ik}\frac{\partial g_{jk}}{\partial y^{c}}%
\doteqdot 0  \label{aux02cc}
\end{equation}%
because $g_{jk}=g_{jk}\left( x^{i}\right) $ for the considered ansatz.

The values
\begin{equation*}
\widehat{C}_{bc}^{a}=\frac{1}{2}h^{ad}\left( \frac{\partial h_{bd}}{\partial
y^{c}}+\frac{\partial h_{cd}}{\partial y^{b}}-\frac{\partial h_{bc}}{%
\partial y^{d}}\right)
\end{equation*}%
for $h_{bd}=[h_{4},h_{5}]$ from the ansatz (\ref{ansatz5}) have nontrivial
components
\begin{equation}
\widehat{C}_{44}^{4}=\frac{h_{4}^{\ast }}{2h_{4}}\doteqdot \beta _{4}^{\ast
},\widehat{C}_{55}^{4}=-\frac{h_{5}^{\ast }}{2h_{4}},\widehat{C}_{45}^{5}=%
\frac{h_{5}^{\ast }}{2h_{5}}\doteqdot \beta _{5}^{\ast }.  \label{aux02d}
\end{equation}

The set of formulas (\ref{aux02})--(\ref{aux02d}) define the nontrivial
coefficients of the canonical d--connection $\widehat{\mathbf{\Gamma }}_{\
\alpha \beta }^{\gamma }=\left( \widehat{L}_{jk}^{i},\widehat{L}_{bk}^{a},%
\widehat{C}_{jc}^{i},\widehat{C}_{bc}^{a}\right) $ (\ref{candcon}) for the
5D ansatz (\ref{ans4d}).

\subsection{Calculation of torsion coefficients}

We should put the nontrivial values (\ref{aux02})-- (\ref{aux02d}) into the
formulas for d--torsion \ (\ref{dtorsb}).

One holds $T_{.jk}^{i}=0$ and $T_{.bc}^{a}=0,$ because of symmetry of
coefficients $L_{jk}^{i}$ and $C_{bc}^{a}.$

We have computed the nontrivial values of $\Omega _{.ji}^{a},$ see \ (\ref%
{omega}) resulting in%
\begin{eqnarray}
T_{23}^{4} &=&\Omega _{23}^{4}=-\Omega _{23}^{4}=w_{2}^{\prime
}-w_{3}^{\bullet }-w_{3}w_{2}^{\ast }-w_{2}w_{3}^{\ast },  \label{aux11} \\
T_{23}^{5} &=&\Omega _{23}^{5}=-\Omega _{23}^{5}=n_{2}^{\prime
}-n_{3}^{\bullet }-w_{3}n_{2}^{\ast }-w_{2}n_{3}^{\ast }.  \notag
\end{eqnarray}

One follows
\begin{equation*}
T_{ja}^{i}=-T_{aj}^{i}=C_{.ja}^{i}=\widehat{C}_{jc}^{i}=\frac{1}{2}g^{ik}%
\frac{\partial g_{jk}}{\partial y^{c}}\doteqdot 0,
\end{equation*}
see (\ref{aux02cc}).

For the components
\begin{equation*}
T_{.bi}^{a}=-T_{.ib}^{a}=P_{.bi}^{a}=\frac{\partial N_{i}^{a}}{\partial y^{b}%
}-L_{.bj}^{a},
\end{equation*}
i. e. for
\begin{equation*}
\widehat{P}_{.bi}^{4}=\frac{\partial N_{i}^{4}}{\partial y^{b}}-\widehat{L}%
_{.bj}^{4}=\partial _{b}w_{i}-\widehat{L}_{.bj}^{4}\mbox{ and }\widehat{P}%
_{.bi}^{5}=\frac{\partial N_{i}^{5}}{\partial y^{b}}-\widehat{L}%
_{.bj}^{5}=\partial _{b}n_{i}-\widehat{L}_{.bj}^{5},
\end{equation*}
we have the nontrivial values%
\begin{eqnarray}
\widehat{P}_{.4i}^{4} &=&w_{i}^{\ast }-\frac{1}{2h_{4}}\left( \partial
_{i}h_{4}-w_{i}h_{4}^{\ast }\right) =w_{i}^{\ast }-\delta _{i}\beta _{4},\
\widehat{P}_{.5i}^{4}=\frac{h_{5}}{2h_{4}}n_{i}^{\ast },  \notag \\
\ \widehat{P}_{.4i}^{5} &=&\frac{1}{2}n_{i}^{\ast },\ \widehat{P}_{.5i}^{5}=-%
\frac{1}{2h_{5}}\left( \partial _{i}h_{5}-w_{i}h_{5}^{\ast }\right) =-\delta
_{i}\beta _{5}.  \label{aux12}
\end{eqnarray}

The formulas\ (\ref{aux11}) and (\ref{aux12}) state the \ nontrivial
coefficients of the canonical d--connection for the chosen ansatz (\ref%
{ans4d}).

\subsection{Calculation of the Ricci tensor}

Let us compute the value $R_{ij}=R_{\ ijk}^{k}$ as in (\ref{dricci}) for
\begin{equation*}
R_{\ hjk}^{i}=\frac{\delta L_{.hj}^{i}}{\delta x^{k}}-\frac{\delta
L_{.hk}^{i}}{\delta x^{j}}%
+L_{.hj}^{m}L_{mk}^{i}-L_{.hk}^{m}L_{mj}^{i}-C_{.ha}^{i}\Omega _{.jk}^{a},
\end{equation*}%
from (\ref{dcurv}). It should be noted that $C_{.ha}^{i}=0$ for the ansatz
under consideration, see (\ref{aux02cc}). We compute

\begin{equation*}
\frac{\delta L_{.hj}^{i}}{\delta x^{k}}=\partial
_{k}L_{.hj}^{i}+N_{k}^{a}\partial _{a}L_{.hj}^{i}=\partial
_{k}L_{.hj}^{i}+w_{k}\left( L_{.hj}^{i}\right) ^{\ast }=\partial
_{k}L_{.hj}^{i}
\end{equation*}%
because $L_{.hj}^{i}$ do not depend on variable $y^{4}=v.$

Derivating (\ref{aux02}), we obtain
\begin{eqnarray*}
\partial _{2}L_{\ 22}^{2} &=&\frac{g_{2}^{\bullet \bullet }}{2g_{2}}-\frac{%
\left( g_{2}^{\bullet }\right) ^{2}}{2\left( g_{2}\right) ^{2}},\ \partial
_{2}L_{\ 23}^{2}=\frac{g_{2}^{\bullet ^{\prime }}}{2g_{2}}-\frac{%
g_{2}^{\bullet }g_{2}^{^{\prime }}}{2\left( g_{2}\right) ^{2}},\ \partial
_{2}L_{\ 33}^{2}=-\frac{g_{3}^{\bullet \bullet }}{2g_{2}}+\frac{%
g_{2}^{\bullet }g_{3}^{\bullet }}{2\left( g_{2}\right) ^{2}}, \\
\ \partial _{2}L_{\ 22}^{3} &=&-\frac{g_{2}^{\bullet ^{\prime }}}{2g_{3}}+%
\frac{g_{2}^{\bullet }g_{3}^{^{\prime }}}{2\left( g_{3}\right) ^{2}}%
,\partial _{2}L_{\ 23}^{3}=\frac{g_{3}^{\bullet \bullet }}{2g_{3}}-\frac{%
\left( g_{3}^{\bullet }\right) ^{2}}{2\left( g_{3}\right) ^{2}},\ \partial
_{2}L_{\ 33}^{3}=\frac{g_{3}^{\bullet ^{\prime }}}{2g_{3}}-\frac{%
g_{3}^{\bullet }g_{3}^{^{\prime }}}{2\left( g_{3}\right) ^{2}}, \\
\partial _{3}L_{\ 22}^{2} &=&\frac{g_{2}^{\bullet ^{\prime }}}{2g_{2}}-\frac{%
g_{2}^{\bullet }g_{2}^{^{\prime }}}{2\left( g_{2}\right) ^{2}},\partial
_{3}L_{\ 23}^{2}=\frac{g_{2}^{ll}}{2g_{2}}-\frac{\left( g_{2}^{l}\right) ^{2}%
}{2\left( g_{2}\right) ^{2}},\partial _{3}L_{\ 33}^{2}=-\frac{g_{3}^{\bullet
^{\prime }}}{2g_{2}}+\frac{g_{3}^{\bullet }g_{2}^{^{\prime }}}{2\left(
g_{2}\right) ^{2}}, \\
\ \partial _{3}L_{\ 22}^{3} &=&-\frac{g_{2}^{^{\prime \prime }}}{2g_{3}}+%
\frac{g_{2}^{\bullet }g_{2}^{^{\prime }}}{2\left( g_{3}\right) ^{2}}%
,\partial _{3}L_{\ 23}^{3}=\frac{g_{3}^{\bullet ^{\prime }}}{2g_{3}}-\frac{%
g_{3}^{\bullet }g_{3}^{^{\prime }}}{2\left( g_{3}\right) ^{2}},\partial
_{3}L_{\ 33}^{3}=\frac{g_{3}^{ll}}{2g_{3}}-\frac{\left( g_{3}^{l}\right) ^{2}%
}{2\left( g_{3}\right) ^{2}}.
\end{eqnarray*}

For these values and (\ref{aux02}), there are only 2 nontrivial components,

\begin{eqnarray*}
R_{\ 323}^{2} &=&\frac{g_{3}^{\bullet \bullet }}{2g_{2}}-\frac{%
g_{2}^{\bullet }g_{3}^{\bullet }}{4\left( g_{2}\right) ^{2}}-\frac{\left(
g_{3}^{\bullet }\right) ^{2}}{4g_{2}g_{3}}+\frac{g_{2}^{ll}}{2g_{2}}-\frac{%
g_{2}^{l}g_{3}^{l}}{4g_{2}g_{3}}-\frac{\left( g_{2}^{l}\right) ^{2}}{4\left(
g_{2}\right) ^{2}} \\
R_{\ 223}^{3} &=&-\frac{g_{3}^{\bullet \bullet }}{2g_{3}}+\frac{%
g_{2}^{\bullet }g_{3}^{\bullet }}{4g_{2}g_{3}}+\frac{\left( g_{3}^{\bullet
}\right) ^{2}}{4(g_{3})^{2}}-\frac{g_{2}^{ll}}{2g_{3}}+\frac{%
g_{2}^{l}g_{3}^{l}}{4(g_{3})^{2}}+\frac{\left( g_{2}^{l}\right) ^{2}}{%
4g_{2}g_{3}}
\end{eqnarray*}%
with%
\begin{equation*}
R_{22}=-R_{\ 223}^{3}\mbox{ and }R_{33}=R_{\ 323}^{2},
\end{equation*}%
or%
\begin{equation*}
R_{2}^{2}=R_{3}^{3}=-\frac{1}{2g_{2}g_{3}}\left[ g_{3}^{\bullet \bullet }-%
\frac{g_{2}^{\bullet }g_{3}^{\bullet }}{2g_{2}}-\frac{\left( g_{3}^{\bullet
}\right) ^{2}}{2g_{3}}+g_{2}^{\prime \prime }-\frac{g_{2}^{l}g_{3}^{l}}{%
2g_{3}}-\frac{\left( g_{2}^{l}\right) ^{2}}{2g_{2}}\right]
\end{equation*}%
which is (\ref{ricci1a}).

Now, we consider
\begin{eqnarray*}
P_{\ bka}^{c} &=&\frac{\partial L_{.bk}^{c}}{\partial y^{a}}-\left( \frac{%
\partial C_{.ba}^{c}}{\partial x^{k}}+L_{.dk}^{c%
\,}C_{.ba}^{d}-L_{.bk}^{d}C_{.da}^{c}-L_{.ak}^{d}C_{.bd}^{c}\right)
+C_{.bd}^{c}P_{.ka}^{d} \\
&=&\frac{\partial L_{.bk}^{c}}{\partial y^{a}}%
-C_{.ba|k}^{c}+C_{.bd}^{c}P_{.ka}^{d}
\end{eqnarray*}%
from (\ref{dcurv}). Contracting indices, we have%
\begin{equation*}
R_{bk}=P_{\ bka}^{a}=\frac{\partial L_{.bk}^{a}}{\partial y^{a}}%
-C_{.ba|k}^{a}+C_{.bd}^{a}P_{.ka}^{d}
\end{equation*}%
Let us denote $C_{b}=C_{.ba}^{c}$ and write%
\begin{equation*}
C_{.b|k}=\delta _{k}C_{b}-L_{\ bk}^{d\,}C_{d}=\partial
_{k}C_{b}-N_{k}^{e}\partial _{e}C_{b}-L_{\ bk}^{d\,}C_{d}=\partial
_{k}C_{b}-w_{k}C_{b}^{\ast }-L_{\ bk}^{d\,}C_{d}.
\end{equation*}%
We express%
\begin{equation*}
R_{bk}=\ _{[1]}R_{bk}+\ _{[2]}R_{bk}+\ _{[3]}R_{bk}
\end{equation*}%
where%
\begin{eqnarray*}
\ _{[1]}R_{bk} &=&\left( L_{bk}^{4}\right) ^{\ast }, \\
\ _{[2]}R_{bk} &=&-\partial _{k}C_{b}+w_{k}C_{b}^{\ast }+L_{\ bk}^{d\,}C_{d},
\\
\ _{[3]}R_{bk} &=&C_{.bd}^{a}P_{.ka}^{d}=C_{.b4}^{4}P_{.k4}^{4}
+C_{.b5}^{4}P_{.k4}^{5}+C_{.b4}^{5}P_{.k5}^{4}+C_{.b5}^{5}P_{.k5}^{5}
\end{eqnarray*}%
and
\begin{eqnarray}
C_{4} &=&C_{44}^{4}+C_{45}^{5}=\frac{h_{4}^{\ast }}{2h_{4}}+\frac{%
h_{5}^{\ast }}{2h_{5}}=\beta _{4}^{\ast }+\beta _{5}^{\ast },  \label{aux6a}
\\
C_{5} &=&C_{54}^{4}+C_{55}^{5}=0  \notag
\end{eqnarray}%
see(\ref{aux02d}) .

We compute%
\begin{equation*}
R_{4k}=\ _{[1]}R_{4k}+\ _{[2]}R_{4k}+\ _{[3]}R_{4k}
\end{equation*}%
with
\begin{eqnarray*}
\ _{[1]}R_{4k} &=&\left( L_{4k}^{4}\right) ^{\ast }=\left( \delta _{k}\beta
_{4}\right) ^{\ast } \\
\ _{[2]}R_{4k} &=&-\partial _{k}C_{4}+w_{k}C_{4}^{\ast }+L_{\
4k}^{4\,}C_{4},L_{\ 4k}^{4\,}=\delta _{k}\beta _{4}\mbox{ see
(\ref{aux02a}} \\
&=&-\partial _{k}\left( \beta _{4}^{\ast }+\beta _{5}^{\ast }\right)
+w_{k}\left( \beta _{4}^{\ast }+\beta _{5}^{\ast }\right) ^{\ast }+L_{\
4k}^{4\,}\left( \beta _{4}^{\ast }+\beta _{5}^{\ast }\right) \\
\ _{[3]}R_{4k}
&=&C_{.44}^{4}P_{.k4}^{4}+C_{.45}^{4}P_{.k4}^{5}+C_{.44}^{5}P_{.k5}^{4}+C_{.45}^{5}P_{.k5}^{5}
\\
&=&\beta _{4}^{\ast }\left( w_{k}^{\ast }-\delta _{k}\beta _{4}\right)
-\beta _{5}^{\ast }\delta _{k}\beta _{5}
\end{eqnarray*}%
Summarizing, we get%
\begin{equation*}
R_{4k}=w_{k}\left[ \beta _{5}^{\ast \ast }+\left( \beta _{5}^{\ast }\right)
^{2}-\beta _{4}^{\ast }\beta _{5}^{\ast }\right] +\beta _{5}^{\ast }\partial
_{k}\left( \beta _{4}+\beta _{5}\right) -\partial _{k}\beta _{5}^{\ast }
\end{equation*}%
or, for
\begin{equation*}
\beta _{4}^{\ast }=\frac{h_{4}^{\ast }}{2h_{4}},\partial _{k}\beta _{4}=%
\frac{\partial _{k}h_{4}}{2h_{4}},\beta _{5}^{\ast }=\frac{h_{5}^{\ast }}{%
2h_{5}},\beta _{5}^{\ast \ast }=\frac{h_{5}^{\ast \ast }h_{5}-\left(
h_{5}^{\ast }\right) ^{2}}{2\left( h_{5}\right) ^{5}},
\end{equation*}%
we can write%
\begin{equation*}
2h_{5}R_{4k}=w_{k}\left[ h_{5}^{\ast \ast }-\frac{\left( h_{5}^{\ast
}\right) ^{2}}{2h_{5}}-\frac{h_{4}^{\ast }h_{5}^{\ast }}{2h_{4}}\right] +%
\frac{h_{5}^{\ast }}{2}\left( \frac{\partial _{k}h_{4}}{h_{4}}+\frac{%
\partial _{k}h_{5}}{h_{5}}\right) -\partial _{k}h_{5}^{\ast }
\end{equation*}%
which is equivalent to (\ref{ricci3a})

In a similar way, we compute%
\begin{equation*}
R_{5k}=\ _{[1]}R_{5k}+\ _{[2]}R_{5k}+\ _{[3]}R_{5k}
\end{equation*}%
with
\begin{eqnarray*}
\ _{[1]}R_{5k} &=&\left( L_{5k}^{4}\right) ^{\ast }, \\
\ _{[2]}R_{5k} &=&-\partial _{k}C_{5}+w_{k}C_{5}^{\ast }+L_{\ 5k}^{4\,}C_{4},
\\
\ _{[3]}R_{5k}
&=&C_{.54}^{4}P_{.k4}^{4}+C_{.55}^{4}P_{.k4}^{5}+C_{.54}^{5}P_{.k5}^{4}+C_{.55}^{5}P_{.k5}^{5}.
\end{eqnarray*}%
We have
\begin{eqnarray*}
R_{5k} &=&\left( L_{5k}^{4}\right) ^{\ast }+L_{\
5k}^{4\,}C_{4}+C_{.55}^{4}P_{.k4}^{5}+C_{.54}^{5}P_{.k5}^{4} \\
&=&\left( -\frac{h_{5}}{h_{4}}n_{k}^{\ast }\right) ^{\ast }-\frac{h_{5}}{%
h_{4}}n_{k}^{\ast }\left( \frac{h_{4}^{\ast }}{2h_{4}}+\frac{h_{5}^{\ast }}{%
2h_{5}}\right) +\frac{h_{5}^{\ast }}{2h_{5}}\frac{h_{5}}{2h_{4}}n_{k}^{\ast
}-\frac{h_{5}^{\ast }}{2h_{4}}\frac{1}{2}n_{k}^{\ast }
\end{eqnarray*}%
which can be written%
\begin{equation*}
2h_{4}R_{5k}=h_{5}n_{k}^{\ast \ast }+\left( \frac{h_{5}}{h_{4}}h_{4}^{\ast }-%
\frac{3}{2}h_{5}^{\ast }\right) n_{k}^{\ast }
\end{equation*}%
i. e. (\ref{ricci4a})

For the values
\begin{equation*}
P_{\ jka}^{i}=\frac{\partial L_{.jk}^{i}}{\partial y^{k}}-\left( \frac{%
\partial C_{.ja}^{i}}{\partial x^{k}}%
+L_{.lk}^{i}C_{.ja}^{l}-L_{.jk}^{l}C_{.la}^{i}-L_{.ak}^{c}C_{.jc}^{i}\right)
+C_{.jb}^{i}P_{.ka}^{b}
\end{equation*}%
from (\ref{dcurv}), we obtain zeros because $C_{.jb}^{i}=0$ and $L_{.jk}^{i}$
do not depend on $y^{k}.$ So,
\begin{equation*}
R_{ja}=P_{\ jia}^{i}=0.
\end{equation*}

Taking
\begin{equation*}
S_{\ bcd}^{a}=\frac{\partial C_{.bc}^{a}}{\partial y^{d}}-\frac{\partial
C_{.bd}^{a}}{\partial y^{c}}+C_{.bc}^{e}C_{.ed}^{a}-C_{.bd}^{e}C_{.ec}^{a}.
\end{equation*}%
from (\ref{dcurv}) and contracting the indices in order to obtain the Ricci
coefficients,%
\begin{equation*}
R_{bc}=\frac{\partial C_{.bc}^{d}}{\partial y^{d}}-\frac{\partial C_{.bd}^{d}%
}{\partial y^{c}}+C_{.bc}^{e}C_{.ed}^{d}-C_{.bd}^{e}C_{.ec}^{d}
\end{equation*}%
with $C_{.bd}^{d}=C_{b}$ already computed, see (\ref{aux6a}), we obtain
\begin{equation*}
R_{bc}=\left( C_{.bc}^{4}\right) ^{\ast }-\partial
_{c}C_{b}+C_{.bc}^{4}C_{4}-C_{.b4}^{4}C_{.4c}^{4}
-C_{.b5}^{4}C_{.4c}^{5}-C_{.b4}^{5}C_{.5c}^{4}-C_{.b5}^{5}C_{.5c}^{5}.
\end{equation*}%
There are nontrivial values,
\begin{eqnarray*}
R_{44} &=&\left( C_{.44}^{4}\right) ^{\ast }-C_{4}^{\ast
}+C_{44}^{4}(C_{4}-C_{44}^{4})-\left( C_{.45}^{5}\right) ^{2} \\
&=&\beta _{4}^{\ast \ast }-\left( \beta _{4}^{\ast }+\beta _{5}^{\ast
}\right) ^{\ast }+\beta _{4}^{\ast }\left( \beta _{4}^{\ast }+\beta
_{5}^{\ast }-\beta _{4}^{\ast }\right) -\left( \beta _{5}^{\ast }\right)
^{\ast } \\
R_{55} &=&\left( C_{.55}^{4}\right) ^{\ast }-C_{.55}^{4}\left(
-C_{4}+2C_{.45}^{5}\right) \\
&=&-\left( \frac{h_{5}^{\ast }}{2h_{4}}\right) ^{\ast }+\frac{h_{5}^{\ast }}{%
2h_{4}}\left( 2\beta _{5}^{\ast }+\beta _{4}^{\ast }-\beta _{5}^{\ast
}\right)
\end{eqnarray*}%
Introducing
\begin{equation*}
\beta _{4}^{\ast }=\frac{h_{4}^{\ast }}{2h_{4}},\beta _{5}^{\ast }=\frac{%
h_{5}^{\ast }}{2h_{5}}
\end{equation*}%
we get%
\begin{equation*}
R_{4}^{4}=R_{5}^{5}=\frac{1}{2h_{4}h_{5}}\left[ -h_{5}^{\ast \ast }+\frac{%
\left( h_{5}^{\ast }\right) ^{2}}{2h_{5}}+\frac{h_{4}^{\ast }h_{5}^{\ast }}{%
2h_{4}}\right]
\end{equation*}%
which is just (\ref{ricci2a}).

Theorem \ref{t5dr} is proven.

\section{Reductions from 5D to 4D}

\label{ss4d}To construct a $5D\rightarrow 4D$ reduction for the ansatz (\ref%
{ansatz5}) and (\ref{ansatzc}) is to eliminate from formulas the variable $%
x^{1}$ and to consider a 4D space (parametrized by local coordinates $\left(
x^{2},x^{3},v,y^{5}\right) )$ being trivially embedded into 5D space
(parametrized by local coordinates $\left( x^{1},x^{2},x^{3},v,y^{5}\right) $
with $g_{11}=\pm 1,g_{1\widehat{\alpha }}=0,\widehat{\alpha }=2,3,4,5)$ with
possible \ 4D conformal and anholonomic transforms depending only on
variables $\left( x^{2},x^{3},v\right) .$ We suppose that the 4D metric $g_{%
\widehat{\alpha }\widehat{\beta }}$ could be of arbitrary signature. In
order to emphasize that some coordinates are stated just for a such 4D space
we put ''hats'' on the Greek indices, $\widehat{\alpha },\widehat{\beta }%
,... $ \ and on the Latin indices from the middle of alphabet, $\widehat{i},%
\widehat{j},...=2,3,$ where $u^{\widehat{\alpha }}=\left( x^{\widehat{i}%
},y^{a}\right) =\left( x^{2},x^{3},y^{4},y^{5}\right) .$

In result, the Theorems \ref{t5dr} and \ref{t5dra}, Corollaries \ref{ceint}
and \ref{cors} and Theorem \ref{texs} can be reformulated for 4D gravity
with mixed holonomic--anholonomic variables. We outline here the most
important properties of a such reduction.

\begin{itemize}
\item The metric (\ref{metric5}) with ansatz (\ref{ansatz5}) and metric (\ref%
{cmetric}) with (\ref{ansatzc}) are respectively transformed on 4D spaces to
the values:

The first type 4D off--diagonal metric is taken
\begin{equation}
\mathbf{g}=\mathbf{g}_{\widehat{\alpha }\widehat{\beta }}\left( x^{\widehat{i%
}},v\right) du^{\widehat{\alpha }}\otimes du^{\widehat{\beta }}
\label{metric4}
\end{equation}%
with the metric coefficients\textbf{\ }$g_{\widehat{\alpha }\widehat{\beta }%
} $ parametrized

{%%\footnotesize
\begin{equation}
\left[
\begin{array}{cccc}
g_{2}+w_{2}^{\ 2}h_{4}+n_{2}^{\ 2}h_{5} & w_{2}w_{3}h_{4}+n_{2}n_{3}h_{5} &
w_{2}h_{4} & n_{2}h_{5} \\
w_{2}w_{3}h_{4}+n_{2}n_{3}h_{5} & g_{3}+w_{3}^{\ 2}h_{4}+n_{3}^{\ 2}h_{5} &
w_{3}h_{4} & n_{3}h_{5} \\
w_{2}h_{4} & w_{3}h_{4} & h_{4} & 0 \\
n_{2}h_{5} & n_{3}h_{5} & 0 & h_{5}%
\end{array}%
\right] ,  \label{ansatz4}
\end{equation}%
} where the coefficients are some necessary smoothly class functions of
type:
\begin{eqnarray}
g_{2,3} &=&g_{2,3}(x^{2},x^{3}),h_{4,5}=h_{4,5}(x^{\widehat{k}},v),  \notag
\\
w_{\widehat{i}} &=&w_{\widehat{i}}(x^{\widehat{k}},v),n_{\widehat{i}}=n_{%
\widehat{i}}(x^{\widehat{k}},v);~\widehat{i},\widehat{k}=2,3.  \notag
\end{eqnarray}

The anholonomically and conformally transformed 4D off--diagonal metric is
\begin{equation}
\mathbf{g}=\omega ^{2}(x^{\widehat{i}},v)\hat{\mathbf{g}}_{\widehat{\alpha }%
\widehat{\beta }}\left( x^{\widehat{i}},v\right) du^{\widehat{\alpha }%
}\otimes du^{\widehat{\beta }},  \label{cmetric4}
\end{equation}%
were the coefficients $\hat{\mathbf{g}}_{\widehat{\alpha }\widehat{\beta }}$
are parametrized by the ansatz %{\scriptsize
\begin{equation}
\left[
\begin{array}{cccc}
g_{2}+(w_{2}^{\ 2}+\zeta _{2}^{\ 2})h_{4}+n_{2}^{\ 2}h_{5} &
(w_{2}w_{3}++\zeta _{2}\zeta _{3})h_{4}+n_{2}n_{3}h_{5} & (w_{2}+\zeta
_{2})h_{4} & n_{2}h_{5} \\
(w_{2}w_{3}++\zeta _{2}\zeta _{3})h_{4}+n_{2}n_{3}h_{5} & g_{3}+(w_{3}^{\
2}+\zeta _{3}^{\ 2})h_{4}+n_{3}^{\ 2}h_{5} & (w_{3}+\zeta _{3})h_{4} &
n_{3}h_{5} \\
(w_{2}+\zeta _{2})h_{4} & (w_{3}+\zeta _{3})h_{4} & h_{4} & 0 \\
n_{2}h_{5} & n_{3}h_{5} & 0 & h_{5}+\zeta _{5}h_{4}%
\end{array}%
\right]  \label{ansatzc4}
\end{equation}%
%
%
%}
where $\zeta _{\widehat{i}}=\zeta _{\widehat{i}}\left( x^{\widehat{k}%
},v\right) $ and we shall restrict our considerations for $\zeta _{5}=0.$

\item We obtain a quadratic line element
\begin{equation}
\delta s^{2}=g_{2}(dx^{2})^{2}+g_{3}(dx^{3})^{2}+h_{4}(\delta
v)^{2}+h_{5}(\delta y^{5})^{2},  \label{dmetric4}
\end{equation}%
written with respect to the anholonomic co--frame $\left( dx^{\widehat{i}%
},\delta v,\delta y^{5}\right) ,$ where
\begin{equation}
\delta v=dv+w_{\widehat{i}}dx^{\widehat{i}}\mbox{ and }\delta
y^{5}=dy^{5}+n_{\widehat{i}}dx^{\widehat{i}}  \label{ddif4}
\end{equation}%
is the dual of $\left( \delta _{\widehat{i}},\partial _{4},\partial
_{5}\right) ,$ where
\begin{equation}
\delta _{\widehat{i}}=\partial _{\widehat{i}}+w_{\widehat{i}}\partial
_{4}+n_{\widehat{i}}\partial _{5}.  \label{dder4}
\end{equation}

\item If the conditions of the 4D variant of the Theorem \ref{t5dr} are
satisfied, we have the same equations (\ref{ep1a}) --(\ref{ep4a}) were we
substitute $h_{4}=h_{4}\left( x^{\widehat{k}},v\right) $ and $%
h_{5}=h_{5}\left( x^{\widehat{k}},v\right) .$ As a consequence we have $%
\alpha _{i}\left( x^{k},v\right) \rightarrow \alpha _{\widehat{i}}\left( x^{%
\widehat{k}},v\right) ,\beta =\beta \left( x^{\widehat{k}},v\right) $ and $%
\gamma =\gamma \left( x^{\widehat{k}},v\right) $ resulting in $w_{\widehat{i}%
}=w_{\widehat{i}}\left( x^{\widehat{k}},v\right) $ and $n_{\widehat{i}}=n_{%
\widehat{i}}\left( x^{\widehat{k}},v\right) .$

\item The 4D line element with conformal factor (\ref{dmetric4}) subjected
to an anhlonomic map with $\zeta _{5}=0$ transforms into
\begin{equation}
\delta s^{2}=\omega ^{2}(x^{\widehat{i}%
},v)[g_{2}(dx^{2})^{2}+g_{3}(dx^{3})^{2}+h_{4}(\hat{{\delta }}%
v)^{2}+h_{5}(\delta y^{5})^{2}],  \label{cdmetric4}
\end{equation}%
given with respect to the anholonomic co--frame $\left( dx^{\widehat{i}},%
\hat{{\delta }}v,\delta y^{5}\right) ,$ where
\begin{equation}
\delta v=dv+(w_{\widehat{i}}+\zeta _{\widehat{i}})dx^{\widehat{i}}%
\mbox{ and
}\delta y^{5}=dy^{5}+n_{\widehat{i}}dx^{\widehat{i}}  \label{ddif24}
\end{equation}%
is dual to the frame $\left( \hat{{\delta }}_{\widehat{i}},\partial _{4},%
\hat{{\partial }}_{5}\right) $ with
\begin{equation}
\hat{{\delta }}_{\widehat{i}}=\partial _{\widehat{i}}-(w_{\widehat{i}}+\zeta
_{\widehat{i}})\partial _{4}+n_{\widehat{i}}\partial _{5},\hat{{\partial }}%
_{5}=\partial _{5}.  \label{dder24}
\end{equation}

\item The formulas (\ref{conf1}) and (\ref{confeq}) from Theorem \ref{t5dra}
must be modified into a 4D form
\begin{equation}
\hat{{\delta }}_{\widehat{i}}h_{4}=0\mbox{\ and\  }\hat{{\delta }}_{\widehat{%
i}}\omega =0  \label{conf14}
\end{equation}%
and the values $\zeta _{\widetilde{{i}}}=\left( \zeta \widehat{_{{i}}},\zeta
_{{5}}=0\right) $ are found as to be a unique solution of (\ref{conf1}); for
instance, if
\begin{equation*}
\omega ^{q_{1}/q_{2}}=h_{4}~(q_{1}\mbox{ and }q_{2}\mbox{ are
integers}),
\end{equation*}%
$\zeta _{\widehat{{i}}}$ satisfy the equations \
\begin{equation}
\partial _{\widehat{i}}\omega -(w_{\widehat{i}}+\zeta _{\widehat{{i}}%
})\omega ^{\ast }=0.  \label{confeq4}
\end{equation}

\item One holds the same formulas (\ref{p2})-(\ref{n}) from the Theorem \ref%
{texs} on the general form of exact solutions with that difference that
their 4D analogs are to be obtained by reductions of holonomic indices, $%
\widehat{i}\rightarrow i,$ and holonomic coordinates, $x^{i}\rightarrow x^{%
\widehat{i}},$ i. e. in the 4D solutions there is not contained the variable
$x^{1}.$

\item The formulae (\ref{einstdiag}) for the nontrivial coefficients of the
Einstein tensor in 4D stated by the Corollary \ref{ceint} are \ written
\begin{equation}
G_{2}^{2}=G_{3}^{3}=-S_{4}^{4},G_{4}^{4}=G_{5}^{5}=-R_{2}^{2}.
\label{einstdiag4}
\end{equation}

\item For symmetries of the Einstein tensor (\ref{einstdiag4}), \ we can
introduce a matter field source with a diagonal energy momentum tensor,
like\ it is stated in the Corollary \ref{cors} by the conditions (\ref%
{emcond}), which in 4D are transformed into
\begin{equation}
\Upsilon _{2}^{2}=\Upsilon _{3}^{3}=\Upsilon _{2}(x^{2},x^{3},v),\ \Upsilon
_{4}^{4}=\Upsilon _{5}^{5}=\Upsilon _{4}(x^{2},x^{3}).  \label{emcond4}
\end{equation}
\end{itemize}

The 4D dimensional off--diagonal ansatz may model certain generalized
Lagrange configurations and Lagrange--affine solutions. They can also
include certain 3D Finsler or Lagrange metrics but with 2D frame transforms
of the corresponding quadratic forms and N--connections.

\section{Generalized Lagrange--Affine Spaces}

We outline and give a brief characterization of five classes of generalized
Finsler--affine spaces (contained in the Table 1 from Ref. \cite{vp1}; see
also in that work the details on classification of such geometries). We note
that the N--connection curvature is computed following the formula $\Omega
_{ij}^{a}=\delta _{\lbrack i}N_{j]}^{a},$ see (\ref{ncurv}), for any
N--connection $N_{i}^{a}.$ A d--connection $\mathbf{D}=[\mathbf{\Gamma }%
_{\beta \gamma }^{\alpha }]=[L_{\ jk}^{i},L_{\ bk}^{a},C_{\ jc}^{i},C_{\
bc}^{a}]$ defines nontrivial d--torsions $\mathbf{T}_{\ \beta \gamma
}^{\alpha }=[L_{[\ jk]}^{i},C_{\ ja}^{i},\Omega _{ij}^{a},T_{\ bj}^{a},C_{\
[bc]}^{a}]$ and d--curvatures $\mathbf{R}_{\ \beta \gamma \tau }^{\alpha
}=[R_{\ jkl}^{i},R_{\ bkl}^{a},P_{\ jka}^{i},P_{\ bka}^{c},S_{\
jbc}^{i},S_{\ dbc}^{a}]$ adapted to the N--connection structure (see,
respectively, the formulas (\ref{dtorsb})\ and (\ref{dcurv})). Any generic
off--diagonal metric $g_{\alpha \beta }$ is associated to a N--connection
structure and reprezented as a d--metric $\mathbf{g}_{\alpha \beta
}=[g_{ij},h_{ab}]$ (see formula (\ref{block2})). The components of a
N--connection and a d--metric define the canonical d--connection $\mathbf{D}%
=[\widehat{\mathbf{\Gamma }}_{\beta \gamma }^{\alpha }]=[\widehat{L}_{\
jk}^{i},\widehat{L}_{\ bk}^{a},\widehat{C}_{\ jc}^{i},\widehat{C}_{\
bc}^{a}] $ (see (\ref{candcon})) with the corresponding values of
d--torsions $\widehat{\mathbf{T}}_{\ \beta \gamma }^{\alpha }$ and
d--curvatures $\widehat{\mathbf{R}}_{\ \beta \gamma \tau }^{\alpha }.$ The
nonmetricity d--fields are computed by using formula $\mathbf{Q}_{\alpha
\beta \gamma }=-\mathbf{D}_{\alpha }\mathbf{g}_{\beta \gamma
}=[Q_{ijk},Q_{iab},Q_{ajk},Q_{abc}],$ see (\ref{nmf}).

The Table \ref{tablegs} outlines five classes of geometries modeled in the
framework of metric--affine geometry as spaces with nontrivial N--connection
structure (for simplicity, we omitted the Berwald configurations, see Ref. %
\cite{vp1}).

\begin{enumerate}
\item Metric--affine spaces (in brief, MA) are those stated as certain
manifolds $V^{n+m}$ of necessary smoothly class\ provided with arbitrary
metric, $g_{\alpha \beta },$ and linear connection, $\Gamma _{\beta \gamma
}^{\alpha },$ structures. For generic off--diagonal metrics, a MA\ space
always admits nontrivial N--connection structures. Nevertheless, in general,
only the metric field $g_{\alpha \beta }$ can be transformed into a
d--metric one $\mathbf{g}_{\alpha \beta }=[g_{ij},h_{ab}],$ but \ $\Gamma
_{\beta \gamma }^{\alpha }$ can be not adapted to the N--connection
structure. As a consequence, the general strength fields $\left( T_{\ \beta
\gamma }^{\alpha },R_{\ \beta \gamma \tau }^{\alpha },Q_{\alpha \beta \gamma
}\right) $ can be also not N--adapted.

\item Distinguished metric--affine spaces (DMA) are defined as manifolds $%
\mathbf{V}^{n+m}$ $\ $provided with N--connection structure $N_{i}^{a},$
d--metric field (\ref{block2}) and arbitrary d--connection $\mathbf{\Gamma }%
_{\beta \gamma }^{\alpha }.$ In this case, all strenghs $\left( \mathbf{T}%
_{\ \beta \gamma }^{\alpha },\mathbf{R}_{\ \beta \gamma \tau }^{\alpha },%
\mathbf{Q}_{\alpha \beta \gamma }\right) $ are N--adapted.

\item Generalized Lagrange--affine spaces (GLA), $\mathbf{GLa}%
^{n}=(V^{n},g_{ij}(x,y),$ $\ ^{[a]}\mathbf{\Gamma }_{\ \beta }^{\alpha })$ $%
, $ are modeled as distinguished metric--affine spaces of odd--dimension, $%
\mathbf{V}^{n+n},$ provided with generic off--diagonal metrics with
associated N--connection inducing a tangent bundle structure. The d--metric $%
\mathbf{g}_{[a]}$ and the d--connection $\ \ ^{[a]}\mathbf{\Gamma }_{\
\alpha \beta }^{\gamma }$ $=\left( \ ^{[a]}L_{jk}^{i},\
^{[a]}C_{jc}^{i}\right) $ are similar to those for the usual Lagrange spaces
but with distorsions $\ ^{[a]}\ \mathbf{Z}_{\ \ \beta }^{\alpha }$ inducing
general nontrivial nonmetricity d--fields $^{[a]}\mathbf{Q}_{\alpha \beta
\gamma }.$

\item Lagrange--affine spaces (LA), $\mathbf{La}%
^{n}=(V^{n},g_{ij}^{[L]}(x,y),$ $\ ^{[b]}\mathbf{\Gamma }_{\ \beta }^{\alpha
}),$ are provided with a Lagrange quadratic form $g_{ij}^{[L]}(x,y)=\frac{1}{%
2}\frac{\partial ^{2}L^{2}}{\partial y^{i}\partial y^{j}}$ inducing the
canonical N--connection structure $^{[cL]}\mathbf{N}=\{\ ^{[cL]}N_{j}^{i}\}$
for a Lagrange space $\mathbf{L}^{n}=\left( V^{n},g_{ij}(x,y)\right) $ but
with a d--connection structure $\ ^{[b]}\mathbf{\Gamma }_{\ \ \alpha
}^{\gamma }=\ ^{[b]}\mathbf{\Gamma }_{\ \alpha \beta }^{\gamma }\mathbf{%
\vartheta }^{\beta }$ distorted by arbitrary torsion, $\mathbf{T}_{\beta },$
and nonmetricity d--fields,$\ \mathbf{Q}_{\beta \gamma \alpha },$ when $%
^{[b]}\mathbf{\Gamma }_{\ \beta }^{\alpha }=\ ^{[L]}\widehat{\mathbf{\Gamma }%
}_{\beta }^{\alpha }+\ \ ^{[b]}\ \mathbf{Z}_{\ \ \beta }^{\alpha }.$ This is
a particular case of GLA spaces with prescribed types of N--connection $%
^{[cL]}N_{j}^{i}$ and d--metric to be like in Lagrange geometry.

\item Finsler--affine spaces (FA), $\mathbf{Fa}^{n}=\left( V^{n},F\left(
x,y\right) ,\ ^{[f]}\mathbf{\Gamma }_{\ \beta }^{\alpha }\right) ,$ in their
turn are introduced by further restrictions of $\mathbf{La}^{n}$ to a
quadratic form $g_{ij}^{[F]}=\frac{1}{2}\frac{\partial ^{2}F^{2}}{\partial
y^{i}\partial y^{j}}$ constructed from a Finsler metric $F\left(
x^{i},y^{j}\right) .$ It is induced the canonical N--connection structure $\
^{[F]}\mathbf{N}=\{\ ^{[F]}N_{j}^{i}\}$ \ as in the Finsler space $\mathbf{F}%
^{n}=\left( V^{n},F\left( x,y\right) \right) $ but with a d--connection
structure $\ ^{[f]}\mathbf{\Gamma }_{\ \alpha \beta }^{\gamma }$ distorted
by arbitrary torsion, $\mathbf{T}_{\beta \gamma }^{\alpha },$ and
nonmetricity, $\mathbf{Q}_{\beta \gamma \tau },$ d--fields, $\ ^{[f]}\mathbf{%
\Gamma }_{\ \beta }^{\alpha }=\ ^{[F]}\widehat{\mathbf{\Gamma }}_{\ \beta
}^{\alpha }\ +\ \ ^{[f]}\ \mathbf{Z}_{\ \ \beta }^{\alpha },$where $\ ^{[F]}%
\widehat{\mathbf{\Gamma }}_{\ \beta \gamma }^{\alpha }$ is the canonical
Finsler d--connection.
\end{enumerate}

%{\footnotesize
\begin{table}[h]
\begin{center}
\begin{tabular}{|l|l|l|l|}
\hline\hline
Space &
\begin{tabular}{l}
\begin{tabular}{l}
N--connection/ \\
N--curvature%
\end{tabular}
\\
\begin{tabular}{l}
metric/ \\
d--metric%
\end{tabular}%
\end{tabular}
&
\begin{tabular}{l}
(d--)connection/ \\
(d-)torsion%
\end{tabular}
&
\begin{tabular}{l}
(d-)curvature/ \\
(d--)nonmetricity%
\end{tabular}
\\ \hline\hline
1. MA &
\begin{tabular}{l}
$N_{i}^{a},\Omega _{ij}^{a}$ \\
\begin{tabular}{l}
off.d.m. $g_{\alpha \beta },$ \\
$\mathbf{g}_{\alpha \beta }=[g_{ij},h_{ab}]$%
\end{tabular}%
\end{tabular}
&
\begin{tabular}{l}
$\Gamma _{\beta \gamma }^{\alpha }$ \\
$T_{\ \beta \gamma }^{\alpha }$%
\end{tabular}
&
\begin{tabular}{l}
$R_{\ \beta \gamma \tau }^{\alpha }$ \\
$Q_{\alpha \beta \gamma }$%
\end{tabular}
\\ \hline
2. DMA &
\begin{tabular}{l}
$N_{i}^{a},\Omega _{ij}^{a}$ \\
$\mathbf{g}_{\alpha \beta }=[g_{ij},h_{ab}]$%
\end{tabular}
&
\begin{tabular}{l}
$\mathbf{\Gamma }_{\beta \gamma }^{\alpha }$ \\
$\mathbf{T}_{\ \beta \gamma }^{\alpha }$%
\end{tabular}
&
\begin{tabular}{l}
$\mathbf{R}_{\ \beta \gamma \tau }^{\alpha }$ \\
$\mathbf{Q}_{\alpha \beta \gamma }$%
\end{tabular}
\\ \hline
3. GLA &
\begin{tabular}{l}
\begin{tabular}{l}
$\dim i=\dim a$ \\
$N_{i}^{a},\Omega _{ij}^{a}$%
\end{tabular}
\\
\begin{tabular}{l}
off.d.m. $g_{\alpha \beta },$ \\
$\mathbf{g}_{[a]}=[g_{ij},h_{kl}]$%
\end{tabular}%
\end{tabular}
&
\begin{tabular}{l}
$^{\lbrack a]}\mathbf{\Gamma }_{\ \alpha \beta }^{\gamma }$ \\
$^{\lbrack a]}\mathbf{T}_{\ \beta \gamma }^{\alpha }$%
\end{tabular}
&
\begin{tabular}{l}
$^{\lbrack a]}\mathbf{R}_{\ \beta \gamma \tau }^{\alpha }$ \\
$^{\lbrack a]}\mathbf{Q}_{\alpha \beta \gamma }$%
\end{tabular}
\\ \hline
4. LA &
\begin{tabular}{l}
\begin{tabular}{l}
$\dim i=\dim a$ \\
$^{\lbrack cL]}N_{j}^{i},\ ^{[cL]}\Omega _{ij}^{a}$%
\end{tabular}
\\
d--metr.$\mathbf{g}_{\alpha \beta }^{[L]}$%
\end{tabular}
&
\begin{tabular}{l}
$^{\lbrack b]}\mathbf{\Gamma }_{\ \alpha \beta }^{\gamma }$ \\
$^{\lbrack b]}\mathbf{T}_{\ \beta \gamma }^{\alpha }$%
\end{tabular}
&
\begin{tabular}{l}
$^{\lbrack b]}\mathbf{R}_{\ \beta \gamma \tau }^{\alpha }$ \\
$^{\lbrack b]}\mathbf{Q}_{\alpha \beta \gamma }=-~^{[b]}\mathbf{D}_{\alpha }%
\mathbf{g}_{\beta \gamma }^{[L]}$%
\end{tabular}
\\ \hline
5. FA &
\begin{tabular}{l}
\begin{tabular}{l}
$\dim i=\dim a$ \\
$^{\lbrack F]}N_{j}^{i};\ ^{[F]}\Omega _{ij}^{k}$%
\end{tabular}
\\
d--metr.$\mathbf{g}_{\alpha \beta }^{[F]}$%
\end{tabular}
&
\begin{tabular}{l}
$^{\lbrack f]}\mathbf{\Gamma }_{\ \alpha \beta }^{\gamma }$ \\
$^{\lbrack f]}\mathbf{T}_{\ \beta \gamma }^{\alpha }$%
\end{tabular}
&
\begin{tabular}{l}
$^{\lbrack f]}\mathbf{R}_{\ \beta \gamma \tau }^{\alpha }$ \\
$^{\lbrack f]}\mathbf{Q}_{\alpha \beta \gamma }=-~^{[f]}\mathbf{D}_{\alpha }%
\mathbf{g}_{\beta \gamma }^{[F]}$%
\end{tabular}
\\ \hline\hline
\end{tabular}%
\end{center}
\caption{Generalized Finsler/Lagrange--affine spaces}
\label{tablegs}
\end{table}
%}


\newpage

\begin{thebibliography}{99}
\bibitem{sgr} P. Deligne, P. Etingof, D. S. Freed et all (eds.), \textit{%
Quanum Fields and Strings: A Course for Mathematicians}, Vols 1 and 2,
Institute for Adavanced Study (American Mathematical Society, 1994); J.
Polchinski, \textit{String Theory,} Vols 1 and 2 (Cambrdge University Press,
1998).

\bibitem{ncg} A. Connes and J. Lott, Nucl. Phys. B \ (Proc. Suppl) \textbf{%
18, }29 (1990); A.\ Connes, \textit{Noncommutative Geometry} (Academic
Press, 1994); A. H. Chamseddine, G. Felder and J. Frohlich, Commun. Math.
Phys. \textbf{155,} 205 (1993); A. H. Chamseddine, J. Frohlich and O.
Grandjean, J. Math. Phys. \textbf{36,} 6255 (1995) ; A. H. Chamseddine,
Commun. Math. Phys. \textbf{218,} 283 (2001); E. Hawkins, Commun. Math.
Phys. \textbf{187,} 471 (1997); I. Vancea, Phys. Rev. Lett. \textbf{79,}
3121 (1997); S. Majid, Int. J. Mod. Phys.\textbf{\ B 14,} 2427 (2000)\ ; \
J. W. Moffat, Phys. Lett. B 491 (2000) 345; A. H. Chamsedine, Phys. Lett.
\textbf{B} \textbf{504} (2001) 33; J. Madore, \textit{An Introduction to
Noncommutative Geometry and its Physical Applications}, LMS lecture note
Series 257, 2nd ed. (Cambridge University Press, 1999); F. Ardalan, H.
Arfaei, M. R. Garousi and A. Ghodsi, Int. J. Mod. Phys. \textbf{A 18, } 1051
(2003); V. Sahakian, JHEP \textbf{0106,} 037 (2001); M. C. B. Abdalla, M. A.
De Anrade, M. A. Santos and I. V. Vancea, Phys. Lett. \textbf{B 548,} 88
(2002); H.\ Nishino and S. Rajpoot, Noncommutative Nonlinear Supersymmetry,
hep-th/0212329; H. Garcia--Compean, O. Obregon, C. Ramirez and M.\ Sabido,
Noncommutative Self--Dual Gravity, hep--th/03021180; M. A. Cardella and D.
Zanon, Class. Quant. Grav. \textbf{20} (2003) L95.

\bibitem{majid} S. Majid, \textit{Foundations of Guantum Group Theory, (}%
Cambridge University Press, Cambridge, 1995); \textit{A Quantum Group Primer,%
} L. M.\ S. Lect. Notes. Series. \textbf{292} (2002); in: Springer Lecture
Notes in Physics, \textbf{541}, 227 (2000); Phys. Trans. Roy. Soc. \textbf{A
358,} 89 (2000).

\bibitem{mag} F. M. Hehl, \ J. D. Mc Grea, E. W. Mielke, \ and Y. Ne'eman.\
Phys. Rep.\ \textbf{258}, 1 (1995); F. Gronwald and F. W. Hehl, in: \textit{%
Quantum Gravity}, Proc. 14th School of Cosmology and Gravitation, May 1995
in Erice,\ Italy. Eds.: P. G. Bergmann, V. de Sabbata and H. -J. Treder
(World Sci. Pubishing, River Edge NY, 1996), 148--198; T.\ Dereli and R.\ W.
Tucker, Class. Quantum Grav. \textbf{12, }L31 (1995); T. Dereli, M. Onder
and R. W. Tucker, Class. Quantum Grav. \textbf{12, }L251 (1995).

\bibitem{oveh} T. Dereli, M. Onder, J. Schray, R. W. Tucker and C. Wang,
Class. Quantum Grav. \textbf{13, }L103 (1996); Yu. N. Obukhov, E. J.
Vlachynsky, W. Esser and F. W. Hehl, Phys. Rev. D \textbf{56,} 7769 (1997).

\bibitem{vp1} S.\ Vacaru, Generalized Finsler Geometry in Einstein, String and
Metric--Affine Gravity, hep-th/0310132.

\bibitem{v1} S. Vacaru, JHEP \textbf{04,} 009 (2001);\ \ \ S. Vacaru and D.
Singleton, J. Math. Phys. \textbf{43,} 2486 (2002); \ \ \ S. Vacaru and D.
Singleton, Class. Quant. Gravity \textbf{19,} 3583 (2002); S. Vacaru and F.
C. Popa, Class. Quant. Gravity \textbf{18,} 4921 (2001); \ Vacaru and D.
Singleton, Class. Quant. Gravity \textbf{19,} 2793 (2002).

\bibitem{v1a} S. Vacaru, D. Singleton, V. Botan and D. Dotenco, Phys. Lett.
\textbf{B} \textbf{519,} 249 (2001); S. Vacaru and O. Tintareanu-Mircea,
Nucl. Phys. \textbf{B} \textbf{626,} 239 (2002); S.\ Vacaru, Int. J. Mod.
Phys. D \textbf{12,} 461 (2003); S.\ Vacaru, Int. J. Mod. Phys. D \textbf{12,%
} 479 (2003); S. Vacaru and H. Dehnen, Gen. Rel. Grav. \textbf{35,} 209
(2003); H. Dehnen and S.\ Vacaru, S. Gen. Rel. Grav. \textbf{35,} 807
(2003); Vacaru and Yu. Goncharenko, \ Int. J. Theor. Phys. \textbf{34,} 1955
(1995).

\bibitem{v2} S.\ Vacaru, Ann. Phys. (NY) 256, 39 (1997); S. Vacaru,\ Nucl.
Phys. B\ \textbf{424 } 590 (1997); S. Vacaru, \ J. Math. Phys. \ \textbf{37,}
508 (1996); S. Vacaru, \ JHEP\ \textbf{09,}\ 011 (1998); S. Vacaru and P.
Stavrinos, \textit{Spinors and Space-Time Anisotropy} (Athens University
Press, Athens, Greece, 2002), 301 pages, gr-qc/0112028; \ S. Vacaru and
Nadejda Vicol, Nonlinear Connections and Clifford Structures,
math.DG/0205190; S. Vacaru, Phys. Lett. B \textbf{498,} 74 (2001) ; S.
Vacaru, Noncommutative Finsler Geometry, Gauge Fields and Gravity,
math--ph/0205023. S. Vacaru, (Non) Commutative Finsler Geometry from String/
M--Theory, hep--th/0211068.\

\bibitem{vnces} S. Vacaru, Exact Solutions with Noncommutative Symmetries in
Einstein and Gauge Gravity, gr--qc/0307103.

\bibitem{fin} P. Finsler,\ $\ddot{U}$\textit{ber Kur\-ven und Fl}$\ddot{a}$%
\textit{chen in All\-gemeiner R}$\ddot{a}$\textit{men,} Disser\-ta\-ti\-on (G%
$\ddot{o}$ttingen, 1918); reprinted\ \ (Basel: Birkh$\ddot{a}$user, 1951);
E. Cartan, \textit{\ Les Espaces de Finsler}\ (Paris: Hermann, 1935); H.
Rund, \ \textit{The Differential Geometry of Finsler Spaces} (Berlin:
Springer--Verlag, 1959); G. Yu. Bogoslovsky, \ Nuov. Cim. \textbf{B40,}\ 99;
116, (1977) ; G., Yu. Bogoslovsky, \ \textit{Theory of Locally Anisotropic
Space--Time }(Moscow, Moscow State University Publ., 1992) [in Russian]; H.,
F. Goenner, and G. Yu. Bogoslovsky, Ann. Phys. (Leipzig) \textbf{9} Spec.
Issue,\ 54; G. S. Asanov, \textit{Finsler Geometry, Relativity and Gauge
Theories} (Boston: Reidel, 1985); A. K. Aringazin and G. S. Asanov, Rep.
Math. Phys. \textbf{25,} 35 (1988); G. S. Asanov, Rep. Math. Phys. \textbf{%
28,} 263 (1989); G. S. Asanov, Rep. Math. Phys. \textbf{42,} 273 (1989); G.
S. Asanov and S. F. Ponomarenko, \textit{Finslerovo Rassloenie nad
Prostranstvom--Vremenem, assotsiiruemye kalibrovochnye polya i sveaz\-nos\-ti%
} \textit{Finsler Bun\-dle on Space--Time. Associated Gauge Fields and
Con\-nec\-tions} (\c{S}tiin\c{t}a, Chi\c{s}in\u{a}u, 1988) [in Russian,
''Shtiintsa, Kishinev, 1988]; G. S. Asanov, Rep. Math. Phys. \textbf{26,}
367 (1988); G. S. Asanov, \textit{Fibered Generalization of the Gauge Field
Theory. Finslerian and Jet Gauge Fields} (Moscow University, 1989) [in
Russian]; \ M. Matsumoto, \ \textit{Foundations of Finsler Geometry and
Special Finsler Spaces }(Kaisisha: Shigaken, 1986); A. Bejancu, \textit{%
Finsler Geometry and Applications } (Ellis Horwood, Chichester, England,
1990; S. Vacaru, \ Nucl. Phys. B\ \textbf{424, } 590 (1997); S. Vacaru,
Interactions, Strings and Isotopies in Higher Order Anisotropic Superspaces
(Hadronic Press, FL, USA, 1998), math--ph/0112056; D. Bao, S.-S. Chern and
Z. Shen, \textit{An Introduction to Riemann-Finsler Geometry.} Graduate
Texts in Mathematics, 200 (Springer-Verlag, New York, 2000).

\bibitem{ma} R. Miron and M. Anastasiei, \textit{The Geometry of Lagrange
Spaces: Theory and Applications} (Kluwer Academic Publishers, Dordrecht,
Boston, London, 1994); R. Miron and M. Anastasiei, \textit{Vector Bundles.
Lagrange Spaces. Application in Relativity} (Academiei, Romania, 1987) [in
Romanian]; [English translation] no. 1 (Geometry Balkan Press, Bucharest,
1997).

\bibitem{mhss} R. Miron, D.\ Hrimiuc, H. Shimada and V. S. Sabau, \textit{%
The Geometry of Hamilton and Lagrange Spaces} (Kluwer Academic Publishers,
Dordrecht, Boston, London, 2000); R. Miron, \textit{The Geometry of
Higher--Order Lagrange Spaces, Application to Mechanics and Physics}, FTPH
no. 82 (Kluwer Academic Publishers, Dordrecht, Boston, London, 1997); R.
Miron, \textit{The Geometry of Higher--Order Finsler Spaces} (Hadronic
Press, Palm Harbor, USA, 1998); R. Miron and Gh. Atanasiu, in: {\ Lagange
geometry, Finsler spaces and noise applied in biology and physics}, \textsl{%
Math. Comput. Modelling.} \textbf{20} (Pergamon Press, 1994), pp. 41--56l;
R. Miron and Gh. Atanasiu, \textit{Compendium sur les Espaces La\-gran\-ge
D'ordre Sup\'{e}rieur,}\ \emph{Seminarul de Mecanic\v{a}. Universitatea din
Ti\-mi\-\c{s}oa\-ra.\ Facultatea de Matematic\v{a}, 1994} \textbf{40} p. 1;\
\emph{Revue Roumaine de Ma\-the\-ma\-tiques Pures et Appliquee}\ \textbf{XLI}%
, $N^{of}$ 3--4 (Timisoara, 1996) pp. 205; 237; 251.

\bibitem{exsolmag} S.\ Vacaru and E. Gaburov, Noncommutative Symmetries and Stability of
Black Ellipsoids in String and Metric--Affine Gravity,
gr-qc/0310134.

\bibitem{vmethod} S.\ Vacaru, \textit{A New Method of Constructing Black
Hole Solutions in Einstein and 5D Dimension Gravity,} hep-th/0110250.

\bibitem{esolmag} R. Tresguerres, Z. Phys. \textbf{C65,} 347 (1995); R.
Tresguerres, Phys. Lett. \textbf{A200,} 405 (1995); R. W. Tucker and C.
Wang, Class. Quant. Grav. \textbf{12,} 2587 (1995); Yu. N. Obukhov, E. J.
Vlachynsky, W. Esser, R. Tresguerres, and F. W. Hehl, Phys. Let. \textbf{%
A220,} 1 (1996); E.\ J. Vlachynsky, R.\ Tresguerres, Yu. N. Obukhov and F.
W. Hehl, Class. Quantum Grav. \textbf{13,} 3253 (1996); R. A. Puntingam, C.
Lammerzahl, F. W. Hehl, Class. Quantum Grav. \textbf{14,} 1347 (1997); A.
Macias, E. W. Mielke and J. Socorro, Class. Quantum Grav. \textbf{15,} 445
(1998); J. Socorro, C. Lammerzahl, A. Macias, and E. W. Mielke, Phys. Lett.
\textbf{A244,} 317 (1998); A. Garcia, F. W. Hehl, C. Lammerzahl, A. Macias
and J. Socorro, Class. Quantum Grav. \textbf{15,} 1793 (1998); A. Macias and
J.\ Socorro, Class. Quantum Grav. \textbf{16,} 1999 (1998); A. Garcia, C.
Lammerzahl, A.\ Macias, E. W. Mielke and J.\ Socorro, Phys. Rev. D \textbf{%
57,} 3457 (1998); A. Garcia, F. W. Hehl, C. Lammerzahl, A. Macias and J.
Socorro, Class. Quantum Grav. \textbf{15}, 1793 (1998); A.\ Garcia, A.
Macias and J.\ Socorro, Class. Quantum Grav. \textbf{16}, 93 (1999); A.
Macias, C. Lammerzahl, and A.\ Garcia, J. Math. Phys. 41, 6369 (2000); E.
Ayon-Beato, A.\ Garcia, A. Macias and H. Quevedo, Phys. Rev. D \textbf{64,}
024026 (2001).

\bibitem{pen} R. Penrose, \textit{Structure of Space-time,} in: \emph{%
Battelle Rencontres, Lectures in Mathematics and Physics,} eds. C. M. DeWitt
and J. A. Wheeler (Benjamin, New York, 1967); R. Penrose and W. Rindler,
\textit{Spinors and Space-Time, vol. 1, Two-Spinor Calculus and Relativistic
Fields} (Cambridge University Press, Cambridge, 1984); R. Penrose and W.
Rindler, \textit{Spinors and Space-Time, vol. 2, Spinor and Twistor Methods
in Space-Time Geometry} (Cambridge University Press, Cambridge, 1986).

\bibitem{barthel} W. Barthel,\ J. Angew. Math.\ \textbf{212,}\ 120 (1963).

\bibitem{rcg} F. W. Hehl, P. von der Heyde and G. D. Kerlick, Phys. Rev.
\textbf{10, }1066 (1974); F. W. Hehl, P. von der Heyde and G. D. Kerlick,
Rev. Mod. Phys. \textbf{48,} 393 (1976).

\bibitem{obet2} R. Tucker and C. Wang, \textit{Non--Riemannian Gravitational
Interactions,} gr--qc/9608055; A. Macias, E. W. Mielke, and J. Socorro,
Class. Quant. Grav. \textbf{15,} 445 (1998).

\bibitem{hm} F. W. Hehl and A. Macias, Int. J. Mod. Phys. D \textbf{8}, 339
(1999).

\bibitem{ggrav} R. Utiama, Phys. Rev. \textbf{101,} 1597 (1956); T. W.
Kibble, J. Math. Phys. \textbf{2,} 212 (1961); D.\ E. Sciama, in \textit{%
Recent Developments in General Relativity} (Pergamon, Oxford, England,
1962); S.\ W. MacDowell and F. Mansourri, Phys. Rev. Lett. \textit{38,} 739
(1977); A. A. Tseytlin, Phys. Rev. \textbf{D 26 }(1982) 3327; \ H. Dehnen
and F. Ghaboussi, Phys. Rev. \textbf{D 33,} 2205 (1986); V. N. Ponomarev, A.
Barvinsky and Yu. N. Obukhov, \textit{Geometrodynamical Methods and the
Gauge Approach to the Gravitational Interactions} (Energoatomizdat, Moscow,
1985) [in Russian]; E. W. Mielke, \textit{Geometrodynamics of Gauge Fields
--- On the Geometry of Yang--Mills and Gravitational Gauge Theories}
(Akademie--Verlag, Berlin, 1987); L. O' Raifeartaigh and N. Straumann, Rev.
Mod. Phys. \textbf{72,} 1 (2000).

\bibitem{cartanmacros} H. H. Soleng, Cartan 1.01 for Unix, gr--qc-9502035.

\bibitem{kir} E.\ Kiritsis, \textit{\ Introduction to Superstring Theory},
Leuven Notes in Mathematical and Theoretical Physics. Series B: Theoretical
Particle Physics, 9. (Leuven University Press, Leuven, 1998); J. Scherk and
J. Schwarz, \ Nucl. Phys. \textbf{B153,} 61 (1979); J. Maharana and J.\
Schwarz, Nucl. Phys. \textbf{B390,} 3 (1993) .

\bibitem{kdp} B. B. Kadomtsev and V. I. Petviashvili, Dokl. Akad. Nauk SSS,
\textbf{192,} 753 (1970).

\bibitem{soliton} V. A. Belinski and V. E. Zakharov, Sov. Phys.--JETP 48,
985 (1978) [translated from Zh. Exsp. Teor. Piz. 75, 1953 (1978), in
Russian]; V. Belinski and E. Verdaguer, Graviational Solitons (Cambridge:
Cambridge University Press, 2001); R. Rajaraman, Solitons and Instantons
(Amsterdam: North--Holland, 1989).

\bibitem{will} C.\ M. Will, \textit{Theory and Experiments in Gravitational
Physics} (Cambridge Universtity Press, 1993).
\end{thebibliography}
\end{document}